\documentclass[10pt, journal, twocolumn]{IEEEtran}

\usepackage[utf8]{inputenc}		
\usepackage[scr=boondoxo,frak=euler,bb=ams]{mathalfa}
\usepackage{amsmath,amsthm,amsfonts,amssymb} 	
\usepackage{graphicx,graphics}
\usepackage{float}
\usepackage{tikz}
\usetikzlibrary{shapes, snakes, patterns}
\usepackage{xcolor}		

\usepackage[export]{adjustbox}
\usepackage[justification=centering]{caption}
\usepackage{subcaption}
\DeclareCaptionLabelSeparator{periodspace}{.\quad}
\usepackage{epstopdf}
\usepackage{enumerate,enumitem} 
\usepackage{cite}
\usepackage{suffix}
\usepackage{mathtools}
\usepackage{tabularx}
\usepackage{booktabs}		
\usepackage{boldline,multirow,diagbox,hhline} 
\usepackage{authblk}
\usepackage{slashbox}
\usepackage{algorithm,algpseudocode}
\usepackage{relsize}
\usepackage{cuted}
\usepackage{titlesec}
\usepackage{chngcntr}
\theoremstyle{plain}
\newtheorem{theorem}{Theorem}[]
\newtheorem{lemma}[theorem]{Lemma}  
\newtheorem{proposition}[theorem]{Proposition}

\newtheorem{model}[]{Model}

\newtheorem{definition}[]{Definition}

\newtheorem{remark}[]{Remark}

\newcommand{\PreserveBackslash}[1]{\let\temp=\\#1\let\\=\temp}
\newcolumntype{C}[1]{>{\PreserveBackslash\centering}p{#1}} 
\newcolumntype{R}[1]{>{\PreserveBackslash\raggedleft}p{#1}} 
\newcolumntype{L}[1]{>{\PreserveBackslash\raggedright}p{#1}} 

\newcolumntype{Y}{>{\centering\arraybackslash}b{0.7cm}}
\newcolumntype{M}{>{\centering\arraybackslash}b{2.1cm}}
\newcolumntype{Z}{>{\centering\arraybackslash}b{1.75cm}}
\newcolumntype{G}{>{\centering\arraybackslash}m{1.5cm}}
\newcolumntype{Q}{>{\centering\arraybackslash}b{3.75cm}}
\newcolumntype{A}{>{\centering\arraybackslash}m{1.5cm}}
\newcolumntype{B}{>{\centering\arraybackslash}m{0.6667cm}}
\newcolumntype{T}{>{\centering\arraybackslash}b{6.5ex}}
\newcolumntype{S}{>{\centering\arraybackslash}b{13.25ex}}
\newcolumntype{?}{!{\vrule width 1pt}}
\newcolumntype{+}{!{\vrule width 2pt}}

\DeclareFontFamily{U}{mathx}{}
\DeclareFontShape{U}{mathx}{m}{n}{<-> mathx10}{}
\DeclareSymbolFont{mathx}{U}{mathx}{m}{n}
\DeclareMathAccent{\widecheck}{0}{mathx}{"71}

\definecolor{bluee}{RGB}{0, 82, 200}
\definecolor{redd}{RGB}{245, 30, 30}
\definecolor{greenn}{RGB}{0, 150, 62}

\newcommand{\norm}[1]{\left\lVert#1\right\rVert}
\newcommand{\abs}[1]{\left\lvert#1\right\rvert}

\newcommand{\psp}{\hspace{0.1em}}
\newcommand{\pspp}{\hspace{0.05em}}
\newcommand{\nsp}{\hspace{-0.1em}}
\newcommand{\nspp}{\hspace{-0.05em}}
\newcommand{\sgn}{\operatorname{sgn}}
\newcommand{\var}{\operatorname{Var}}

\newcommand{\erfi}{\operatorname{erfi}}

\newcommand{\summ}{\textstyle\sum\limits}

\newcommand{\comp}{^\textup{\texttt{c}}}
\newcommand{\idist}{\stackrel{\textup{\texttt{d}}}{=}}
\newcommand{\uso}{\leq_{\textup{\texttt{st}}}^{}}

\newcommand{\neighb}{{\rotatebox[origin=c]{90}{\footnotesize{$\rangle\nsp\langle\!$}}}_{\nsp_\mathcal{X}}\nspp}
\newcommand{\neighbb}{{\rotatebox[origin=c]{90}{\scriptsize{$\rangle\nsp\langle\!$}}}_{\nsp_\mathcal{X}}\nspp}
\newcommand{\neighbdsets}{\mathcal{D}\psp  \neighb \widecheck{\mathcal{D}}}
\newcommand{\neighbdsetss}{\mathcal{D}\psp  \neighbb \widecheck{\mathcal{D}}}
\newcommand{\supoverneighb}{\begin{array}{c} \sup\\[-1.25em] {}_{\neighbdsetss} \end{array}\!\nsp}
\newcommand{\plossrv}{\mathfrak{L}^{\mathcal{D}\nsp,\widecheck{\mathcal{D}}}_{_\mathcal{M}}}
\newcommand{\plossrvv}{\mathfrak{L}^{\widecheck{\mathcal{D}}\nsp,\mathcal{D}}_{_\mathcal{M}}}
\newcommand{\plossup}{{\zeta}_{\pspp \mathbf{d}}^{_{(\nspp u\nspp)}}\nsp} 
\newcommand{\plossupp}{{\zeta}_{\pspp {d}}^{_{(\nspp u\nspp)}}\nsp} 

\newcounter{asubsection}[section]
\titleclass{\asubsection}{straight}[\part]
\titleformat{\asubsection}[hang]{\normalfont\itshape}{\theasubsection.}{0.5em}{}
\titlespacing{\asubsection}{0pt}{2.45ex plus 0.5ex minus .2ex}{0.75ex plus .2ex}
\renewcommand{\theasubsection}{\textup{\Alph{section}-\Roman{asubsection}}}
\makeatletter
\@addtoreset{subsubsection}{asubsection}
\makeatother 

\title{Grafting Laplace and Gaussian distributions: A new noise mechanism for differential privacy}

\author{
	\IEEEauthorblockN{
		Gokularam Muthukrishnan\textsuperscript{*}, Sheetal Kalyani\textsuperscript{*}
		\thanks{\noindent\textsuperscript{*}The authors are with the Department of Electrical Engineering, Indian Institute	of Technology Madras, Chennai 600036, India (e-mail: \texttt{ee17d400@smail.iitm.ac.in}; \texttt{skalyani@ee.iitm.ac.in})}}
}

\begin{document}
	
	\maketitle	
	
	\begin{abstract} 
		
		The framework of differential privacy protects an individual's privacy while publishing query responses on congregated data. 
		In this work, a new noise addition mechanism for differential privacy 
		is introduced where the noise added is sampled from a hybrid density that resembles 
		Laplace in the centre and Gaussian in the tail.
		With a sharper centre and light, sub-Gaussian tail, this density has 
		the best characteristics of both distributions. 
		We theoretically analyze the proposed mechanism, and
		we derive the necessary and sufficient condition in one dimension and a sufficient condition in high dimensions for the mechanism to guarantee $(\epsilon,\delta)$-differential privacy. Numerical simulations corroborate the efficacy of the proposed mechanism compared to other existing mechanisms in achieving 
		a better trade-off between 
		privacy and accuracy. 
	\end{abstract}
	
	
	\section{Introduction}\label{sec:intro}

	\IEEEPARstart{D}{ifferential} privacy (DP)  is a mathematical framework to obscure an individual's presence in a sensitive dataset while publishing responses to queries on that dataset \cite{dwork2014algorithmic}. 
	It preserves the privacy of an individual while still allowing inferences and determining patterns from the entire data. 
	The privacy constraints are captured in the parameters $\epsilon$ and $\delta\pspp$, which are colloquially termed the privacy budget and privacy leakage, respectively.

	Mechanisms that warrant DP are essentially randomized, and 
	in machine learning models, randomization 
	is typically achieved by perturbing either the output \cite{dwork2006calibrating}, input \cite{kang2020input}, 
	objective \cite{zhang2012functional,chaudhuri2011differentially} 
	or  gradient updates 
	\cite{song2013stochastic,bassily2014private,abadi2016deep} 
	by adding noise sampled from a distribution. 
	Some of the well-known DP mechanisms are Laplacian \cite{dwork2006calibrating}, Gaussian \cite{dwork2006our,balle2018improving} and exponential \cite{mcsherry2007mechanism} mechanisms. 
	We defer the formal definitions of DP and its mechanisms to Section \ref{sec:definitions}. 
	%
	%
	Differential privacy has wide-ranging applications; it is 
	used to provide privacy guarantees on learning problems such as empirical risk minimization \cite{chaudhuri2011differentially}, clustering \cite{shechner2020private}, 
	data mining \cite{mohammed2011differentially}, low-rank matrix completion \cite{chien2021private} and many more. It has also been deployed in the 2020 US census \cite{us2021disclosure}. 
	
	DP involves a three-way trade-off between privacy, accuracy and the number of data records in the dataset. Of these, the first two are of interest as they are directly controllable. For the additive noise mechanism, accuracy translates to the variance of added noise and attaining a good privacy-accuracy trade-off 
	will be attributed to the proper choice of noise distribution; i.e., with the appropriate selection of noise density, accuracy can be improved for a given privacy constraint and vice-versa. 
	
	In this article, we propose a new mechanism for differential privacy called the flipped Huber mechanism, which adds flipped Huber noise. 
	Formed as a hybrid of Laplace and Gaussian densities, this new noise density offers a better privacy-accuracy trade-off than existing noise mechanisms in a wide range of scenarios. 
	
	
	\subsection{Prior works}
	\label{sec:rel_works}
	Ever since the advent of the differential privacy framework, several noise mechanisms have been proposed,  
	and their privacy guarantees have been well-documented.  
	Among those, 
	Laplace and Gaussian mechanisms are the popular ones; 
	the Laplace mechanism \cite{dwork2006calibrating} is capable of offering pure DP (with $\delta=0$), while the Gaussian mechanism is not. The Gaussian mechanism \cite{dwork2006our,dwork2014algorithmic,balle2018improving}  
	is the most extensively studied noise mechanism among all. 
	%
	Subbotin or Generalized-Gaussian distribution 
	has Gaussian and Laplace as special cases. It involves a power parameter, $p\pspp$, which determines how the tail decays. Recently, \cite{liu2018generalized} and \cite{vinterbo2022differential} have suggested the use of Subbotin noise for differential privacy.
	
	In \cite{canonne2020discrete}, the authors proposed to add noise from discretized Gaussian distribution to avoid loss of privacy due to finite precision errors.
	Almost every existing noise mechanism 
	adds noise 
	sampled from a log-concave distribution, i.e., a distribution whose density function is log-concave \cite{boyd2004convex}; the recent work \cite{vinterbo2022differential} provides the necessary and sufficient conditions for $(\epsilon,\delta)$-DP for scalar queries
	%
	when the noise density is log-concave, 
	%
	Recently proposed Offset Symmetric Gaussian Tails (OSGT) mechanism \cite{sadeghi2022offset} adds noise sampled from a novel distribution whose density function 
	is proportional to the product of Gaussian and Laplace densities. 
	
	Several works \cite{geng2016optimalstaircase,geng2015staircase,
		geng2016optimaluniflap,geng2019optimal0del,geng2020tighttrunclapl} have studied optimal noise mechanisms under various  regimes and constraints. In one dimension, the staircase mechanism is the optimal $\epsilon$-DP additive noise mechanism for real-valued and discrete queries, and Laplace is the optimal $\epsilon$-DP mechanism for small $\epsilon$ \cite{geng2016optimalstaircase,geng2015staircase}. However, guaranteeing $\epsilon$-DP in higher dimensions results in huge perturbations and lesser utility \cite{mironov2017renyi}. For one-dimensional discrete queries,  uniform and discrete Laplace are nearly optimal approximate DP mechanisms in the high-privacy regime \cite{geng2016optimaluniflap}. The truncated Laplace mechanism \cite{geng2020tighttrunclapl} is the optimal $(\epsilon,\delta)$ mechanism for a single real-valued query in the high-privacy regime, and it provides better utility than the Gaussian mechanism.
	
	However, to our best knowledge, the optimal noise distribution for $(\epsilon,\delta)$-DP has not been studied for arbitrary dimensions. In this work, we propose a new distribution for additive noise
	that offers 
	a better privacy-utility trade-off for multi-dimensional queries. We motivate the design of the noise distribution next.
	
	
	\subsection{Motivation}
	\label{sec:motiv}
	Though the Laplace mechanism offers a stricter notion of privacy than the Gaussian, it typically tends to add a larger amount of noise when employed on high dimensional queries, while the Gaussian mechanism does not \cite{mironov2017renyi,steinke2022composition}. This could be attributed to the light tail of the Gaussian density, which reduces the probability of outliers \cite{canonne2020discrete,sadeghi2022offset}. However, in very low dimensions, Laplace outperforms Gaussian by a large margin.  
	We believe that it is important to explore if there exist other noise mechanisms and to analyze how they fare against well-established ones. A noise mechanism combining the advantages of both of these 
	would significantly contribute to the DP literature.
	%
	{In the case of 
		DP, unlike the applications 
		where the noise is added by the environment, the designer is free to choose the noise to be added; therefore},  restricting only to well-known density limits the scope for improvement.

	We attribute the better performance of Laplace in low dimensions to its \textit{centre}, which is `sharp'. In estimation literature, the densities which are sharper tend to result in measurements that are more informative of the location parameter. This is because the Fisher information, which measures the curvature of the density at the location parameter, is inversely related to the variance of the estimate, and hence noise densities that are sharper give more accuracy \cite{hogg2019introduction}. As mentioned before, Gaussian gives better accuracy in higher dimensions because of the lighter tail. Motivated by these, we design a hybrid noise density, which we name the flipped Huber\footnote{We name it 
		so because the negative log density, with absolute valued centre and quadratic tails, 
		looks like the reverse of the Huber loss\cite{huber2009robust}. 
	} density, by splicing together the Laplace centre and Gaussian tail.

	\subsection{Contributions}
	Our major contributions are as follows. 
	We introduce a new noise distribution 
	for the additive noise DP mechanism. 
	This distribution is sub-Gaussian 
	and has larger Fisher information compared to Gaussian. 
	We then propose the flipped Huber mechanism that adds noise sampled from this new distribution. 
	Privacy guarantees for the proposed mechanism are theoretically derived. Necessary and sufficient in one dimension and sufficient condition in $K$ dimensions are obtained. 
	We also show 
	that the proposed 
	mechanism outperforms the existing noise mechanisms by requiring lesser noise variance for the given privacy constraints for a wide range of scenarios. 
	We further demonstrate that the flipped Huber noise offers better performance than Gaussian in private empirical risk minimization through coordinate descent for learning logistic and linear regression models on several real-world datasets.

\subsection{Basic notations} 

Let ${a }\psp \wedge\psp {b } = \min\pspp({a },{b })$ and ${a }\psp \vee\psp {b }=\max\pspp({a },{b })\pspp$; these operators have the
least precedence over other arithmetic operators. The unary operator $[\,\cdot\,]_{\nspp +}$ provides the positive part of its argument, i.e., $[{a }]_{\nspp +}={a }\vee 0\psp$. 
$\sgn(\cdot)$ denotes the signum (or sign) function 
and
$\log(\cdot)$ indicates the natural logarithm. 
We use $\mathbb{R}_{++}^{}$ to denote the set of positive real numbers $(0,\infty)\pspp$, and with the inclusion of 0, we write $\mathbb{R}_{+}^{}\pspp$. 
The vectors are denoted with bold-face 
letters, 
and the $\ell_p^{}$-norm of a vector is denoted as $\norm{\cdot}_p^{}\pspp$.

$\mathbb{P}\{\cdot\}$ denotes the probability measure, and $\mathbb{E}[\pspp\cdot\pspp]$ is the expectation. Let ${g}_{\pspp T}^{}(\cdot)$ and ${G}_{\pspp T}^{}(\cdot)$ respectively denote the probability density function (PDF) 
and cumulative distribution function (CDF) 
of the random variable $T\pspp$; $\overline{{G}}_{\pspp T}^{}(\cdot)$ and ${G}_{\pspp T}^{-1}(\cdot)$ are respectively the survival (complementary CDF) and the quantile (inverse distribution) functions. 
$\mathbb{M}_{\pspp T}^{}({s})=\mathbb{E}\big[e^{{s}T}\big]$ is the moment generating function (MGF) of $T\pspp$. If the 
random variables $X$ and $Y$ are distributed identically, we write $X \idist Y\nspp\pspp$.
The Laplace (or bilateral exponential) distribution centred at ${\upsilon}$ with scale parameter $\beta$ is denoted by $\mathcal{L}({\upsilon},\beta)\pspp$, and $\mathcal{N}({\upsilon},\sigma^2)$ denotes normal distribution with mean ${\upsilon}$ and variance $\sigma^2\nspp\pspp$.  
Also, let $\Phi(\cdot)$ and ${Q}(\cdot)$ respectively denote the CDF and survival function of the standard normal distribution $\mathcal{N}(0,1)$ and let $\erfi(\cdot)$ denote the imaginary error function,  $\erfi({a })=\frac{2}{\sqrt{\pi}}\! \int_{0}^{{a }}\! e^{\pspp{r }^2} \mathrm{d}{r }\pspp$.
$\mathcal{SG}({\sigma}^2)$ denotes the class of sub-Gaussian distributions 
with proxy variance ${\sigma}^2_{}\nspp\pspp$.

\subsection{Organization of the paper}
The remainder of this article is organized as follows. 	In Section \ref{sec:definitions}, the 
definitions pertaining to DP 
are provided. 
We introduce the flipped Huber density and mechanism 
in Section \ref{sec:flip_hub_mech}. Section \ref{sec:theor_anal} provides a theoretical analysis of the proposed mechanism, 
and Section \ref{sec:emp_result} numerically validates the derived analytical results. 
In Section \ref{sec:appn_discussion}, the efficacy of the flipped Huber mechanism in a machine learning setup is demonstrated. We discuss the strengths and limitations of the proposed mechanism in the same section. 
We provide the concluding remarks in Section \ref{sec:conc}. 

\section{Background and essential definitions}\label{sec:definitions}

We now provide some definitions from differential privacy literature, in part to establish 
notations and conventions that will be followed in this article. 
We also highlight the shortcomings of popular 
additive noise mechanisms for differential privacy, 
viz., Laplace and Gaussian, and {introduce the statistical estimation viewpoint} that will lead to the development of a new 
mechanism in the next section.


The dataset $\mathcal{D}$ is a collection of data records from several individuals. 
Let $\mathcal{X}$ be the space of datasets, and $N$ be the number of data records 
in each dataset. The query function ${f}:\mathcal{X}\to\mathcal{Y}$ acts on dataset $\mathcal{D}$ and provides the query response ${f}(\mathcal{D})\pspp$. 
The goal of DP 
is to conceal any individual's presence 
in $\mathcal{D}$ while publishing ${f}(\mathcal{D})\pspp$.
The following notion of neighbouring datasets is crucial to define DP. 

\begin{definition}[Neighbouring datasets]
	Any pair of datasets that differ by only a single data record is said to be neighbouring (or adjacent) datasets. 
	Let $\neighb$ be the binary relation that denotes 
	neighbouring datasets in the space $\mathcal{X}\pspp$; 
	we write $\neighbdsets$ when $\mathcal{D}$ and $\widecheck{\mathcal{D}}$ are neighbouring datasets from $\mathcal{X}\pspp$.
\end{definition}

\subsection{Differential privacy}
As mentioned earlier, DP 
makes it hard 
to detect the presence of any individual in a dataset from the query responses on that dataset.
This is ensured by \textit{randomizing} the query responses 
such that the responses on neighbouring datasets are probabilistically `similar'.
A private mechanism	$\mathcal{M}$ is an algorithm that, on input with a dataset, provides a randomized output to a query. 

\begin{definition}[$(\epsilon,\delta)$-DP \cite{dwork2014algorithmic}]\label{defn:DP}
	The randomized mechanism $\mathcal{M}:\mathcal{X}\to\mathcal{Y}$ is said to guarantee $(\epsilon,\delta)$-differential privacy \textup{(}$(\epsilon,\delta)$-DP in short\textup{)} if for every 
	pair of 
	neighbouring datasets $\neighbdsets$ and every 
	measurable set $\mathcal{E}$ in $\mathcal{Y}\pspp$, 
	\begin{equation} \label{eq:DP}
		\mathbb{P}\{\mathcal{M}(\mathcal{D})  \in \mathcal{E}\} 
		\leq 
		e^{\epsilon}_{}\psp \mathbb{P}\{\mathcal{M}(\widecheck{\mathcal{D}})  \in \mathcal{E}\} + \delta
		\psp,
	\end{equation}
	where $\epsilon\in\mathbb{R}_+^{}$ is the privacy budget parameter and $\delta\in[0,1]$ is the privacy leakage parameter. 
\end{definition}
The privacy budget $\epsilon$ quantifies the tolerance limit 
on the dissimilarity between the distributions of outputs from neighbouring datasets, and the leakage parameter $\delta$ is the bound on 
stark deviations from this limit. 
Smaller $\epsilon$ essentially means that the privacy guarantee is stronger and smaller $\delta$ assures the reliability of this guarantee. 
%
%
%
%
Conventionally, $\delta$ has been seen as the probability with which the mechanism fails to ensure 
$\epsilon$-DP \cite{dwork2014algorithmic,le2013differentially}. However, 
this \textit{probabilistic} DP (pDP) viewpoint is not equivalent to $(\epsilon,\delta)$-DP in Definition 2, but just a sufficient condition for the same. 
Though the pDP framework makes the analysis simpler, it usually results in excess perturbation 
\cite{balle2018improving}. 


\subsection{Privacy loss and alternate characterizations of privacy}
%
%
Privacy loss encapsulates 
the distinction	between the mechanism's outputs on neighbouring datasets 
as a 
univariate random variable (RV), thereby facilitating 
the interpretation of $(\epsilon,\delta)$-DP guarantee through the extreme (tail) events of this RV. 
Let ${\mu}$ and $\widecheck{{\mu}}$ be, respectively, the probability measures associated with $\mathcal{M}(\mathcal{D})$ and $\mathcal{M}(\widecheck{\mathcal{D}})$ 
and let
${\mu}$ be absolutely continuous with respect to $\widecheck{{\mu}}\pspp$. 
The function $\log \frac{\mathrm{d}{\mu}}{\mathrm{d}\widecheck{{\mu}}}$ is known as the \textit{privacy loss function}.  Consequently, the \textit{privacy loss random variable} \cite{dwork2010boosting,dwork2016concentrated,bun2016concentrated} of the mechanism $\mathcal{M}$ on the neighbouring datasets $\neighbdsets$ is defined as $\plossrv =\log\frac{\mathrm{d}{\mu}}{\mathrm{d}\widecheck{{\mu}}}(\mathbf{{V}})\pspp$, where $\mathbf{{V}} \sim {\mu}\pspp$.

Privacy profile \cite{balle2018improving,balle2020privacy} 
of the randomized mechanism $\mathcal{M}$, $\delta_{\mathcal{M}}^{}(\epsilon)$ 
is the curve of the lowest possible $\delta$ for which $\mathcal{M}$ is differentially private for each $\epsilon\in\mathbb{R}_+^{}$. 
Hence, $\mathcal{M}$ is $(\epsilon,\delta)$-DP if $\delta_{\mathcal{M}}^{}(\epsilon)\leq \delta\pspp$.
The following expression from 
\cite{balle2018improving} 
highlights the 
connection between the privacy profile and the privacy losses, 
thereby providing an equivalent condition for 
$(\epsilon,\delta)$-DP in terms of privacy losses.
\vspace{-0.85em}
\begin{equation}\label{eq:balle_profile}
\delta_{\mathcal{M}}^{}(\epsilon)=
\supoverneighb
\, \mathbb{P}\big\{\plossrv > \epsilon\big\}-e^{\epsilon}_{}\psp \mathbb{P}\big\{\plossrvv < -\epsilon\big\} 
\psp.
\end{equation}
This characterization facilitates easier analysis whenever the privacy losses are adequately simple. We witness one such instance in the 
following subsection for the 
case of additive noise mechanisms, where we redefine the privacy loss in terms of noise density to make our analysis simpler.


Several alternate definitions of differential privacy \cite{bun2016concentrated,mironov2017renyi,dwork2016concentrated,dong2022gaussian} have been introduced 
to facilitate 
simpler characterization of 
the graceful degradation of privacy 
when several 
mechanisms are composed. One such variant, which we use 
in this work, is defined below.
\begin{definition}[$({\xi},{\eta})$-zCDP \cite{bun2016concentrated}]
\label{defn:zCDP}
The randomized mechanism $\mathcal{M}:\mathcal{X}\to\mathcal{Y}$ is said to satisfy $({\xi},{\eta})$-zero concentrated differential privacy \textup{(}$({\xi},{\eta})$-zCDP\textup{)} if
\begin{equation} \label{eq:zCDP}
\mathfrak{D}_{\pspp{\Lambda}}^{(\textup{\texttt{R}})}\nsp ({\mu}\!\parallel\!\widecheck{{\mu}})
\leq
{\xi} + {\Lambda}\psp{\eta} \ \,\ \forall\, {\Lambda} \in (1,\infty)
\ \text{ and } \ \neighbdsets
\psp,
\end{equation}
%
where 
$\mathfrak{D}_{\pspp{\Lambda}}^{(\textup{\texttt{R}})}\nsp ({\mu}\!\parallel\!\widecheck{{\mu}})$ is the $\Lambda$-R\'{e}nyi divergence \cite{renyi1961measures} 
between the distributions of $\mathcal{M}({\mathcal{D}})$ and $\mathcal{M}(\widecheck{\mathcal{D}})$ 
(i.e., 
${\mu}$ and $\widecheck{{\mu}}$). 
When ${\xi}=0\pspp$, the mechanism is said to guarantee ${\eta}$-zCDP. 
\end{definition}
The zCDP guarantee in \eqref{eq:zCDP} can be alternatively interpreted as the imposed bound on the MGF of the privacy loss RV, 
$\mathbb{E}\Big[\exp\!\Big(\nsp{{s}\psp\plossrv}\Big)\nsp\Big]\nspp\pspp \leq 
e^{\pspp s\pspp({\xi}+  (s+1)\pspp{\eta})}_{} \,\ \forall s>0\pspp$. 
%

\subsection{Additive noise mechanism}
When the query response is a numeric vector, the straightforward way to randomize it is by adding noise. 

\begin{definition}[Additive noise mechanism]\label{defn:add_noise}
Consider a {$K$-dimensional, real-valued} query function ${f}: \mathcal{X} \rightarrow \mathbb{R}^{K}_{}\pspp$. The additive noise mechanism (noise mechanism in short) perturbs the query output for the dataset $\mathcal{D}$ as
$	\mathcal{M}( \mathcal{D} ) = {f}(\mathcal{D}) + \mathbf{t}
\pspp$, 
where $\mathbf{t}=[{t}_1\ {t}_2\ \cdots\ {t}_{K}]^\top_{} \in \mathbb{R}^{K}_{}$ is the noise sample added to impart differential privacy that is drawn from a distribution with CDF ${G}_{\pspp\mathbf{T}}^{}$ and PDF ${g}_{\pspp \mathbf{T}}^{}\pspp$.
\end{definition}
Conventionally, ${t}_i,\ i=1,\,2,\,\ldots,\,K$ are the independent and identically distributed (i.i.d.) noise samples obtained from some known univariate distribution with density function ${g}_{\pspp T}^{}\pspp$. 
Several 
noise mechanisms have been proposed and studied 
in the past, the most popular being 
Laplace 
and Gaussian. 
%
%
%
%
%
%
%
Though the 
framework in Definition \ref{defn:add_noise} pertains to output perturbation, it can be generalized by appropriately  specifying 
${f}\pspp$. For e.g., \cite{zhang2012functional} provides an objective perturbation scheme, which is prominent in private regression analysis \cite{dwork2014analyze}; while training machine learning models through gradient descent algorithms, DP is usually ensured by jittering the gradient updates 
\cite{song2013stochastic,bassily2014private,abadi2016deep} 
(see Section \ref{sec:appn_dpcd}). 
%


The accuracy and hence, the utility of the noise-perturbed 
output is dictated by the `amount' of  
noise 
added to the true response. The privacy parameters $\epsilon$ and $\delta$ have a direct impact on 
the noise added
. 
Apart from these, 
the noise \textit{variance} %
is determined by the sensitivity of the query function. 

\begin{definition}[Sensitivity]
The $\ell_p$-sensitivity of the {real-valued, $K$-dimensional} query function ${f}: \mathcal{X} \rightarrow \mathbb{R}^{K}_{}$ is evaluated as 
\vspace{-0.85em}
\begin{equation}\label{eq:sensitivity}
\Delta_p^{} =
\supoverneighb 
\, \big\|{f}(\mathcal{D})-{f}(\widecheck{\mathcal{D}})\big\|_{p}^{}
\,,\quad p\in[1,\infty]
\psp.
\end{equation}
If $K=1$ or $p=\infty\pspp$, we simply denote the sensitivity by $\Delta\pspp$.
\end{definition}

Thus, sensitivity 
indicates how sensitive the true response is to the replacement of a single entry in the dataset.
In order to 
ensure privacy, the noise added should be able to mask this variation 
and thus, sensitivity plays a crucial role in determining the noise variance. 
Using the 
equivalence of norms \cite{goldberg1987equivalence}, 
we write
\begin{equation}\label{eq:norm_equiv}
\Delta_{{q}}^{} \leq {K}^{{\left[\pspp1\nspp/\nspp {q}\pspp-\pspp1\nspp/\nspp {r}\pspp\right]_{\nspp +}^{}}}_{\stackrel{}{}}\! \times \nsp \Delta_{{r}}^{}
\,,\quad \,\ \forall\, {q},\, {r} \in[1,\infty]
\end{equation} 
and hence, it is evident that $\Delta_{p}^{}$ is monotonic decreasing in ${p}\pspp$, i.e., $\Delta_{{q}}^{} \leq \Delta_{{r}}^{} \,\ \forall\, {r}\geq {q}\pspp$. 
%
%
%
%
\\

\subsubsection{Equivalent characterization of privacy loss:}
We now revisit the concepts that were defined earlier in the context of additive noise mechanism, which 
in turn would make the analysis simpler. The following model captures a few 
notations that we will use.

\begin{model}\label{model:mod_1}
For the {neighbouring} datasets $\neighbdsets\pspp$, 
$\mathbf{{z}}={f}(\mathcal{D})$ and $\widecheck{\mathbf{{z}}}={f}(\widecheck{\mathcal{D}})$ denote the true responses
and $\mathbf{d}=\mathbf{{z}}-\widecheck{\mathbf{{z}}} 
\pspp$ 
denotes the difference between 
them. 
Also, let $\mathbf{{V}}=\mathbf{{z}}+\mathbf{T}$ and $\widecheck{\mathbf{{V}}}=\widecheck{\mathbf{{z}}}+\mathbf{T}$ be the random vectors modelling the 
mechanism's outputs, where $\mathbf{T}$ 
corresponds to the 
noise whose coordinates are i.i.d. RVs 
from a distribution with density 
${g}_{\pspp T}^{}$ and $\psi=-\log {g}_{\pspp T}^{}$ is 
the negative log-density function. 
\end{model}



Now, the density of the output $\mathcal{M}(\mathcal{D})$ is ${g}_{\pspp\mathbf{{V}}}^{}\nspp(\mathbf{{v}})
={g}_{\pspp\mathbf{T}}^{}(\mathbf{t})
$ and also, ${g}_{\pspp\widecheck{\mathbf{{V}}}}\nspp(\mathbf{{v}})
={{g}_{\pspp\mathbf{T}}^{}(\mathbf{t}+\mathbf{d})}\pspp$, 
where $\mathbf{t}=\mathbf{{v}}-\mathbf{{z}}\pspp$. Since the densities of $\mathbf{{V}}$ and $\widecheck{\mathbf{{V}}}$ are merely the translations of noise density, the privacy loss 
can be equivalently represented in terms of the noise PDF alone; 
alongside rendering simpler analysis, this 
also offers more intuition to construct a better noise PDF. 


The \textit{equivalent privacy loss function} in one dimension is given by
${\zeta}_{\pspp d}^{}\nsp(t)=\log\frac{{g}_{\pspp T}^{}(t)}{{g}_{\pspp T}^{}(t+d)}{=\psi(t+d)-\psi(t)}\pspp$. In $K$-dimensions, the privacy loss is \textit{additive} under i.i.d. noise, i.e., ${\zeta}_{\pspp \mathbf{d}}^{}\nsp(\mathbf{t}) 
=\sum_{i=1}^K {\zeta}_{\pspp d_i}^{}\nsp(t_i)\pspp$. 
%
%
Furthermore, the random variable ${\zeta}_{\pspp \mathbf{d}}^{}\nsp(\mathbf{T}),\ \mathbf{T}\sim {G}_{\pspp\mathbf{T}}^{}$ is 
probabilistically equivalent to 
$\plossrv$, and any event in $\plossrv$ and $\plossrvv$ can respectively be represented with 
${\zeta}_{\pspp \mathbf{d}}^{}\nsp(\mathbf{T})$ and ${\zeta}_{\pspp-\mathbf{d}}^{}\nsp(\mathbf{T})$ in a seamless manner. For e.g., 
the necessary and sufficient condition for the additive noise mechanism to be $(\epsilon,\delta)$-DP can be re-derived in terms of the equivalent privacy loss function (see the supplementary material) as
\vspace{-0.85em}
\begin{equation}\label{eq:balle_K}
\supoverneighb \,
\mathbb{P}\{{\zeta}_{\pspp \mathbf{d}}^{}\nsp(\mathbf{T})\geq\epsilon\}-e^{\epsilon}_{}\psp \mathbb{P}\{{\zeta}_{\pspp -\mathbf{d}}^{}\nsp_{}(\mathbf{T})\leq-\epsilon\} \leq \delta
\psp,	
\end{equation}
which mirrors \eqref{eq:balle_profile}.
%
This reformulation 
of privacy loss in terms of noise density 
facilitates the understanding of 
the role that noise distribution plays in determining accuracy 
for the given privacy parameters
and sheds light on how to 
design the noise density to enhance the accuracy. 

It is sometimes convenient to express the privacy loss function in a symmetric fashion. We define the \textit{centered privacy loss} as 
$\widetilde{{\zeta}}_{\pspp d}^{}\nsp(t)
=\psi\nspp\big(\nspp t\nspp+\nspp\tfrac{d}{2}\nspp\big)\! - \psi\nspp\big(\nspp t\nspp-\nspp\tfrac{d}{2}\nspp\big)\!\pspp\pspp$. 
It is clear that ${\zeta}_{\pspp d}^{}\nsp(t)=\widetilde{{\zeta}}_{\pspp d}^{}\nsp\!\left(t\nspp+\nspp\frac{d}{2}\nspp\right)\!\nspp\psp\pspp$. 
When the noise density is symmetric, $\widetilde{{\zeta}}_{\pspp d}^{}\nsp(t)$ exhibits odd symmetry, i.e., 
$\widetilde{{\zeta}}_{\pspp d}^{}\nsp(-t)=-\widetilde{{\zeta}}_{\pspp d}^{}\nsp(t)\pspp$.
Also, if ${g}_{\pspp T}^{}$ is a log-concave density, 
then $\widetilde{{\zeta}}_{\pspp d}^{}\nsp(t)$ will be a monotonic function in the direction of $\sgn(d)\pspp$, i.e., monotonic increasing for $d>0$ and decreasing for $d<0\pspp$ \cite{saumard2014log}. 
%
%
In the succeeding subsection, 
we examine the noise perturbation in the purview 
of statistical estimation that offers new insights onto designing noise density.

\subsection{Estimation formulation}
By observing Model \ref{model:mod_1}, the mechanism's output 
can be visualized as a single noisy measurement 
of unperturbed query response
. Thus, if the noise density ${g}_{\pspp T}^{}$ has the mode at $t=0\pspp$, $\mathcal{M}(\mathcal{D})$ is the \textit{maximum likelihood estimate} \cite{hogg2019introduction} 
for the `location parameter' ${f}(\mathcal{D})\pspp$. 
%
%
%
%
We desire 
noise such that
$\mathcal{M}(\mathcal{D})$ is \textit{most informative} of ${f}(\mathcal{D})$ and also light-tailed so that large samples of noise do not result in excess perturbation.
In the location model $v_i=z_i+t_i$, where $t_i$ is a 
sample from the distribution with density function ${g}_{\pspp T}^{}=e^{-\psi}_{}\pspp$, the Fisher information
that $v_i$ carries about $z_i$ is 
$ 
\mathcal{I}_{\pspp T}^{}
= \int_{-\infty}^{\infty}\! (\psi'(t)\nspp)^2
\psp {g}_{\pspp T}^{}(t)
\, \mathrm{d}t
\pspp$. 
%
%
Since the Fisher information is dependent on the scale of ${g}_{\pspp T}^{}\pspp$, we define the normalized Fisher information (NFI) as the product of Fisher information corresponding to the location model when the noise is sampled from that distribution and the variance of the distribution, i.e., $\widetilde{\mathcal{I}}_{\pspp T}^{}=\mathcal{I}_{\pspp T}^{}\times \sigma_{T}^2\pspp$, whenever both 
are defined and finite. 
%
%
The 
NFI measures how `concentrated' or `sharp' the density is about zero 
after accounting for the scale of the distribution. 
%
%
Gaussian distribution has the least  
NFI among all the distributions 
($\!\!$\cite{stein2014lower} and references therein). 
However, Gaussian is still a popular noise mechanism 
as it is light-tailed. Despite having higher NFI, the tail of Laplace is relatively heavy, 
so the noise samples take large values frequently.
%
%
We 
seek the distribution whose density is light-tailed like Gaussian but has a larger curvature around zero like Laplace. Such a distribution would have more NFI than Gaussian while generating large noise samples less frequently. 

\section{Proposed mechanism}\label{sec:flip_hub_mech}

In order to leverage the benefits of both Laplace and Gaussian mechanisms, we first introduce the flipped Huber distribution, which has Gaussian tails and a Laplacian centre. We then present the flipped Huber mechanism for differential privacy, which adds flipped Huber noise to the query result.

\subsection{Flipped Huber distribution}\label{sec:flip_hub_dist}
\begin{definition}
The flipped Huber loss function is defined as
\begin{equation}\label{eqn:flip_hub_loss}
\rho_{\alpha}^{}\nspp(t) = 
\begin{cases}
\\[-3.3em]
\alpha |t| \, , &  |t| \leq \alpha \\[-0.5em]
({t^2+\alpha^2})/{2}  \, , & |t| > \alpha
\end{cases} 
\psp,
\end{equation} 
where $\alpha\in\mathbb{R}_{+}^{}$  is the transition parameter which indicates the transition of $\rho_{\alpha}^{}\nspp(t)$ from an absolute-valued function to a quadratic function.
\end{definition} 
Note that the flipped Huber loss function is absolute-valued in the middle and quadratic in the tails; this is 
the reverse of 
the Huber loss function, and hence the name `flipped' Huber. 
The flipped Huber loss function is symmetric and  convex\footnote{$\rho_{\alpha}^{}\nspp(t)$ is a convex function since 
it always lies above the tangents at every $t\in \mathbb{R}$ \cite{boyd2004convex}.}. 

We now devise a distribution based on the flipped Huber loss function. Inspired from the  statistical physics, we define the flipped Huber distribution as the \textit{Boltzmann-Gibbs distribution} \cite{landau2013statistical} 
corresponding to \textit{energy function} $\mathcal{E}(t)=\rho_{\alpha}^{}\nspp(t)\pspp$, \textit{thermal energy} 
$\mathcal{E}_{0}=\gamma^2$ and \textit{partition function} ${\kappa}\pspp$.
\begin{figure*}[h!]
\centering
\begin{minipage}{0.45\linewidth}
\centering
\vspace{-0.82em}
\includegraphics[width=0.85\linewidth]{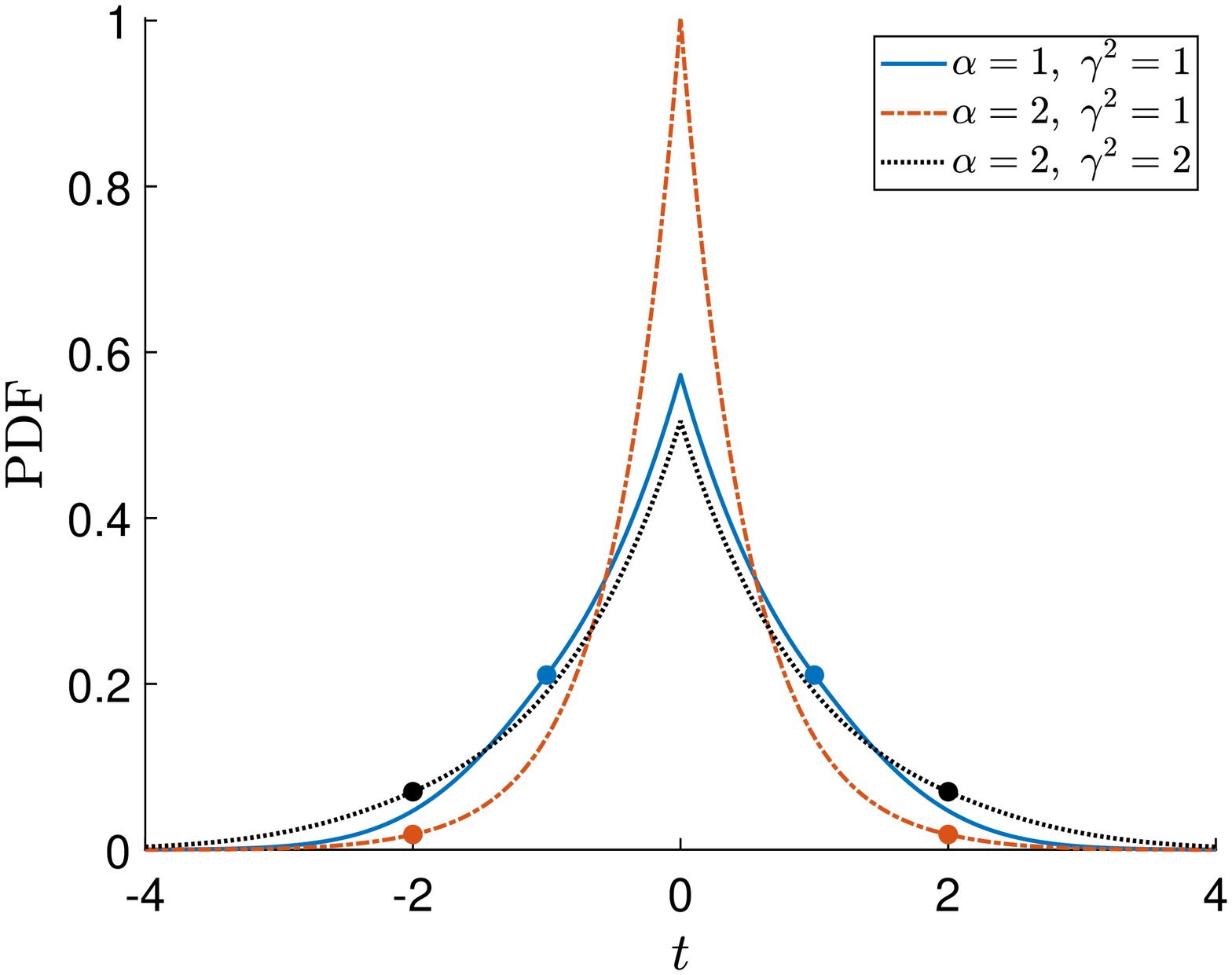}
\vspace{0.645em}
\caption{Flipped Huber density functions for various choices of $\alpha$ and $\gamma\pspp$. The transition from  Laplace to Gaussian segment is marked by `$\bullet$'.}
\label{fig:flip_hub_pdf}
\end{minipage}%
\hspace{2em}
\begin{minipage}{0.45\linewidth}
\centering
\includegraphics[width=\linewidth]{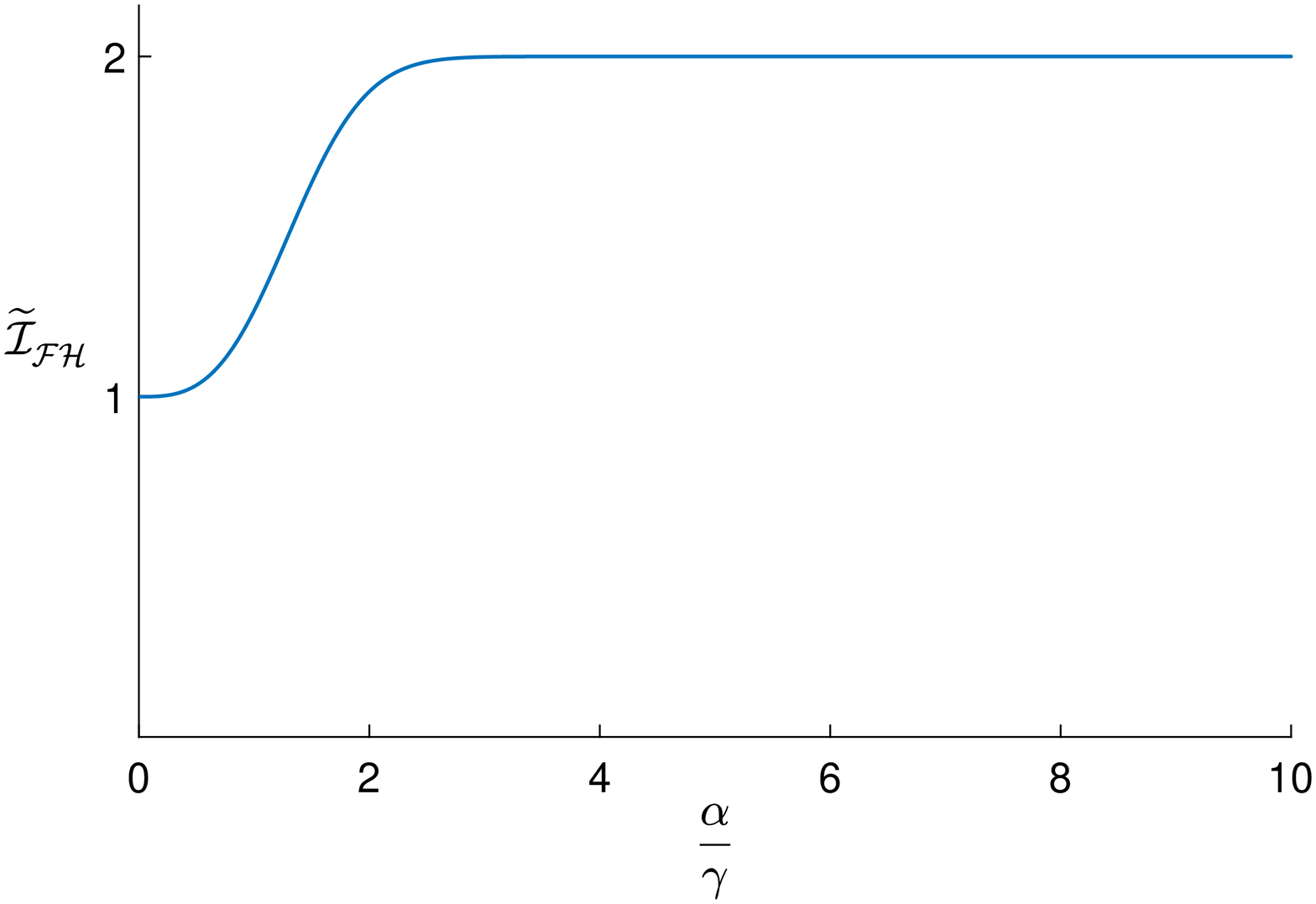}
\caption{Normalized Fisher information of the flipped Huber distribution as a function of 
${\alpha}/{\gamma}\pspp$.}
\label{fig:flip_hub_norm_fish_inf}
\end{minipage}
\end{figure*}
\begin{definition}
The flipped Huber distribution $\mathcal{FH}(\alpha,\gamma^2)$ is specified by the flipped Huber density function, 
\begin{equation}
\label{eqn:pdf_flip_hub}
{g}_{\mathcal{FH}}^{}(t;\alpha,\gamma^2) = \tfrac{1}{{\kappa}} \exp\!\left(\nsp-\tfrac{\rho_{\alpha}^{}\nspp(t)}{\gamma^2_{}}\nsp\right)\! 
\psp,
\end{equation}
where $\alpha\in\mathbb{R}_{+}^{}$  is the transition parameter, $\gamma\in \mathbb{R}_{++}^{}$ is the scale parameter 
and ${\kappa}$ is the normalization constant, which is given by 
%
${\kappa}= \gamma \pspp {\omega} \pspp e^{-\alpha^2/2\gamma^2_{}}\nspp\pspp$, where
\begin{equation}\label{eq:omega}
{\omega}=2\nspp\!\left[\nspp\sqrt{2\pi}\psp{Q}\!\left(\nspp\tfrac{\alpha}{\gamma}\nspp\right)\! +\tfrac{2\gamma}{\alpha}\sinh\!\left(\nspp\tfrac{\alpha^2}{2\gamma^2_{}}\nspp\right)\!\right]\!
\psp.
\end{equation}
A random variable $T$ from the flipped Huber distribution is termed as the \textit{flipped Huber RV} and is denoted as $T\sim \mathcal{FH}(\alpha,\gamma^2)\pspp$.
\end{definition}
%
%
%
%

Flipped Huber densities 
for various choices of $\alpha$ and $\gamma$ are demonstrated in Fig.  \ref{fig:flip_hub_pdf}. 
%
The parameter $\alpha$ can be adjusted to achieve the desired combination of Laplace and Gaussian densities.
The flipped Huber density is the same as Gaussian when $\alpha=0$ and resembles Laplace density as $\alpha\to \infty\pspp$.
The scale parameter $\gamma$ controls how the flipped Huber density is concentrated around zero. 
%
%
%
%
The CDF 
of the flipped Huber distribution ${\mathcal{FH}}(\alpha,\gamma^2)$ is 
\begin{equation}\label{eq:flip_hub_cdf}
{G}_{\mathcal{FH}}^{}(t;\alpha,\gamma^2) 
\!
= 
\!
\begin{cases}
\\[-3em]
\hspace{-0.225em}
\tfrac{1}{2} {+} \tfrac{2\gamma}{\alpha\pspp{\omega}}\nspp \exp\nspp\!\left(\nsp\tfrac{\alpha}{2\gamma^2_{}} (\alpha {-} |t|)\!\right)\nsp\sinh\!\left(\nsp\tfrac{\alpha t}{2\gamma^2_{}}\nsp\right)\!\nsp
, & \hspace{-0.85em}
|t| \nspp\leq\nspp \alpha\\[-0.5em]
\hspace{-0.225em}
\tfrac{1}{2} {+} \sgn(t)\nsp\!\left[\tfrac{1}{2} {-} \tfrac{\sqrt{2\pi}}{{\omega}}\psp {Q}\!\left(\nspp\tfrac{|t|}{\gamma}\nspp\right)\!\right]\!
, & \hspace{-0.85em}
|t| \nspp>\nspp \alpha \\
\end{cases}
\hspace{-0.12em}.
\end{equation}

The factor $\omega$ in \eqref{eq:omega} 
depends on the parameters $\alpha$ and $\gamma$ only through their ratio $\tfrac{\alpha}{\gamma}\pspp$. The following observation will be useful in our analyses. 

\begin{proposition}\label{prop:omeg_2pi} 
The factor ${\omega}$ has the property that ${\omega} \geq \sqrt{2\pi}\pspp$. 
\end{proposition}
\begin{proof} 
Let ${b}=\frac{\alpha}{\gamma}\nspp\pspp$. Thus, ${b}\in\mathbb{R}_{+}^{}$ and 
$Q({b})
\geq \frac{1}{2}-\frac{{b}}{\sqrt{2\pi}}\nspp\pspp$. Now, 	
${\omega}=2\sqrt{2\pi}\psp{Q}({b}) + \frac{4}{{b}}\nspp \sinh\!\big(\nsp\frac{{b}^2}{2}\nsp\big)\!
\geq \nspp\sqrt{2\pi} \psp-\psp 2{b} \psp + \frac{4}{{b}}\nsp\big(\nsp\frac{{b}^2}{2}\nsp\big)\!
=\nspp\sqrt{2\pi}\pspp$, 
where 
the fact that $\sinh({a})\geq {a} \,\ \forall\, {a}\in\mathbb{R}_{+}^{}$ 
is used, which is evident from the Taylor expansion of $\sinh({a})\pspp$. 
%
\end{proof}

\subsection{Features of flipped Huber distribution}\label{sec:flip_hub_feat}
%
Since the flipped Huber loss is symmetric and convex, the flipped Huber distribution is a symmetric log-concave distribution \cite{vinterbo2022differential}. 
%
%
%
In Laplace and Gaussian distributions, the scale parameter is the sole determinant of both accuracy and privacy, which is a disadvantage in the differential privacy setup. But, for the flipped Huber distribution, multiple sets of parameters $(\alpha,\gamma)$ could result in the same variance, as we show below. 
This additional degree of freedom of the flipped Huber 
is what 
makes it 
well-suited noised for DP.  

Now, we investigate the aspects in which Flipped Huber is better than Gaussian and Laplace. We first demonstrate that flipped Huber distribution is better than Gaussian in terms of variance and Fisher information. 
\\

\subsubsection{Variance and Fisher information:}

The variance of the flipped Huber random variable $T\sim \mathcal{FH}(\alpha,\gamma^2)$ is given by
\begin{equation}\label{eq:var_flip_hub}
\sigma_{\mathcal{FH}}^2 
\nspp=\nspp
\gamma^2_{} \nspp\!\left[ 1\nspp-\nspp \tfrac{1}{{\omega}}\big(\tfrac{2\gamma}{\alpha}\big)^{\!3}_{}\pspp 
\!\left(\nspp\tfrac{\alpha^2}{2\gamma^2_{}}\cosh\!\left(\nspp\tfrac{\alpha^2}{2\gamma^2_{}}\nspp\right)\! -\sinh\!\left(\nspp\tfrac{\alpha^2}{2\gamma^2_{}}\nspp\right)\!\nspp\right)\!
\right]\!
,
\end{equation}
and the Fisher information of $\mathcal{FH}(\alpha,\gamma^2)$ is
\begin{equation}\label{eq:fisher_flip_hub}
\mathcal{I}_{\mathcal{FH}}^{}
=\tfrac{1}{\gamma^2_{}}
\nsp\!\left[1+\tfrac{4\gamma}{\alpha\pspp{\omega}}\!\left(\nspp\tfrac{\alpha^2}{2\gamma^2_{}}e^{\psp\alpha^2/2\gamma^2_{}}\!-\sinh\!\left(\nspp\tfrac{\alpha^2}{2\gamma^2_{}}\nspp\right)\!\nspp\right)\!
\right]\!  
\psp.
\end{equation}

It is observed that $\sigma_{\mathcal{FH}}^2 \leq \gamma^2_{}$ and $\mathcal{I}_{\mathcal{FH}}^{} \geq	\tfrac{1}{\gamma^2_{}}$; therefore, $\mathcal{FH}(\alpha,\gamma^2)$ has less variance and more Fisher information  than the Gaussian with the same 
scale parameter\footnote{The proofs for these are provided in Appendix \ref{appx:var_fish_FH}, along with the derivations of variance and Fisher information of $\mathcal{FH}(\alpha,\gamma^2)\pspp$.}, i.e., $\mathcal{N}(0,\gamma^2)\pspp$. 
Fig.  \ref{fig:flip_hub_norm_fish_inf} plots the normalized Fisher information of the flipped Huber distribution, $\widetilde{\mathcal{I}}_{\mathcal{FH}}^{}=\mathcal{I}_{\mathcal{FH}}^{}\times\sigma^2_{\mathcal{FH}}$ as a function of $\frac{\alpha}{\gamma}\pspp$. It is evident that $\widetilde{\mathcal{I}}_{\mathcal{FH}}^{}$ is always between $1$ and $2\pspp$, 
i.e., between the normalized Fisher information of Gaussian and Laplace distributions.


Now, we highlight the feature of the flipped Huber distribution, which makes it superior to Laplace.
\\

\subsubsection{MGF and sub-Gaussianity:}

Because of the log-concavity, we know that the flipped Huber distribution is sub-exponential \cite{saumard2014log}. However, we further expect it to be sub-Gaussian since it has Gaussian tails by construction.
%
The moment-generating function of the flipped Huber distribution 
is given as
$
\mathbb{M}_{\mathcal{FH}}^{}({s})
=\tfrac{\sqrt{2\pi}}{{\omega}}\nsp
\big(\nspp{\mathcal{P}_{\alpha,\gamma}^{}({s})\psp+\, \mathcal{P}_{\alpha,\gamma}^{}(-{s})}\nspp\big)\!\psp
e^{\pspp\gamma^2{s}^2/2}_{}
\pspp$,
where $\mathcal{P}_{\alpha,\gamma}^{}({s}) 
= 
\big(\nspp e^{\pspp {m}_{\nspp {s}}^{}\nspp \alpha/\gamma}-1\nspp\big) \frac{e^{-{{m}_{\nspp {s}}^{2}\nspp }/{2}}}{{m}_{\nspp {s}}^{}\nspp \sqrt{2\pi}}+{Q}({m}_{\nspp {s}}^{}\nspp)$ 
and ${m}_{\nspp {s}}^{}\nspp =\tfrac{\alpha}{\gamma}\nspp +\nspp \gamma {s}\pspp$.
%

\begin{lemma}[Sub-Gaussianity of flipped Huber] %
\label{lem:sub_gau}
The Flipped Huber distribution $\mathcal{FH}(\alpha,\gamma^2)$ is sub-Gaussian with proxy variance $\gamma^2_{}\pspp$, i.e., $\mathcal{FH}(\alpha,\gamma^2) \in \mathcal{SG}(\gamma^2_{})\pspp$. 
\end{lemma}

\begin{proof}
Proving the sub-Gaussianity in a direct way, i.e., showing $\mathbb{M}_{\mathcal{FH}}^{}({s}) \leq e^{\pspp\gamma^2 {s}^2/2} \,\ \forall\, {s}\in\mathbb{R}\pspp$, is 
difficult. Hence, we prove the sub-Gaussianity using the Orlicz condition. 
From Theorem 2.1.$\pspp$(IV) in \cite{wainwright2019high}, the necessary and sufficient condition for sub-Gaussianity with proxy variance $\gamma^2_{}$ is
$
\mathbb{E}\!\left[\exp\!\left(\nsp\frac{{s}T^2_{}}{2\gamma^2_{}}\nsp\right)\!\right]\! \leq \frac{1}{\sqrt{1-{s}}}
\,\ \forall\, {s}\in[\pspp 0,1)
\pspp$. 
%
%
When $T\sim\mathcal{FH}(\alpha,\gamma^2)\pspp$, we shall obtain 
$
\mathbb{E}\!\left[\exp\!\left(\nsp\frac{{s}T^2_{}}{2\gamma^2_{}}\nsp\right)\!\right]\! = \frac{1}{\sqrt{1-{s}}}\psp\mathcal{C}_{\alpha,\gamma}^{}({s})\pspp$, where 
$\mathcal{C}_{\alpha,\gamma}^{}({s})=\tfrac{\sqrt{2\pi}}{{\omega}}
\nsp\Big[\nspp
\sqrt{\tfrac{1}{{s}}\nsp-\nsp 1} \times 
\exp\!\left(\nsp-\!\left(\nspp\tfrac{1}{{s}}\nsp - \nsp 1\nspp\right)\nsp\! \tfrac{\alpha^2}{2\gamma^2_{}}\nsp\right)\!
\!\Big(\nsp\erfi\nspp\!\left(\nspp\tfrac{1}{\sqrt{2{s}}}\tfrac{\alpha}{\gamma}\nspp\right)\! -\erfi\nspp\!\left(\nspp\tfrac{(1-{s})}{\sqrt{2{s}}}\tfrac{\alpha}{\gamma}\nspp\right)\!\!\Big)\!
\nspp +\nspp 
2{Q}\!\left(\nsp\sqrt{1\nsp-\nsp{s}}\,\tfrac{\alpha}{\gamma}\nsp\right)\!
\!\Big]\!\pspp\pspp$. 
Note that $\mathcal{C}_{\alpha,\gamma}^{}({s})$ 
depends on $\alpha$ and $\gamma$ only through 
$\tfrac{\alpha}{\gamma}\nspp\pspp$. It can be verified that $\forall\, \tfrac{\alpha}{\gamma}\in\mathbb{R}_{+}^{}\pspp$, $\mathcal{C}_{\alpha,\gamma}^{}({s})$ is a decreasing function in ${s}\in [\pspp 0,1)$ and $\lim\limits_{{s}\rightarrow 0^+}\mathcal{C}_{\alpha,\gamma}^{}({s})=1\pspp$. 
Thus, $\mathcal{C}_{\alpha,\gamma}^{}({s})\leq 1 \,\ \forall\, \alpha\in\mathbb{R}_{+}^{}$ and $\gamma \in \mathbb{R}_{++}^{}$ 
and hence, $\mathcal{FH}(\alpha,\gamma^2) \in \mathcal{SG}(\gamma^2_{})\pspp$.
\end{proof}
This result is also used later in this article in the derivation of a sufficient condition in higher dimensions in Section \ref{sec:priv_K_dim} and the characterization of zCDP in Section \ref{sec:appn_discussion_zcdp}.
Since the flipped Huber distribution is sub-Gaussian, unlike the Laplace, the chances of extreme perturbations of query results are very less. This feature of sub-Gaussian tails (like Gaussian) that limits the outliers and the sharper centre (like Laplace) that offers more Fisher information make the flipped Huber mechanism, presented formally below, a well-suited additive noise mechanism. 


\subsection{Flipped Huber mechanism for differential privacy}

Flipped Huber mechanism imparts privacy by adding i.i.d. noise samples from the flipped Huber distribution 
to each coordinate of the real-valued query response vector. 
Flipped Huber noise samples can be obtained using Smirnov transform; Appendix \ref{appx:sampling} describes how to obtain i.i.d. samples from the distribution $\mathcal{FH}(\alpha,\gamma^2)\pspp$. 
The 
mechanism is presented formally in Algorithm \ref{alg:flip_hub}.

\begin{algorithm}[h]
\caption{Flipped Huber Mechanism.} \label{alg:flip_hub}
\textbf{Input}: Dataset $\mathcal{D}\pspp$, query function ${f}:\mathcal{X}\rightarrow\mathbb{R}^{K}_{}\nspp\pspp$, flipped Huber parameters $\alpha$ and $\gamma\pspp$. 
\begin{algorithmic}[1]
\State Evaluate the query response ${f}(\mathcal{D})\pspp\pspp$.
\State Obtain i.i.d. noise samples $t_i^{},\ i=1,\,2,\,\ldots,\,K$ from $\mathcal{FH}(\alpha,\gamma^2)\pspp$. 
\State Compute $\mathcal{M}_{\mathcal{FH}}^{}( \mathcal{D}) 
= {f}(\mathcal{D}) + [\,t_1^{} \ t_2 \ \cdots \ t_{K}^{}\,]^{\top}_{}\nspp\pspp$. 
\end{algorithmic}
\textbf{Output}: Perturbed query response $\mathcal{M}_{\mathcal{FH}}^{}( \mathcal{D})\pspp$. 
\end{algorithm}

Since our preliminary analysis of flipped Huber distribution in Section \ref{sec:flip_hub_feat} corroborates its advantage over Gaussian and Laplace distributions, we expect the flipped Huber mechanism to achieve a better privacy-accuracy trade-off than Gaussian and Laplace mechanisms. 
Note that the parameters of the distribution, viz., $\alpha$ and $\gamma\pspp$, are not specified directly and have to be determined from the privacy parameters. We need conditions that indicate what $\alpha$ and $\gamma$ are valid for the given $\epsilon$ and $\delta\pspp$.
In the 
sequel, we derive the privacy guarantees of the proposed flipped Huber mechanism that enables us to determine the noise parameters.

\section{Theoretical analysis}\label{sec:theor_anal}
\renewcommand{\arraystretch}{1.1}
\begin{table*}[h!] 
\begin{center}
\caption{Cases to be considered for one dimensional necessary and sufficient condition.}
\label{tab:thm1}
\setlength{\tabcolsep}{2.8pt}
\begin{tabular}{?c?c|c?c|c?c|c?}

\clineB{2-7}{2}
\multicolumn{1}{c?}{$\hspace{1.9em}$} & \multirow{2}{*}{Range of $\epsilon$} &  \multirow{2}{*}{\begin{tabular}{c} Non-empty\\[-0.5em]only when \end{tabular}} & \multirow{2}{*}{$t_1^{}$} & \multirow{2}{*}{$\hspace{5.5em} t_2^{} \hspace{5.5em} $} &\multicolumn{2}{c?}{Intervals}\\
\cline{6-7} 
\multicolumn{1}{c?}{} & & &  &  & $t_1^{}$ & $ \hspace{1.8em} t_2^{} \hspace{1.8em} $\\[0.25em]
\clineB{2-7}{2}
\specialrule{1pt}{2pt}{0pt}
(i) & 
$\bigg[\psp 0, \dfrac{(\Delta-2\alpha)\Delta}{2\gamma^2_{}}\nsp\bigg)$ &  
$\alpha < \frac{\Delta}{2}$ &
$\dfrac{\gamma^2_{}\epsilon}{\Delta}-\dfrac{\Delta}{2}$ &
$\dfrac{\gamma^2_{}\epsilon}{\Delta}+\dfrac{\Delta}{2}$ &  
$(-\infty,-\alpha)$ & $(\alpha,\infty)$
\\[0.85em] \hline

(ii) &
$\bigg[\psp 0, \dfrac{((2\alpha-\Delta)\wedge\Delta)\alpha}{\gamma^2_{}}\nsp\bigg)$ &  
$\alpha > \frac{\Delta}{2}$ &
$\dfrac{\gamma^2_{}\epsilon}{2\alpha}-\dfrac{\Delta}{2}$ &
$\dfrac{\gamma^2_{}\epsilon}{2\alpha}+\dfrac{\Delta}{2}$ & 
$(-\alpha,0)$ & $(0,\alpha)$
\\[0.85em] \hline

(iii) &
$\bigg[\nspp \dfrac{(2\alpha \vee \Delta )^{2}-2\alpha\Delta}{2\gamma^2_{}}, \dfrac{\nsp\left(\nspp[\Delta-\alpha]_+\nspp\right)^{\!2}\!+2\alpha\Delta}{2\gamma^2_{}}\nspp \bigg)$ & 
$\alpha < {\Delta}$ &
$-\alpha+\sqrt{2(\gamma^2_{}\epsilon+\alpha\Delta)}-\Delta$ &
$\sqrt{2(\gamma^2_{}\epsilon+\alpha\Delta)}-\alpha$ & 
$[-\alpha,0)$ & $[\psp\alpha,\infty)$
\\[0.85em] \hline

(iv) &
$\bigg[ \dfrac{\nsp\left(\nspp[\Delta-\alpha]_+\nspp\right)^{\nsp 2}\!+2\alpha\Delta}{2\gamma^2_{}}, \dfrac{(\Delta+2\alpha)\Delta}{2\gamma^2_{}}\nsp\nspp\bigg)$ & 
%
$-$ &
$\alpha+\sqrt{2(\gamma^2_{}\epsilon-\alpha\Delta)}-\Delta$ &
$\sqrt{2(\gamma^2_{}\epsilon-\alpha\Delta)}+\alpha$ & 
$[\psp 0,\alpha)$ & $[\psp\alpha,\infty)$
\\[0.85em] \hline

(v) &
$\bigg[\nspp\dfrac{(\Delta+2\alpha)\Delta}{2\gamma^2_{}}, \infty\bigg) $ & 
%
$-$ &
$\dfrac{\gamma^2_{}\epsilon}{\Delta}-\dfrac{\Delta}{2}$ &
$\dfrac{\gamma^2_{}\epsilon}{\Delta}+\dfrac{\Delta}{2}$ & 
$[\psp \alpha,\infty)$ & $[\psp\alpha,\infty)$
\\[0.85em] 

\specialrule{1pt}{0pt}{0pt}

\end{tabular}
\end{center}
\end{table*}
\renewcommand{\arraystretch}{1}
Firstly, we obtain a necessary and sufficient condition for $(\epsilon,\delta)$-DP 
for the single-dimensional case.  In the high dimensional scenario, we observe that it is very difficult to obtain a necessary and sufficient condition because of the heterogeneous nature of the flipped Huber density. Therefore, a sufficient condition is derived using the principles of sub-Gaussianity and stochastic ordering. 

The primary requirement for the derivation of privacy guarantees is the privacy loss function ${\zeta}_{\pspp d}^{}\nsp(t)\pspp$, which can be equivalently characterized by the centered privacy loss function $\widetilde{{\zeta}}_{\pspp d}^{}\nsp(t)\pspp$ since ${\zeta}_{\pspp d}^{}\nsp(t)=\widetilde{{\zeta}}_{\pspp d}^{}\nsp\!\left(t\nspp+\nspp\frac{d}{2}\nspp\right)\nspp\pspp$. Because of the piecewise nature of noise 
density, the derivation of $\widetilde{{\zeta}}_{\pspp d}^{}\nsp(t)$ for the flipped Huber mechanism is algebraically involved
. Appendix \ref{appx:zeta} provides 
the piecewise closed-form expression for $\widetilde{{\zeta}}_{\pspp d}^{}\nsp(t)\pspp$. 

\subsection{Necessary and sufficient condition in one dimension}\label{sec:priv_1_dim}

We deduce the 
necessary and sufficient condition for $(\epsilon,\delta)$-DP in one dimension using the fact that the flipped Huber density is symmetric and log-concave. The following theorem provides the condition.
%
%
\begin{table*}[h]
\centering
\setlength{\tabcolsep}{0pt}
\begin{tabular}{p{\textwidth}}
\begin{equation}\label{eq:del_fh}
\hspace{-0.7em}
\delta_{\mathcal{FH}}^{(1)}(\epsilon)= 
\begin{cases}
\\[-2.75em]
\left(1 \nsp - \nsp \frac{\sqrt{2\pi}}{{\omega}}\right)\! + \frac{\sqrt{2\pi}}{{\omega}}
\nsp\!\left[ {Q}\!\left(\nsp\frac{\gamma\epsilon}{\Delta} \nsp - \nsp \frac{\Delta}{2\gamma}\nsp\right)\!
- e^{\epsilon}_{} {Q}\!\left(\nsp\frac{\gamma\epsilon}{\Delta} \nsp + \nsp \frac{\Delta}{2\gamma}\nsp\right)\! \right]\!\!
\, , & \hspace{-0.7em}
0\leq\nspp  \epsilon \nspp < \nsp \frac{(\Delta-2\alpha)\Delta}{2\gamma^2_{}} \\
\frac{1}{2}(1\nsp - \nsp e^{\epsilon}_{}) + \frac{\gamma}{\alpha\pspp{\omega}} e^{\psp\alpha^2_{}/2\gamma^2_{}} \!\left(1 \nsp + \nsp e^{\epsilon}_{} \nsp - \nsp 2\exp\!\left(\frac{\epsilon}{2} - \frac{\alpha\Delta}{2\gamma^2_{}}\right)\!\right)\!\!
\, , & \hspace{-0.7em}
0\leq\nspp  \epsilon \nspp < \nsp \frac{((2\alpha-\Delta)\wedge\Delta)\alpha}{\gamma^2_{}}\\
\frac{1}{2} \nsp + \nsp \frac{\gamma}{\alpha\pspp{\omega}} e^{\psp\alpha^2_{}/2\gamma^2_{}} \nsp\!\left[1 \nsp - \nsp \exp\!\left(\nsp\frac{\alpha}{\gamma^2_{}}\nsp\big(\! -\alpha \nsp + \nsp \sqrt{2(\gamma^2_{}\epsilon \nsp + \nsp \alpha\Delta)} \nspp-\nsp \Delta \big)\nspp\!\right)\!\right]\! - e^{\epsilon}_{} \frac{\sqrt{2\pi}}{{\omega}} {Q}\!\left(\nsp\frac{\sqrt{2(\gamma^2_{}\epsilon + \alpha\Delta)} - \alpha}{\gamma}\nsp\right)\!\!\nspp
\, , & \hspace{-0.7em}
\frac{\left(2\alpha \vee \Delta\right)^{2}-2\alpha\Delta}{2\gamma^2_{}}
\nsp \leq\nspp \epsilon \nspp < \nsp \frac{\left([\Delta-\alpha]_{\nspp +}\nsp\right)^{\nsp 2}\! +  2\alpha\Delta}{2\gamma^2_{}} \!\!\!\\
\frac{1}{2} \nsp - \nsp  \frac{\gamma}{\alpha\pspp{\omega}} e^{\psp\alpha^2_{}/2\gamma^2_{}} \nsp\!\left[1 \nsp - \nsp \exp\!\left(\nsp\frac{\alpha}{\gamma^2_{}}\nsp\big(\!-\alpha \nsp - \nsp \sqrt{2(\gamma^2_{}\epsilon \nsp - \nsp \alpha\Delta)} \nspp+\nsp \Delta \big)\nspp\!\right)\!\right]\! - e^{\epsilon}_{} \frac{\sqrt{2\pi}}{{\omega}} {Q}\!\left(\nsp\frac{\sqrt{2(\gamma^2_{}\epsilon - \alpha\Delta)} + \alpha}{\gamma}\nsp\right)\!\!\nspp
\, , & \hspace{-0.7em}
\frac{\left([\Delta-\alpha]_{\nspp +}\nsp\right)^{\nsp 2}\! +  2\alpha\Delta}{2\gamma^2_{}}
\nsp \leq \nspp \epsilon \nspp < \nsp  \frac{(\Delta+2\alpha)\Delta}{2\gamma^2_{}} \\
\frac{\sqrt{2\pi}}{{\omega}} \nsp\!\left[{Q}\!\left(\nsp\frac{\gamma\epsilon}{\Delta}\nsp - \nsp\frac{\Delta}{2\gamma}\nsp\right)\! -e^{\epsilon}_{} {Q}\!\left(\nsp\frac{\gamma\epsilon}{\Delta}\nsp + \nsp\frac{\Delta}{2\gamma}\nsp\right)\!\right]\!\!
\, , & \hspace{-0.7em}
\epsilon \nspp \geq \nsp \frac{(\Delta+2\alpha)\Delta}{2\gamma^2_{}} \\
\end{cases}
\end{equation}
\\
\hline
\end{tabular}
\end{table*}
%
%
\begin{theorem}\label{thm:priv_1_dim}
The one-dimensional flipped Huber mechanism guarantees $(\epsilon,\delta)$ differential privacy if and only if
$\delta_{\mathcal{FH}}^{(1)}(\epsilon) \leq \delta
\pspp$,
where $	\delta_{\mathcal{FH}}^{(1)}(\epsilon)$ is the privacy profile given in \eqref{eq:del_fh}.
\end{theorem}

\begin{proof}
Since the flipped Huber density is symmetric and log-concave, we can use the necessary and sufficient condition from \cite{vinterbo2022differential}, which can be expressed in our notation for the flipped Huber mechanism as\footnote{We have re-derived the result with our notational convention in Appendix \ref{appx:ns_cond}.
}
\begin{equation}\label{eq:log_concave}
\overline{{G}}_{\mathcal{FH}}^{}\nsp\!\left(\nspp \widetilde{{\zeta}}^{-1}_{\Delta}\nsp(\epsilon) - \tfrac{\Delta}{2}\nsp\right)\!
- e^{\epsilon}_{}\psp \overline{{G}}_{\mathcal{FH}}^{}\nsp\!\left(\nspp \widetilde{{\zeta}}^{-1}_{\Delta}\nsp(\epsilon) + \tfrac{\Delta}{2}\nsp\right)\! \leq \delta
\psp,
\end{equation}
where $\widetilde{{\zeta}}^{-1}_{\Delta}\nsp({\nu})=\sup\nspp\big\{{t}\,\big|\,\widetilde{{\zeta}}^{}_{\Delta}\nsp({t})\leq {\nu}\big\}\pspp$ is the right-continuous partial inverse of the centered privacy loss function $\widetilde{{\zeta}}^{}_{\Delta}\nsp(t)$ of the flipped Huber mechanism, which can be derived in closed form using \eqref{eq:zeta_d_flip_hub} in Appendix \ref{appx:zeta} as
\begin{equation}\label{eq:zeta_inv_flip_hub}
\widetilde{{\zeta}}_{\pspp \Delta}^{-1}\nsp({\nu}) = 
\begin{cases}
\\[-2.75em]
\frac{\gamma^2_{}{{\nu}}}{2\alpha} \nspp\nsp
\, , & \hspace{-0.8em}
|{\nu}|\! < \!\frac{((2\alpha-\Delta)\wedge\Delta)\alpha}{\gamma^2_{}}\!\\[-0.35em]
\mathcal{J}_{\Delta,\gamma}^{}({\nu},\alpha)\nspp\nsp
\, , & \hspace{-0.8em}
|{\nu}|\! \in \!\Big[\nspp \frac{(2\alpha \vee \Delta)^{2}_{}-2\alpha\Delta}{2\gamma^2_{}},  {\nu}_1^{} \nspp\Big)\!\!\!\\[-0.35em]
\mathcal{J}_{\Delta,\gamma}^{}({\nu},-\alpha)\nspp\nsp
\, , & \hspace{-0.8em}
|{\nu}|\! \in [ {\nu}_1^{}, {\nu}_2^{} )\!\\[-0.35em]
\frac{\gamma^2_{}{{\nu}}}{\Delta}\nspp\nsp
\, , & \hspace{-0.8em}
|{\nu}|\! \in \nsp\mathbb{R}_{+}^{}\!\nsp\left\backslash\nsp\Big[ \frac{(\Delta-2\alpha)\Delta}{2\gamma^2_{}},  {\nu}_2^{} \nspp\Big)\right.\!\!\!\!\!\\
\end{cases}
\psp,
\end{equation}
where ${\nu}_1^{}=\frac{\left(\nspp[\Delta-\alpha]_+\nsp\right)^{\nsp 2}\nsp+2\alpha\Delta}{2\gamma^2_{}}
\nspp\pspp$, ${\nu}_2^{}=\frac{(\Delta+2\alpha)\Delta}{2\gamma^2_{}}\nspp\pspp$ and
\begin{equation*}
\mathcal{J}_{\Delta,\gamma}^{}({\nu},\alpha)=\sgn({{\nu}})\nspp\big(\nspp\sqrt{2(\gamma^2_{}|{\nu}|+\alpha\Delta)}-\tfrac{\Delta}{2}-\alpha\big)\nspp
\psp. \end{equation*}
Now, let us denote  $t_1^{}=	\widetilde{{\zeta}}_{\pspp \Delta}^{-1}\nsp(\epsilon)\nsp-\nsp \tfrac{\Delta}{2}$ and $t_2^{}=	\widetilde{{\zeta}}_{\pspp \Delta}^{-1}\nsp(\epsilon)\nsp+\nsp \tfrac{\Delta}{2}\nspp\pspp$. We can rewrite 
\eqref{eq:log_concave} as 
\begin{equation}\label{eq:priv_prof_1}
\delta_{\mathcal{FH}}^{(1)}(\epsilon)
\triangleq
\overline{{G}}_{\mathcal{FH}}^{}\nsp(t_1^{})
- e^{\epsilon}_{}\psp \overline{{G}}_{\mathcal{FH}}^{}\nsp(t_2^{}) \leq \delta
\psp,
\end{equation} 
where $\delta_{\mathcal{FH}}^{(1)}(\epsilon)$ is the privacy profile in one dimension.
Since the flipped Huber CDF in \eqref{eq:flip_hub_cdf} is piecewise, we now need to characterize in which interval $t_1^{}$ and $t_2^{}$ fall to proceed further with evaluating the condition \eqref{eq:priv_prof_1}. Table \ref{tab:thm1} lists the various possible ranges in which $\epsilon$ can lie and the corresponding 
$t_1^{}$ and $t_2^{}\pspp$, 
along with the interval in which they fall.
Note that depending on the relation between $\alpha$ and $\Delta\pspp$, some of these ranges for $\epsilon$ might 
correspond to empty sets, which are also mentioned in the table. 
Taking the cases 
in Table \ref{tab:thm1} into account,  
after algebraic simplifications of $\overline{{G}}_{\mathcal{FH}}^{}\nsp(t_1^{})
- e^{\epsilon}_{}\psp \overline{{G}}_{\mathcal{FH}}^{}\nsp(t_2^{})\pspp$, 
we shall deduce $\delta_{\mathcal{FH}}^{(1)}(\epsilon)$ as in \eqref{eq:del_fh}. Consequently, 
the necessary and sufficient condition is given as $\delta_{\mathcal{FH}}^{(1)}(\epsilon)\leq\delta\pspp$.
\end{proof}

From the above theorem, it can be shown that the performance of flipped Huber will be better than that of Gaussian in one dimension. We show it through the existence of $(\alpha,\gamma)$ that results in 
lesser variance than Gaussian for the given set of privacy parameters $\epsilon $ and $\delta\pspp$.
\begin{itemize}
\item Let $\sigma$ be the minimum scale parameter of the Gaussian noise to ensure $(\epsilon,\delta)$-DP.
\item Let the parameters of flipped Huber mechanism be $\gamma=\sigma$ and $\alpha=\big[\tfrac{\gamma^2\epsilon}{\Delta}-\tfrac{\Delta}{2}\big]_+^{}$.
We have 
\begin{equation*}
{Q}\!\left(\nspp\tfrac{\gamma\epsilon}{\Delta}\nspp-\nspp\tfrac{\Delta}{2\gamma}\nspp\right)\!
-e^{\epsilon}_{} {Q}\!\left(\nspp\tfrac{\gamma\epsilon}{\Delta}\nspp+\nspp\tfrac{\Delta}{2\gamma}\nspp\right)
\pspp\leq\pspp	\delta 
\pspp\leq\pspp  \delta\cdot\tfrac{{\omega}}{\sqrt{2\pi}}\nspp
\psp,
\end{equation*}
where the first inequality is due to the necessary and sufficient condition for the Gaussian mechanism \cite{balle2018improving}, and the second inequality follows from Proposition \ref{prop:omeg_2pi}.
\item Therefore, we have 
$\delta_{\mathcal{FH}}^{(1)}(\epsilon) \leq \delta$ from \eqref{eq:del_fh}
(corresponding to the last case), 
and hence for this choice $\alpha$ and $\gamma$, flipped Huber mechanism is $(\epsilon,\delta)$-DP.  
\end{itemize} 
Recall from Section  \ref{sec:flip_hub_feat}
that the variance of the flipped Huber distribution $\mathcal{FH}(\alpha,\gamma^2_{})$ can never exceed 
$\gamma^2_{}$, i.e., $\var(\mathcal{FH}(\alpha,\gamma^2_{})) \leq 
\gamma^2_{}=\sigma^2_{}
=\var(\mathcal{N}(0,\sigma^2_{}))$. Also, note that 
this choice of $\alpha$ and $\gamma$ 
need not be the optimal set of parameters resulting in the least possible variance for the given $(\epsilon,\delta)\pspp$ but rather provides an upper bound for the same.
Hence, the variance of the noise added by the one-dimensional flipped Huber mechanism satisfying $(\epsilon,\delta)$-DP is always lesser than that of the corresponding Gaussian mechanism.  

The performance of the one-dimensional flipped Huber mechanism under the necessary and sufficient condition in Theorem \ref{thm:priv_1_dim} is validated through simulations in Section \ref{sec:res_1_D}.

\subsection{Sufficient condition in higher dimensions}\label{sec:priv_K_dim}


Now, we derive a condition for the flipped Huber mechanism to be $(\epsilon,\delta)$ differentially private in $K$ dimensions when $K>1\pspp$, where we assume that  i.i.d. flipped Huber noise	samples are added to each coordinate of the query output. 
%
Deriving a closed-form necessary and sufficient condition is mathematically intractable
because of 
the hybrid, piecewise nature 
of the flipped Huber density. Also, 
the expressions for 
${\zeta}_{\pspp \mathbf{d}}^{}\nsp(\mathbf{T})$  become highly complex in high dimensions, 
making it difficult to characterize the tail probabilities of privacy loss, 
and hence, condition \eqref{eq:balle_K} cannot be used.

In such cases, the privacy profile can be obtained using $K$-fold iterative numerical integration, as suggested  
in \cite[Section IV-B]{sadeghi2022offset}. By using this technique, the performance of flipped Huber mechanism in $K$ dimensions is empirically studied in Section \ref{sec:res_K_D}. 

However, determining suitable noise parameters that give at most accuracy for a given $\epsilon$ and $\delta$ would involve grid search, and it becomes very complex 
since iterative integration has to be performed for every point in the grid. 
Hence, we now derive 
a sufficient condition for $(\epsilon,\delta)$-DP, 
which is obtained from \eqref{eq:balle_K} {under the case of restricted parameters}. 

Due to the symmetry of the flipped Huber density, we have $\mathbf{T}\idist-\mathbf{T}$ and also
${\zeta}_{\pspp -\mathbf{d}}^{}\nsp(\mathbf{t}) =\log\frac{{g}_{\pspp \mathbf{T}}^{}(\mathbf{t})}{{g}_{\pspp \mathbf{T}}^{}(\mathbf{t}-\mathbf{d})}
=\log\frac{{g}_{\pspp \mathbf{T}}^{}(-\mathbf{t})}{{g}_{\pspp \mathbf{T}}^{}(-\mathbf{t}+\mathbf{d})}
={\zeta}_{\pspp \mathbf{d}}^{}\nsp(-\mathbf{t})\pspp$. 
Subsequently, 
${\zeta}_{\pspp -\mathbf{d}}^{}\nsp(\mathbf{T})={\zeta}_{\pspp \mathbf{d}}^{}\nsp(-\mathbf{T})\idist{\zeta}_{\pspp \mathbf{d}}^{}\nsp(\mathbf{T})\pspp$, as the functions of identically distributed random variables are identically distributed. 
Therefore, we shall rewrite the condition for $(\epsilon,\delta)$-DP using \eqref{eq:balle_K}
\begin{equation}\label{eq:balle_K_suff0}
\mathbb{P}\{ {\zeta}_{\pspp \mathbf{d}}^{}\nsp(\mathbf{T})\geq\epsilon \}-e^{\epsilon}_{}\psp \mathbb{P}\{ {\zeta}_{\pspp \mathbf{d}}^{}\nsp(\mathbf{T})\leq-\epsilon\} \leq \delta
\psp. 
\end{equation}
Thus, a sufficient condition can be devised by bounding the tail probabilities of the privacy loss; we will upper bound the upper tail probability $\mathbb{P}\{ {\zeta}_{\pspp \mathbf{d}}^{}\nsp(\mathbf{T})\geq\epsilon \}$ and then use non-trivial 
lower bound in place of the lower tail probability $\mathbb{P}\{ {\zeta}_{\pspp \mathbf{d}}^{}\nsp(\mathbf{T})\leq-\epsilon\}\pspp$.

Note that the privacy loss RV 
${\zeta}_{\pspp \mathbf{d}}^{}\nsp(\mathbf{T})$ is a transformation of flipped Huber random vector by the privacy loss function. 
The following result pertaining to such functionally expressed variables is essential 
for 
deriving our sufficient condition.

\begin{lemma}\label{lem:zosd}
If ${h}_1^{}\!:\mathbb{R}^{K}_{}\rightarrow\mathbb{R}$ and ${h}_2^{}\!:\mathbb{R}^{K}_{}\rightarrow\mathbb{R}$ are functions such that ${h}_1^{}(\mathbf{t})\leq {h}_2^{}(\mathbf{t}), \, \ \forall\, \mathbf{t}\in\mathbb{R}^{K}_{}\nspp\pspp$, then for every ${a}\in \mathbb{R}\pspp$,
%
%
\begin{enumerate}[label=(\roman*)]
\item $\mathbb{P}\{{h}_1^{}(\mathbf{T}) \geq {a}\} \leq \mathbb{P}\{{h}_2^{}(\mathbf{T}) \geq {a}\}$ and
\item $\mathbb{P}\{{h}_1^{}(\mathbf{T}) \leq {a}\} \geq \mathbb{P}\{{h}_2^{}(\mathbf{T}) \leq {a}\}\pspp$. \label{lem:zosd_case_b}
\end{enumerate}
\end{lemma}

\begin{proof}
Since we have ${h}_1^{}(\mathbf{t})\leq {h}_2^{}(\mathbf{t})\pspp$, 
${h}_1^{}(\mathbf{t})\geq {a}$ implies ${h}_2^{}(\mathbf{t}) \geq {a} \, \ \forall\, \mathbf{t}\in\mathbb{R}^{K}_{}, \ {a}\in \mathbb{R}$ and hence the upper level set $\{\mathbf{t}\in\mathbb{R}^{K}_{} | {h}_1^{}(\mathbf{t})\geq {a}\} $ is a subset of $ \{\mathbf{t}\in\mathbb{R}^{K}_{} | {h}_2^{}(\mathbf{t})\geq {a}\}\pspp$. By invoking the monotonicity of probability measure, we obtain $\mathbb{P}\{{h}_1^{}(\mathbf{T}) \geq {a}\} \leq \mathbb{P}\{{h}_2^{}(\mathbf{T}) \geq {a}\}  \,\ \forall\, {a}\in \mathbb{R}\pspp$. Similarly, we can prove the other case in \ref{lem:zosd_case_b}.
\end{proof}
Thus, one can appropriately choose the bounding function in order to obtain bounds on the tail probabilities. We 
apply this result on ${\zeta}_{\pspp \mathbf{d}}^{}\nsp(\mathbf{t})$ to obtain the sufficient condition. Since the linear combinations of i.i.d. random 
variables are relatively easy to handle, 
we look 
for the bounding function $\plossup\nsp(\mathbf{t})\pspp$, 
which is linear or affine in $\mathbf{t}\pspp$; this is facilitated by the linear tails of the privacy loss function. 
Let $\mathcal{R}_{\Delta}^{}\nsp(\alpha) = \alpha^2_{}-(\pspp[\alpha-\Delta]_{\nspp +}^{})^{ 2}_{}\nspp\pspp$. Using the  affine upper bound (derived in Appendix \ref{appx:zeta}) 
\begin{equation}\label{eq:zeta_up_bnd}
\plossup\nsp(\mathbf{t}) = \tfrac{\mathbf{t}^{\top}\mathbf{d}}{\gamma^2_{}}+ \tfrac{\norm{\mathbf{d}}^2_{2}}{2\gamma^2_{}}+\tfrac{K\mathcal{R}_{\Delta}^{}\nsp(\alpha)}{2\gamma^2_{}}
\psp,
\end{equation}
in place of 
${\zeta}_{\pspp \mathbf{d}}^{}\nsp(\mathbf{t})$ in  \eqref{eq:balle_K_suff0}, we arrive at the sufficient condition
\begin{equation}\label{eq:balle_K_suff}
\mathbb{P}\{ \plossup\nsp(\mathbf{T})\geq\epsilon \}-e^{\epsilon}_{}\psp \mathbb{P}\{ \plossup\nsp(\mathbf{T})\leq-\epsilon\} \leq \delta
\psp. 
\end{equation}
However, this condition is still 
difficult to evaluate since the distribution of $\mathbf{T}^{\top}\mathbf{d}$ is difficult to characterize as it involves convolutions of 
flipped Huber densities. In the sequel, we derive an even looser (but tractable) sufficient condition by upper bounding $\mathbb{P}\{\plossup\nsp(\mathbf{T})\geq\epsilon\}$ and lower bounding $\mathbb{P}\{\plossup\nsp(\mathbf{T})\leq-\epsilon\}\pspp$. 
%
\\

\subsubsection{Upper bounding the upper tail probability:} 
First, we upper bound 
$\mathbb{P}\{\plossup\nsp(\mathbf{T})\geq\epsilon\}$ 
using the sub-Gaussianity of the flipped Huber distribution. 

\begin{lemma}\label{lem:up_tail}
Let $T_1,\, T_2,\, \ldots,\, T_K$ be i.i.d. flipped Huber RVs, 
$T_i \sim \mathcal{FH}(\alpha,\gamma^2)$ 
and $\mathbf{T}=[\, T_1 \ T_2 \, \cdots \, T_K\,]^{\top}\nspp\pspp$. If $\mathcal{R}_{\Delta}^{}\nsp(\alpha)= 
\alpha^2_{}-(\pspp[\alpha-\Delta]_{\nspp +}^{})^{ 2}_{} 
\leq ({2\gamma^2_{}\epsilon-\Delta_2^2})/{K}\pspp$,  then
$
\mathbb{P}\{ \plossup\nsp(\mathbf{T}) \geq \epsilon \}
\leq {Q}\!\left(\nsp\frac{\gamma\epsilon}{\Delta_2^{}}-\frac{\Delta_2^{}}{2\gamma}-\frac{K\mathcal{R}_{\Delta}^{}\nsp(\alpha)}{2\gamma\Delta_2^{}}\nsp\right)\nspp
\pspp$.
\end{lemma}

\begin{proof}
We have $\mathbb{M}_{\pspp T_i}^{}({s}) =\mathbb{M}_{\mathcal{FH}}^{}({s}) \leq e^{\pspp\gamma^2_{}{s}^2_{}/2}$$ \,\ \forall\, i=1,\,2,\,\ldots,\,K\nspp\pspp$. Let ${W}=\mathbf{T}^{\top}\mathbf{d}\pspp$. Therefore,  
$
\mathbb{M}_{\pspp W}^{}({s})=\prod_{i=1}^{K}\mathbb{M}_{\mathcal{FH}}^{}(d_i \pspp {s}) \leq \exp\!\left(\nspp\tfrac{\gamma^2_{}{s}^2_{}}{2}\nsp\norm{\mathbf{d}}_2^2\nspp\right)\!
\pspp\pspp$.
Hence, ${W}$ 
is a sub-Gaussian random variable, ${G}_{\pspp W}^{}\in\mathcal{SG}\big(\gamma^2_{}\!\norm{\mathbf{d}}^{2}_{2}\nsp\big)\nspp\pspp$. From the equivalent condition (2.11) in \cite{wainwright2019high} for sub-Gaussianity, 
\begin{equation}\label{eq:sub_G_tail_flip_hub}
\mathbb{P}\!\left\{ \mathbf{T}^{\top}\mathbf{d} > {a}\right\}\! \leq {Q}\!\left(\nspp\tfrac{a}{\gamma \norm{\mathbf{d}}_{2}^{}}\nspp\right)\!
\,\ \forall\,{a}\in\mathbb{R}_{+}^{} 
\psp.
\end{equation}

From \eqref{eq:zeta_up_bnd} and monotonicity of any survival function, 
$	
\mathbb{P}\{\plossup\nsp(\mathbf{T})\geq\epsilon\}
\leq \mathbb{P}\pspp\Big\{\nspp \mathbf{T}^{\top}\mathbf{d} \geq \gamma^2_{}\epsilon - \frac{\Delta^2_{2}}{2} -\frac{K\mathcal{R}_{\Delta}^{}\nsp(\alpha)}{2} \nspp \Big\}
\pspp$.  
The given  condition $\mathcal{R}_{\Delta}^{}\nsp(\alpha)=\alpha^2_{}-(\pspp[\alpha-\Delta]_{\nspp +}^{})^{ 2}_{} \leq ({2\gamma^2_{}\epsilon-\Delta_2^2})/{K}$ implies $\gamma^2_{}\epsilon \geq \frac{\Delta^2_{2}}{2}+\frac{K\mathcal{R}_{\Delta}^{}\nsp(\alpha)}{2}\nspp\pspp$; 
using \eqref{eq:sub_G_tail_flip_hub}, we have 
\begin{align*}
\mathbb{P}\{ \plossup\nsp(\mathbf{T}) \geq \epsilon \}
&
\leq {Q}\!\left(\nsp \tfrac{1}{\gamma\norm{\mathbf{d}}_2^{}}	\!\left(\nsp \gamma^2_{}\epsilon-\tfrac{\Delta^2_{2}}{2} -\tfrac{K\mathcal{R}_{\Delta}^{}\nsp(\alpha)}{2}\nsp\right)\! \nspp \right)\!
\\&
\leq {Q}\!\left(\nsp \tfrac{1}{\gamma\Delta_2^{}}	\!\left(\nsp \gamma^2_{}\epsilon-\tfrac{\Delta^2_{2}}{2} -\tfrac{K\mathcal{R}_{\Delta}^{}\nsp(\alpha)}{2}\nsp\right)\! \nspp \right)\!\pspp
\!\psp, \end{align*}
where the last inequality is because ${Q}(\cdot)$ is a monotonic decreasing function and $\norm{\mathbf{d}}_2^{} \leq \Delta_2^{}\pspp$. Simplifying the bound proves the result.
\end{proof}

The parameter restriction $\mathcal{R}_{\Delta}^{}\nsp(\alpha)\leq ({2\gamma^2_{}\epsilon-\Delta_2^2})/{K}$ in Lemma \ref{lem:up_tail} can be rewritten as $\alpha\leq \mathcal{R}_{\Delta}^{-1}\nsp\Big(\nsp\frac{2\gamma^2_{}\epsilon-\Delta_2^2}{K}\nsp\Big)\nsp\pspp$. 
Note that $\mathcal{R}_{\Delta}^{}\nsp(\alpha)=\alpha^2_{}\nspp\pspp$, if $\alpha<\Delta$ and $\mathcal{R}_{\Delta}^{}\nsp(\alpha)=2\alpha\Delta-\Delta^2_{}\nspp\pspp$, if $\alpha\geq\Delta$
Hence, $\mathcal{R}_{\Delta}^{}\nsp(\alpha)$ is a monotonic increasing function on $\alpha\in\mathbb{R}_{+}^{}\pspp$, for which we can define the inverse
\begin{equation}
\mathcal{R}_{\Delta}^{-1}\nsp({\nu})=\begin{cases}
\\[-3.25em]
\hspace{-0.225em}
\sqrt{{{\nu}}}, & \hspace{-0.85em}
0 < {{\nu}} < \Delta^2_{}\\[-0.5em]
\hspace{-0.225em}
\frac{{{\nu}}+\Delta^2_{}}{2\Delta}, & \hspace{-0.5em}
{{\nu}} \geq \Delta^2_{}\\
\end{cases}
\psp.
\end{equation}
%

\subsubsection{Lower bounding the lower tail probability:} 

Now, we derive a lower bound for the lower tail probability of $\plossup\nsp(\mathbf{T})\pspp$. 
To achieve a non-trivial lower bound, we use the notion of stochastic ordering
\cite{shaked2007stochastic}. 
The \textit{usual stochastic order} relates two random variables based on the ordering of their CDFs, or equivalently, survival functions. 
The random variable $X$ is 
\textit{stochastically smaller than} $Y$ (denoted as $X\uso Y$) if 
${G}_{\pspp X}^{}({a}) \geq {G}_{\pspp Y}^{}({a}) \,\ \forall\, {a}\in\mathbb{R}\pspp$. 

We begin by 
stochastically bounding the univariate flipped Huber variable. We 
want the bounding variable to be from some simple distribution. Since the flipped Huber has Gaussian tails, we stochastically bound the flipped Huber variable by a Gaussian random variable. Also, since the convolution of Gaussian densities is again a Gaussian, it is simple to characterize the bound on the tail probability in high dimensions.

Consider the univariate flipped Huber random variable $T\sim \mathcal{FH}(\alpha,\gamma^2_{})\pspp$. We know that $\mathbb{E}[T]=0\pspp$. Since the usual stochastic ordering implies ordering of means, we require the bounding variable to have a positive mean; we consider the Gaussian variable ${W}\sim\mathcal{N}(\theta,\gamma^2_{})$ with $\theta \geq 0\pspp$. The mean parameter $\theta$ for which $T \uso {W}$ holds is derived next.

\begin{lemma}\label{lem:theta}
The flipped Huber distribution $\mathcal{FH}(\alpha,\gamma^2_{})$ is stochastically upper bounded by the Gaussian distribution $\mathcal{N}(\theta,\gamma^2_{})\pspp$, i.e., $\mathcal{FH}(\alpha,\gamma^2_{})\uso \mathcal{N}(\theta,\gamma^2_{})\pspp$, where 
\begin{equation}\label{eq:theta}
\theta=\gamma\psp {Q}^{-1}\!\nsp\left(\nspp\tfrac{1}{{\omega}}\sqrt{\tfrac{\pi}{2}}\right)\!
\psp.
\end{equation}
\end{lemma}
\begin{proof}
We need to show that the CDF of $\mathcal{FH}(\alpha,\gamma^2_{})$ at $t$ is lower bounded by that of $\mathcal{N}(\theta,\gamma^2_{}) \,\ \forall\, t\in \mathbb{R}\pspp$. From \eqref{eq:flip_hub_cdf}, it can be verified that ${G}_{\mathcal{FH}}^{}(t;\alpha,\gamma^2_{})\geq
\frac{\sqrt{2\pi}}{{\omega}} {Q}\!\left(\nsp-\frac{t}{\gamma}\nsp\right)\! \,\ \forall\, t \in \mathbb{R}\pspp$. 	
Using Esseen's inequality \cite[p. 291]{mitrinovic1970analytic} 
${Q}({a}+{b})\leq 2\psp {Q}({a}) \psp {Q}({b}), \,\ \forall\, {a}\geq0 \pspp$,
we get
$
{G}_{\mathcal{FH}}^{}(t;\alpha,\gamma^2_{})\geq {Q}\!\left(\nsp-\frac{t}{\gamma} + {Q}^{-1}\!\nsp\left(\nsp\frac{1}{{\omega}} \sqrt{\frac{\pi}{2}}\right)\!\nsp\right)\!
=\Phi\!\left(\nsp\frac{t\pspp-\pspp \theta}{\gamma}\nsp\right)\!\pspp
\pspp$, proving the result.
\end{proof}
{Note that ${Q}^{-1}$ is defined only on $[0,1]\pspp$, and the existence of $\theta$ is ensured by Proposition  \ref{prop:omeg_2pi}, from which we know $\tfrac{1}{{\omega}} \sqrt{\tfrac{\pi}{2}} \leq \tfrac{1}{2}$ and hence, $\theta \geq 0$}.

In the high dimensional case, each of the i.i.d. variables $T_i\sim \mathcal{FH}(\alpha,\gamma^2_{}),\ i =1,\,2,\,\ldots,\, K$ is stochastically bounded by corresponding ${W}_i\pspp$, where ${W}_i\sim \mathcal{N}(\theta,\gamma^2_{}),\ i =1,\,2,\,\ldots,\, K$ are i.i.d. Gaussian variables with mean $\theta$ as in \eqref{eq:theta}. 
The following lemma provides a lower bound for $\mathbb{P}\{{\zeta}_{\pspp \mathbf{d}}^{}\nsp(\mathbf{T})\leq-\epsilon\}$ by extending the stochastic bounds on each of $T_i$ to $\mathbf{T}^{\top}\mathbf{d}\pspp$.

\begin{lemma}\label{lem:low_tail}
Let $T_1,\, T_2,\, \ldots,\, T_K$ be i.i.d. flipped Huber random variables,  $T_i \sim \mathcal{FH}(\alpha,\gamma^2)$ 
and $\mathbf{T}=[\, T_1 \ T_2 \, \cdots \, T_K\,]^{\top}\nspp\pspp$. We have 
$
\mathbb{P}\{\plossup\nsp(\mathbf{T})\leq-\epsilon\} \geq
{Q}\!\left(\nsp\tfrac{\gamma\epsilon}{\Delta_2^{}}+\tfrac{\Delta_2^{}}{2\gamma}+\tfrac{K\mathcal{R}_{\Delta}^{}\nsp(\alpha)}{2\gamma\Delta_2^{}}+\tfrac{ \theta\Delta_1^{}}{\gamma\Delta_2^{}}\nsp\right)\nspp
\pspp$, 
where $\theta=\gamma\psp {Q}^{-1}\!\nsp\left(\nsp\tfrac{1}{{\omega}}\sqrt{\tfrac{\pi}{2}}\right)\!\psp
\pspp$.
\end{lemma}

\begin{proof}  
Let ${W}_i\sim\mathcal{N}(\theta,\gamma^2),\ i =1,\,2,\,\ldots,\, K$ be i.i.d. Gaussian variables. 
From Lemma \ref{lem:theta}, $T_i\uso {W}_i \,\ \forall\, i=1,\,2,\,\ldots,\,K\nspp\pspp$. 
Due to the symmetry of the flipped Huber density, $d_i\,T_i \idist|d_i|\psp T_i\pspp$. Since the usual stochastic order is preserved after the application of an increasing function and convolutions \cite[Theorem 1.A.3]{shaked2007stochastic}, we have $|d_i|\psp T_i\uso |d_i|\psp {W}_i\pspp$ 
%
and consequently
$\sum_{i=1}^{K}\! |d_i|\psp T_i  \uso \sum_{i=1}^{K}\! |d_i| \psp {W}_i
\triangleq {W}_0
\pspp$. 

It is clear that 
${W}_0 \sim \mathcal{N}(\theta\!\norm{\mathbf{d}}_1^{},\gamma^2\!\norm{\mathbf{d}}_2^2)\pspp$. Consider a new variable ${W} \sim \mathcal{N}(\theta\Delta_1^{},\gamma^2_{}\Delta_2^2)\pspp$. Clearly, ${W}_0 \uso {W}$ as we know that $\Phi\!\left(\nsp\frac{a-\theta\norm{\mathbf{d}}_1^{}}{\gamma\norm{\mathbf{d}}_2^{}}\nsp\right)\! \leq \Phi\!\left(\nsp\frac{a-\theta\Delta_1^{}}{\gamma\Delta_2^{}}\nsp\right)\! \,\ \forall\, a\in\mathbb{R}$ since $\norm{\mathbf{d}}_1^{} \leq \Delta_1^{}\pspp$, $\norm{\mathbf{d}}_2^{} \leq \Delta_2^{}\pspp$, $\theta \geq 0$ and $\Phi(\cdot)$ is a monotonic increasing function.
By gathering the results, we have $\mathbf{T}^{\top}\mathbf{d}=\sum_{i=1}^{K}\! d_i T_i  \uso {W}$ and hence, 
$
\mathbb{P}\{\mathbf{T}^{\top}\mathbf{d} \leq a\} \geq \mathbb{P}\{{W} \leq a\} \,\ \forall\, a\in\mathbb{R}
\pspp$. 
From \eqref{eq:zeta_up_bnd},
\begin{align*}
\mathbb{P}\{\plossup\nsp(\mathbf{T})\leq-\epsilon\}
& = \mathbb{P}\pspp\Big\{\nspp\mathbf{T}^{\top}\mathbf{d} \leq -\gamma^2_{}\epsilon - \tfrac{\norm{\mathbf{d}}^2_{2}}{2}-\tfrac{K\mathcal{R}_{\Delta}^{}\nsp(\alpha)}{2} \nspp\Big\}
\\&
\geq  \mathbb{P}\pspp\Big\{\nspp\mathbf{T}^{\top}\mathbf{d} \leq -\gamma^2_{}\epsilon - \tfrac{\Delta_2^2}{2}-\tfrac{K\mathcal{R}_{\Delta}^{}\nsp(\alpha)}{2} \nspp\Big\}
\psp,
\end{align*}
where the inequality is because CDF is a monotonic increasing function.
Using the stochastic bound,   we get
\begin{equation*}
\mathbb{P}\{\plossup\nsp(\mathbf{T})\leq-\epsilon\}
\geq
\Phi\!\left(\nsp\tfrac{1}{\gamma\Delta_2^{}}\!\left(\nsp -\gamma^2_{}\epsilon - \tfrac{\Delta_2^2}{2} -\tfrac{K\mathcal{R}_{\Delta}^{}\nsp(\alpha)}{2} - \theta\Delta_1^{} \nsp\right)\!\nspp\right)\nsp
\!\psp, \end{equation*}
which proves the lemma.
\end{proof}
%

\subsubsection{Overall sufficient condition:} 

We now consolidate the results derived so far to provide the sufficient condition. 
\begin{theorem}\label{thm:priv_K_dim}
The $K$-dimensional flipped Huber mechanism guarantees $(\epsilon,\delta)$ differential privacy if $\mathcal{R}_{\Delta}^{}\nsp(\alpha)=
\alpha^2_{}-(\pspp[\alpha-\Delta]_{\nspp +}^{})^{ 2}_{} \leq ({2\gamma^2_{}\epsilon-\Delta_2^2})/{K}$ 
and 
\begin{equation*}
{Q}\nspp\!\left(\nsp\nspp\tfrac{\gamma\epsilon}{\Delta_2^{}}\nsp-\nsp\tfrac{\Delta_2^{}}{2\gamma}\nsp-\nsp\tfrac{K\mathcal{R}_{\Delta}^{}\nsp(\alpha)}{2\gamma\Delta_2^{}}\nsp\right)\!
-e^{\epsilon}_{}\nspp {Q}\nspp\!\left(\nsp\nspp\tfrac{\gamma\epsilon}{\Delta_2^{}}\nsp+\nsp\tfrac{\Delta_2^{}}{2\gamma}\nsp+\nsp\tfrac{K\mathcal{R}_{\Delta}^{}\nsp(\alpha)}{2\gamma\Delta_2^{}}\nsp+\nsp\tfrac{\theta\Delta_1^{}}{\gamma\Delta_2^{}}\nsp\right)\!
\nspp\leq\nspp \delta\nspp
\psp, \end{equation*}
where 
$\theta=\gamma\psp {Q}^{-1}\!\nsp\left(\nsp\frac{1}{{\omega}}\sqrt{\frac{\pi}{2}}\right)\!\psp\pspp$.
\end{theorem}

\begin{proof}
Using the bounds in Lemmata \ref{lem:up_tail} and \ref{lem:low_tail}, we shall write $\mathbb{P}\{ \plossup\nsp(\mathbf{T})\geq\epsilon \}-e^{\epsilon}_{}\psp \mathbb{P}\{ \plossup\nsp(\mathbf{T})\leq-\epsilon\} \leq {Q}\!\left(\nsp\frac{\gamma\epsilon}{\Delta_2^{}}-\frac{\Delta_2^{}}{2\gamma} -\frac{K\mathcal{R}_{\Delta}^{}\nsp(\alpha)}{2\gamma\Delta_2^{}}\nsp\right)\!
-e^{\epsilon}_{} {Q}\!\left(\nsp\frac{\gamma\epsilon}{\Delta_2^{}}+\frac{\Delta_2^{}}{2\gamma} +\frac{K\mathcal{R}_{\Delta}^{}\nsp(\alpha)}{2\gamma\Delta_2^{}} +\frac{\theta\Delta_1^{}}{\gamma\Delta_2^{}}\nsp\right)\nsp\pspp$. 
By using this in \eqref{eq:balle_K_suff}, the theorem is proved.
\end{proof}

The performance of the flipped Huber mechanism under this sufficient condition is studied through simulations in Section \ref{sec:res_K_D}, where its efficacy is compared with the necessary and sufficient condition through the numerical iterative integration technique \cite{sadeghi2022offset}. 
%
\begin{figure*}[h!]
\begin{subfigure}{0.42\linewidth}
\centering
\includegraphics[width=\linewidth]{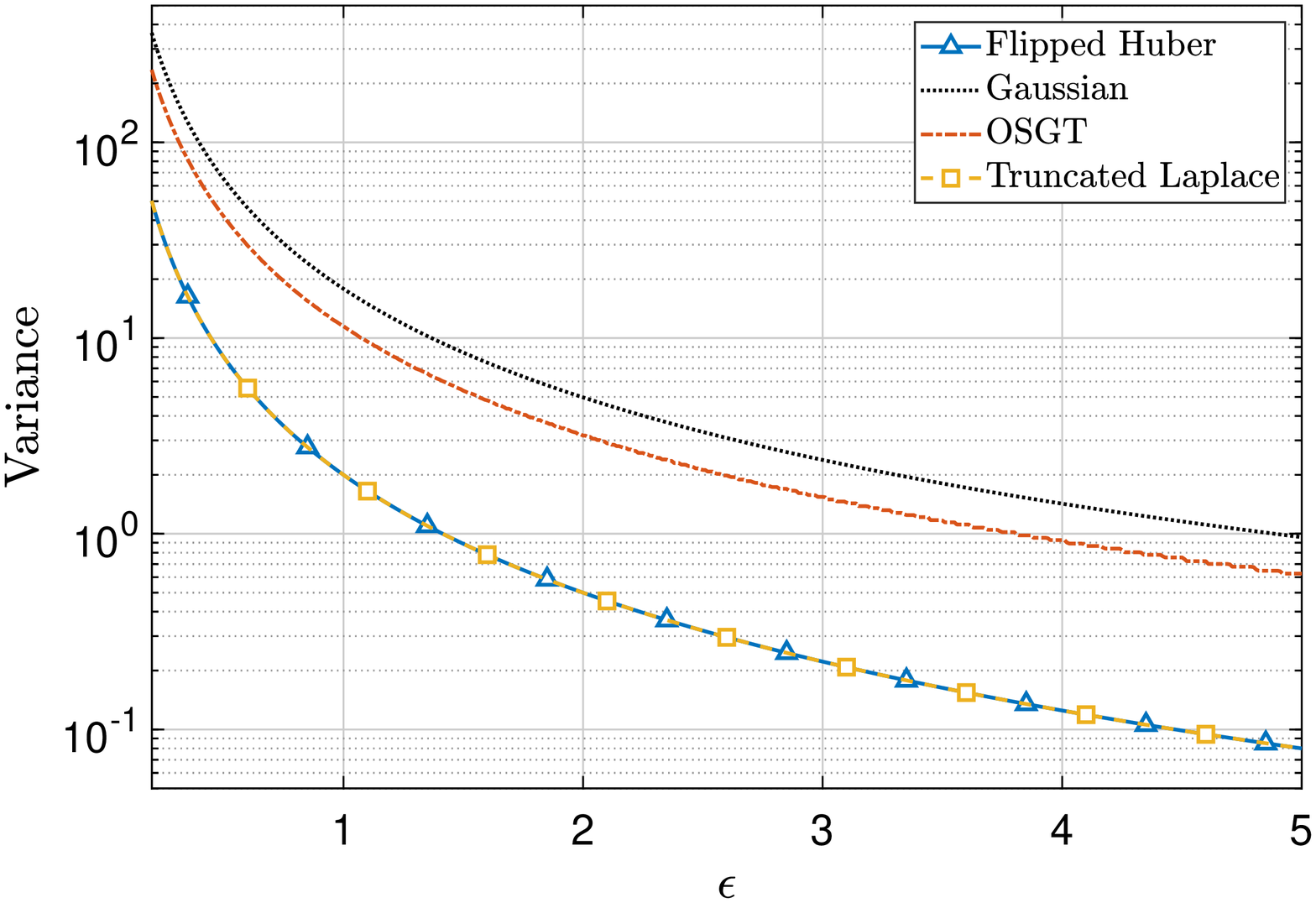}
\caption{}
\label{fig:priv_1_dim_a}
\end{subfigure}
\begin{subfigure}{0.29\linewidth}
\centering
\includegraphics[width=\linewidth]{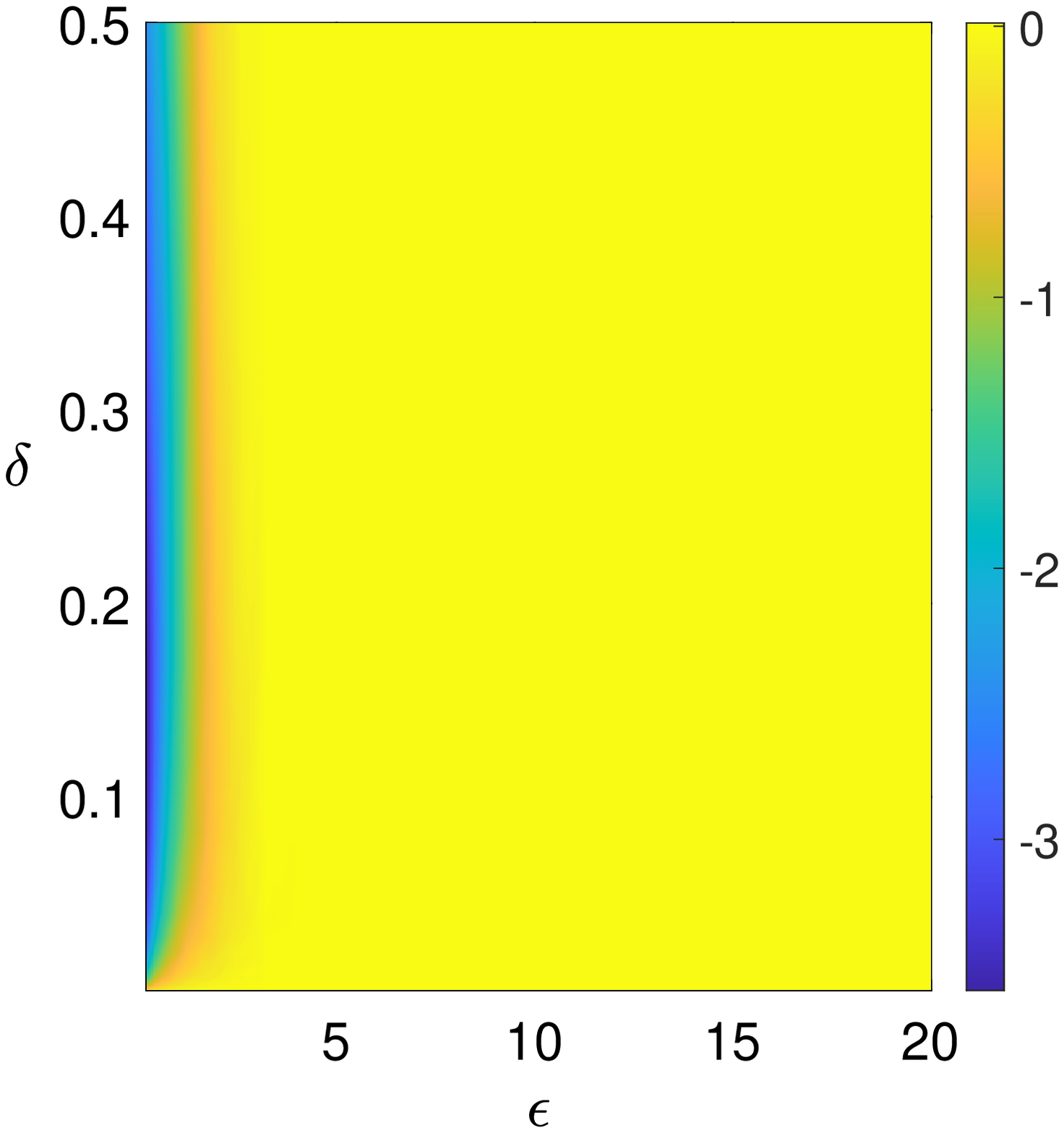}
\caption{}
\vspace{-0.15em}
\label{fig:priv_1_dim_b}
\end{subfigure}
\begin{subfigure}{0.29\linewidth}
\centering
\includegraphics[width=\linewidth]{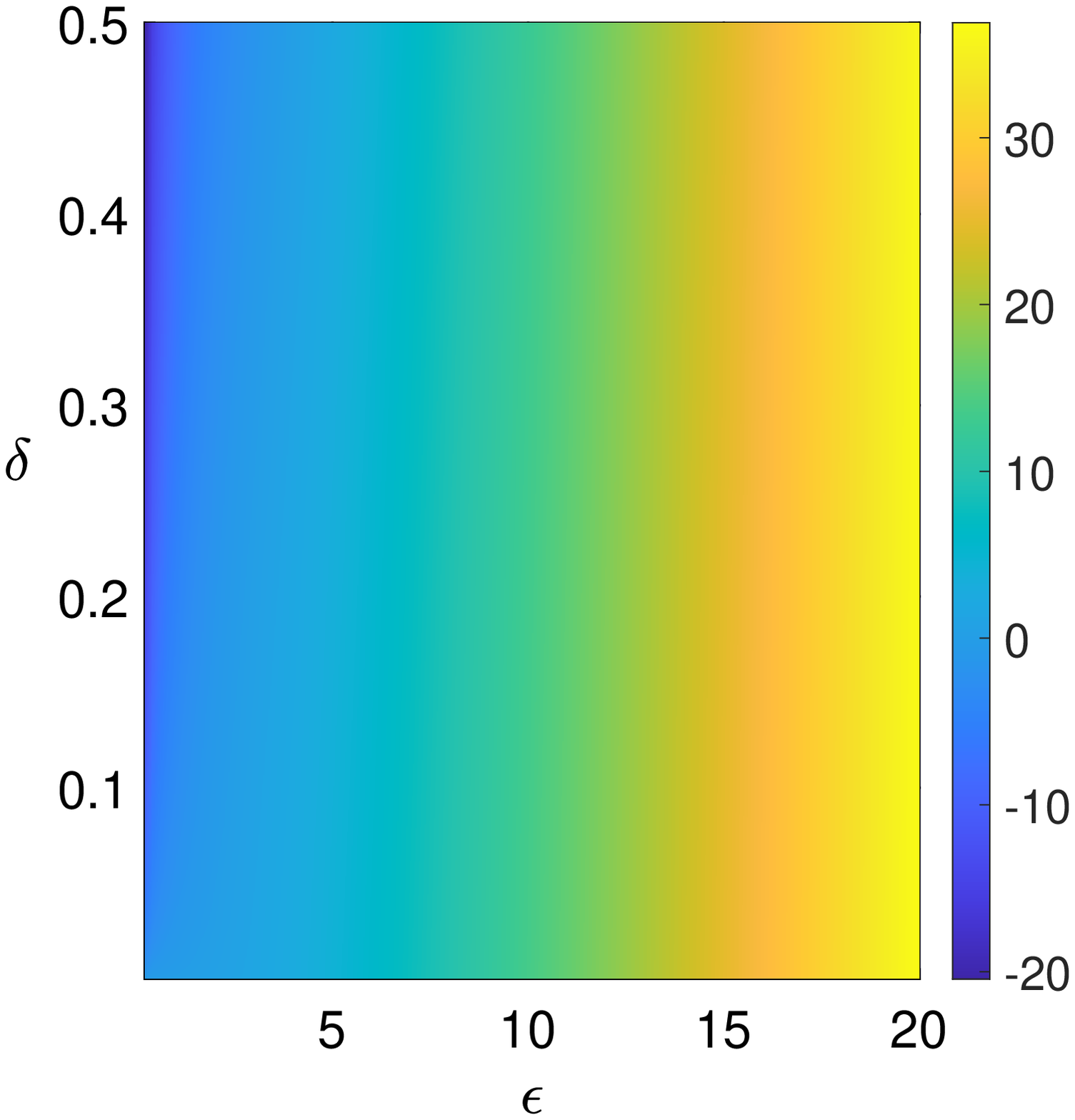}
\caption{}
\vspace{-0.15em}
\label{fig:priv_1_dim_c}
\end{subfigure}
\caption{Performance of various differentially private noise mechanisms in single dimension for $\Delta=1\pspp$: 
(a) Variances of flipped Huber, Gaussian, truncated Laplace and OSGT noises for $\delta=10^{-6}\nspp\pspp$, and the ratio of the variance of flipped Huber noise to that of (b) Laplace and (c) staircase noises in decibel scale.}
\label{fig:priv_1_dim}
\end{figure*}
%
\section{Empirical validation}\label{sec:emp_result}

In this section, we validate the theoretical results derived in the previous section through simulations and clearly show the advantage of the proposed mechanism. 
We compare the performance of our flipped Huber mechanism 
with the Gaussian \cite{balle2018improving}, 
Laplace \cite{dwork2006calibrating,vinterbo2022differential}, 
staircase \cite{geng2016optimalstaircase,geng2015staircase}, 
truncated Laplace \cite{geng2020tighttrunclapl}
and 
Offset Symmetric Gaussian Tails (OSGT) \cite{sadeghi2022offset} mechanisms. 
%
%
A few additional results and analyses are provided 
in Appendix \ref{appx:addn_res}.

\subsection{One dimension}\label{sec:res_1_D}

Firstly, we demonstrate the performance of the single-dimensional flipped Huber mechanism. We consider the following simulation setup. For a given $\delta$ and $\epsilon\pspp$, 
parameters satisfying $(\epsilon,\delta)$-DP are identified for each mechanism, and the minimum variances achieved under such parameters are compared. Such a comparison would reveal which mechanisms are better in terms of the privacy-accuracy trade-off. 

We first compare the proposed mechanism with other sub-Gaussian mechanisms - Gaussian, truncated Laplace and OSGT. 
The Gaussian mechanism is the most popular choice, and truncated Laplace is the optimal $(\epsilon,\delta)$-DP mechanism in one dimension, resulting in the smallest variance in the high-privacy regime \cite[Theorem 7]{geng2020tighttrunclapl}. 
We set $\delta=10^{-6}_{}$ and the sensitivity as $\Delta=1$. For flipped Huber and OSGT mechanisms, we perform grid search for $(\alpha,\gamma)$ and $({\vartheta},{\varrho})$ 
in $[0.02,150]\times[0.02,50]$ to identify the parameters conforming with given $(\epsilon,\delta)$ and for Gaussian mechanism, we use the binary search algorithm presented in \cite{balle2018improving} to determine the scale parameter. 

The results are plotted in Fig.  \ref{fig:priv_1_dim_a}, where we show the variance of noise added by these mechanisms for various $\epsilon$. It can be observed that the flipped Huber mechanism performs identically to the optimal truncated Laplace mechanism. 
We observe that the proposed mechanism outperforms Gaussian 
by a large margin; it also performs better than OSGT. For instance, when $\epsilon=0.3\pspp$, the variances of noises added by the Gaussian and OSGT mechanisms are respectively $168.80$ and $108.94\pspp$, whereas flipped Huber and truncated Laplace mechanisms add noises only of variance $22.21$. This reduction in the variance of flipped Huber noise improves with $\epsilon\pspp$: when $\epsilon=3\pspp$, there is roughly a 71-fold reduction in the variances of Gaussian and OSGT noises compared to the case of $\epsilon=0.3$ (to $2.38$ and $1.54\pspp$, respectively), whereas, for flipped Huber and truncated Laplace, the variances decrease by 100 times. 
This improved gain in the low-privacy regime is in accordance with the observation in \cite[Section 6]{geng2020tighttrunclapl}.
%
The results clearly demonstrate the benefit of improving Fisher information of noise density.
The intuitive explanation of why flipped Huber outperforms Gaussian was provided in Section \ref{sec:priv_1_dim}.  
In Appendix \ref{appx:addn_res_osgt}, we explain why flipped Huber is performing better than OSGT. 

In Figs. \ref{fig:priv_1_dim_b} and \ref{fig:priv_1_dim_c}, we compare flipped Huber mechanism with $(\epsilon,\delta)$-DP Laplace\cite{balle2020privacy,vinterbo2022differential} and $\epsilon$-DP staircase \cite{geng2016optimalstaircase} mechanisms\footnote{Since the $(\epsilon,\delta)$-DP 
condition for staircase mechanism is not 
studied in the literature, we compare $(\epsilon,\delta)$-DP flipped Huber and $\epsilon$-DP staircase mechanisms.}; the figures show the ratio of the variance of flipped Huber noise to those of Laplace and staircase noises for a range of $\epsilon$ and $\delta$ in decibel scale as heat maps. 
The staircase mechanism is the optimal $\epsilon$-DP mechanism in one dimension, and Laplace is optimal in the high-privacy regime 
\cite{geng2016optimalstaircase,geng2016optimaluniflap}. We observe that the flipped Huber noise variance is identical to that of Laplace, except for the region of small $\epsilon$ and 
$\delta>0\pspp$, where flipped Huber noise is better.  
However, our mechanism performs worse 
than the staircase mechanism in the low-privacy regime ($\epsilon \to \infty$). 
In the following subsection, we show that the staircase mechanism, similar to Laplace, performs poorly in high dimensions 
compared to the flipped Huber mechanism.
\begin{figure*}[h!]
\centering
\begin{subfigure}{0.42\linewidth}
\centering
\includegraphics[width=\linewidth]{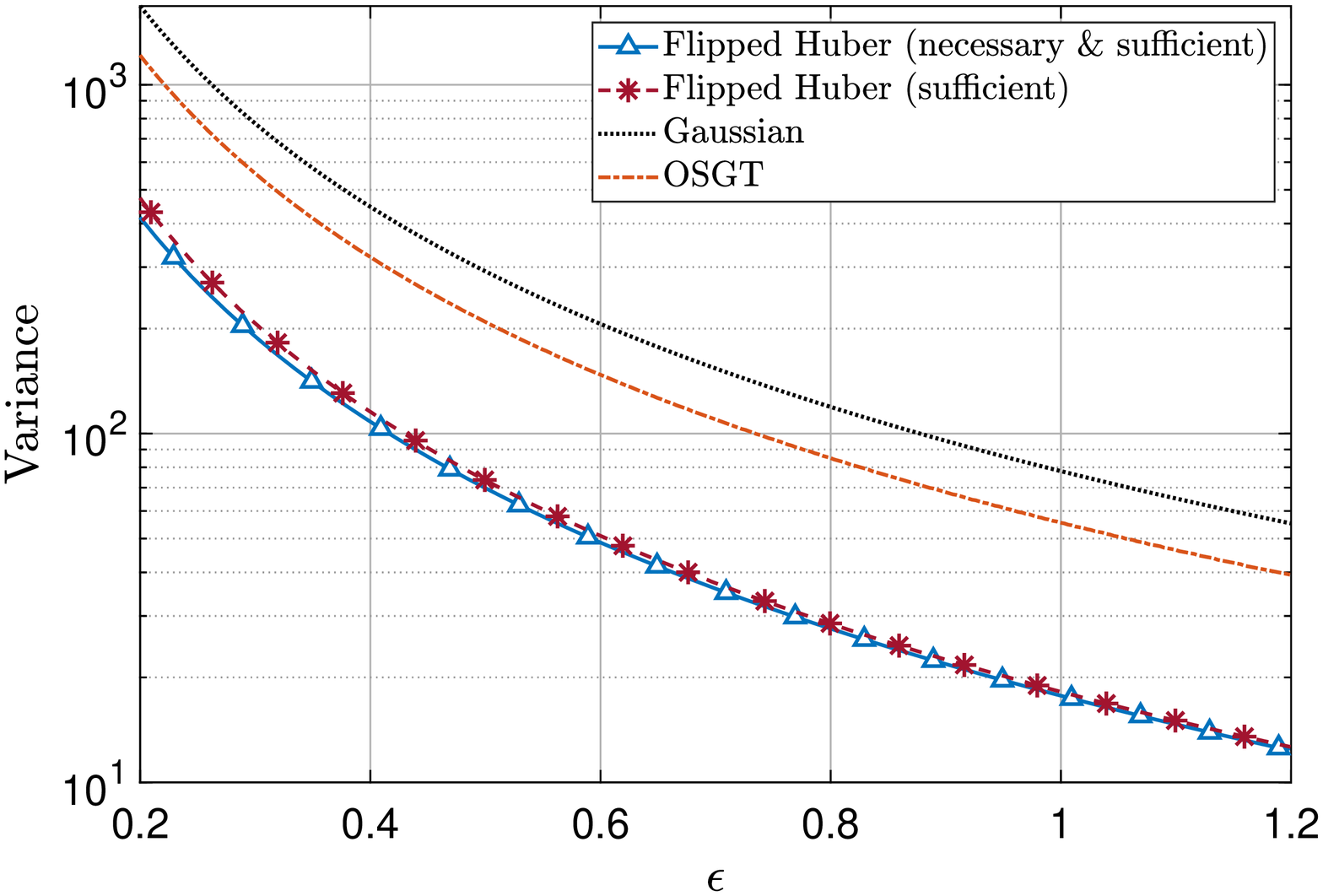}
\caption{$K=3\pspp$.}
\label{fig:priv_3_dim}
\end{subfigure}
\hspace{4em}
\begin{subfigure}{0.42\linewidth}
\centering
\includegraphics[width=\linewidth]{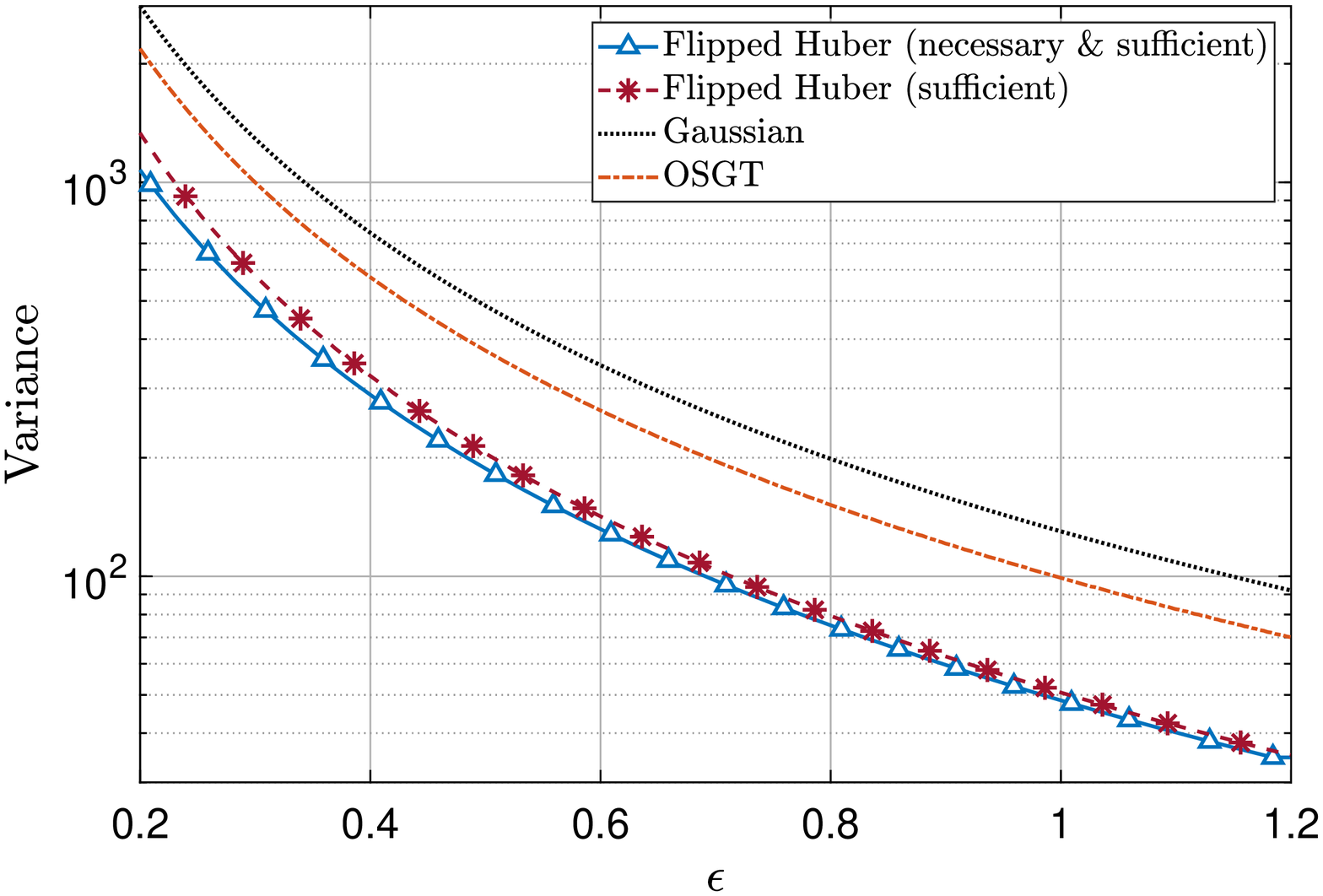}
\caption{$K=5\pspp$.}
\label{fig:priv_5_dim}
\end{subfigure}
\caption{Performance of various differentially private noise mechanisms in $K$ dimensions for $\delta=10^{-8}\nspp\pspp$, $\Delta=1\pspp$, $\Delta_2^{}=\sqrt{K}$ and $\Delta_1^{}=K$. 
}
\label{fig:priv_high_dim}
\end{figure*}
\subsection{$K$-dimensions}\label{sec:res_K_D}
We illustrate the performance of flipped Huber mechanism in $K$-dimensions under the necessary and sufficient condition as well as the sufficient condition derived in Section \ref{sec:priv_K_dim}. 
The simulation setup is similar to that of the one-dimensional case. 
We consider $\delta=10^{-8}_{}$ and set $\Delta_2^{}=\sqrt{K}\Delta$ and $\Delta_1^{}={K}\Delta$ based on the edge cases of the norm equivalence \eqref{eq:norm_equiv}, and the per coordinate sensitivity (or the $\ell_\infty^{}$-sensitivity) is set as $\Delta=1\pspp$. 
When obtaining the results of the flipped Huber 
under the sufficient condition, for each $\gamma$ in the search region, we restrict the search region for $\alpha $ as dictated by Theorem \ref{thm:priv_K_dim}. 


As discussed before, the necessary and sufficient condition for $K>1$ is very difficult to obtain in closed form for flipped Huber. We can perform iterative integration and grid search, as explained in \cite{sadeghi2022offset}, to obtain parameters satisfying privacy constraints. Algorithm \ref{alg:iter_int} presents this method for the flipped Huber mechanism with our notational convention. We exclude small values for $\alpha$ and $\gamma$ in the grid search as they were observed to cause instability in the numerical integration. For the Gaussian case, since the necessary and sufficient condition is available in closed form, we present its results, and for OSGT, we report the results from iterative integration.
%
\begin{algorithm}[h]
\caption{Determining parameters for $K$-dimensional flipped Huber mechanism.}
\label{alg:iter_int}
\textbf{Input}: Set of grid points $\mathcal{G}\subseteq\mathbb{R}_{++}^2$ for $(\alpha,\gamma)\pspp$, set of grid points $\mathscr{E}\subseteq \mathbb{R}$ for $\epsilon$ for numerical integration, per coordinate sensitivity $\Delta\pspp$. 
\begin{algorithmic}[1]
\For{each $(\alpha,\gamma)\in\mathcal{G}\pspp$, ${\nu}\in\mathscr{E}$}
\State $\delta^{( 1)}_{}\nsp({\nu})\nsp\gets\nsp\overline{{G}}_{\mathcal{FH}}^{}\nsp\big(\nspp \widetilde{{\zeta}}^{-1}_{\Delta}\nsp({\nu}) \nsp-\nsp \tfrac{\Delta}{2}\nsp\big)\!
- e^{\pspp {\nu}}_{} \overline{{G}}_{\mathcal{FH}}^{}\nsp\big(\nspp \widetilde{{\zeta}}^{-1}_{\Delta}\nsp({\nu}) \nsp+\nsp \tfrac{\Delta}{2}\nsp\big)\!\psp$. 
\For{$k=2,\,3,\,\ldots,\, K$, \textbf{for} each ${\nu} \in \mathscr{E}$}
\State $\delta^{(k)}_{}\nsp({\nu}) \gets \int_{-\infty}^{\infty} {g}_{\mathcal{FH}}^{}(t)\,\delta^{(k-1)}_{}\!\left(\nsp\epsilon-\widetilde{\zeta}_{\pspp \Delta}^{}\nsp\big( t\nspp+\nspp\tfrac{\Delta}{2}\nsp\big)\!\nspp\right)\! \,\mathrm{d}t\pspp$. 
\EndFor
\EndFor
\State Determine the set of points $\mathcal{G}^* \subseteq \mathcal{G}$ for which $\delta^{(K)}_{}\nsp(\epsilon)\leq\delta\pspp$.
\end{algorithmic}
\textbf{Output}: $(\alpha,\gamma)\in\mathcal{G}^*$ that results in minimum variance.
\end{algorithm}

The results of the simulation are presented in Fig.  \ref{fig:priv_high_dim} for $K=3$ and $K=5$. The flipped Huber results in lesser noise variance for fixed $(\epsilon,\delta)$ compared to Gaussian and OSGT in higher dimensions also. 
However, we observe that the improvement over Gaussian reduces with dimension. For example, compared to $K=3$, the improvement of flipped Huber over Gaussian reduces by $1.7\mathrm{dB}$ at $\epsilon=0.3\pspp$.

The derived sufficient condition for the flipped Huber performs close to the necessary and sufficient condition,  especially when the $\epsilon$ is not very small. For instance, when $K=5$, for $\epsilon=0.3$ and the necessary and sufficient condition results in the variance of $502$ and the noise variance using sufficient condition is $557\pspp$. However, the difference between variance reduces to less than 2 for $\epsilon=1\pspp$. Also, note that the gap between the variances corresponding to sufficient and necessary and sufficient condition 
increases with the dimension.

Now, we provide the comparison of flipped Huber mechanism with Laplace and staircase mechanisms for the same simulation setup but in $K=20$ dimensions. The results are provided in Table \ref{tab:fh_vs_lap_Kd}. Note that since iterative integration is computationally very difficult to perform for such a large $K$, we compare 
only with flipped Huber results corresponding to the sufficient condition. 
Also, for the staircase mechanism, we have considered independent noise 
samples\footnote{Since the noise variance for arbitrary dimensional staircase mechanism is not characterized, we consider independent noise coordinates.} across the coordinates\cite{geng2015staircasearxiv}. 
We see that the variance of noise added by the flipped Huber mechanism is significantly lesser than both these mechanisms, unlike what we observed for the case of $K=1\pspp$.
%
%
\setlength{\tabcolsep}{2.8pt}
\begin{table}[h!] 
\caption{Variance of noise added by flipped Huber and Laplace mechanisms in 20 dimensions.\vspace{-2ex} 
}
\label{tab:fh_vs_lap_Kd}
\begin{center}
\begin{tabular}{?C{6.5em}?C{3.75em}|C{3.75em}|C{3.75em}|C{3.75em}|C{3.75em}?} 
\specialrule{1pt}{0pt}{0pt}
$\epsilon$ 					 & 0.2      & 0.4       & 1       & 2.2       & 5    \\
\specialrule{1pt}{0pt}{0pt}
\begin{tabular}{c}
\\[-2.5em]	Flipped Huber \\[-0.5em] (sufficient)\\[0.1em]
\end{tabular} & {7237.09}  & {1971.36}  & {359.57}   & {87.09}   & {19.49}\\
\hline
\begin{tabular}{c}
\\[-1.75em]	Laplace \\[1em] 
\end{tabular} & {$2\nspp\cdot\nspp10^{4}_{}$}  & {5000}  & {800}   & {165.29}   & {32}\\\hline
\begin{tabular}{c}
\\[-2.5em]	Staircase \\[-0.5em] (independent)\\[0.1em]
\end{tabular}& {19999.92}  & {4999.92}  & {799.92}   & {165.21}   & {31.92}\\
\specialrule{1pt}{0pt}{0pt}
\end{tabular}
\end{center}
\end{table}
%
%
%
%

\section{Application and Discussions} \label{sec:appn_discussion}
In this section, we illustrate the utility 
of the proposed mechanism in a private machine learning setup.
We also discuss the strengths and 
limitations of the flipped Huber mechanism. 
First, we study the composition performance of the flipped Huber mechanism, which extends its scope to iterative algorithms.

\subsection{Flipped Huber under composition} \label{sec:appn_discussion_zcdp}
Several variants of DP have been proposed in the literature for easier analysis of privacy degradation under composition. The zCDP definition involves the sub-Gaussianity constraint on the privacy loss RV \cite{bun2016concentrated}.  
The following theorem characterizes the zCDP of the flipped Huber mechanism.

\begin{table*}[h!] 
	\caption{Performance of Gaussian and Flipped Huber noises on differentially private coordinate descent.}
	\label{tab:dpcd_FH_results}
	\begin{center}
		\begin{tabular}{? C{5em} ? C{9em} | C{7em} ? C{6.5em} | C{6.5em} ? C{6.5em} | C{6.5em} ?} 
			%
			\clineB{2-7}{2}
			\multicolumn{1}{c?}{}   &   \multirow{2}{*}{\textbf{Dataset}}    & \multirow{2}{*}{\begin{tabular}{c} \textbf{Regularization}\\[-0.5em]\textbf{and parameter} \end{tabular}}
			&  \multicolumn{2}{c?}{\textbf{Gaussian}}  
			&  \multicolumn{2}{c?}{\textbf{Flipped Huber}} 	\\
			\clineB{4-7}{1}
			\multicolumn{1}{c?}{} & & & \textbf{NMSE} & \textbf{Test error} & \textbf{NMSE} & \textbf{Test error} 	\\
			\clineB{2-7}{2}
			\specialrule{1pt}{2pt}{0pt}
			%
			\multirow{4}{*}{
				\begin{tabular}{c}
					\textbf{Logistic} \\[-0.5em] \textbf{regression}
			\end{tabular}} 
			& Houses \cite{californiadataset} & $(\ell_2^{},\psp0.1)$ & $0.6371\nsp\cdot\nsp10^{-3}_{}$ & $0.0391$	& $0.6165\nsp\cdot\nsp10^{-3}_{}$	& $0.0389$		\\ 
			\clineB{2-7}{1}
			& Wine quality \cite{winedataset} & $(\ell_2^{},\psp2\nsp\cdot\nsp10^{-4}_{})$ & $0.2250$ & $0.0614$ & $0.1421$	& $0.0513$		\\
			\clineB{2-7}{1}
			& Pumpkin seeds \cite{pumpkinseedsdataset} & $(\ell_2^{},\psp0.1)$  & $0.0255$ & $0.1224$	& $0.0189$	& $0.1152$		\\
			\clineB{2-7}{1}
			& Heart \cite{heartdataset} & $(\ell_2^{},\psp0.1)$ & $0.2384$ & $0.1741$ & $0.1989$ & $0.1556$	\\ 
			%
			\clineB{1-7}{2}
			\specialrule{1pt}{2pt}{0pt}
			%
			\multirow{4}{*}{
				\begin{tabular}{c}
					\textbf{Linear} \\[-0.5em] \textbf{regression}
			\end{tabular}} 
			& California \cite{californiadataset} & $(\ell_1^{},\psp0.01)$ & $0.0479$ & $0.4532$	& $0.0465$	& $0.4298$	\\
			\clineB{2-7}{1}
			& Boston housing \cite{bostonhousingdataset} & $(\ell_1^{},\psp0.01)$ & $0.3579$ & $0.3743$	& $0.3406$	& $0.3253$		\\
			\clineB{2-7}{1}
			& Airfoil \cite{airfoildataset} & $(\ell_1^{},\psp0.01)$ & $0.0206$ & $0.5161$	& $0.0190$	& $0.4558$		\\
			\clineB{2-7}{1}
			& Diabetes \cite{diabetesdataset} & $(\ell_1^{},\psp0.1)$ & $0.2515$ & $0.5741$ & $0.1489$ & $0.4384$		\\ 
			\clineB{1-7}{2}
			%
		\end{tabular}
	\end{center}
\end{table*}
\begin{theorem}\label{thm:zCDP}
The single-dimensional flipped Huber mechanism guarantees 
$({\xi},{\eta})$ zero concentrated differential privacy 
for ${\xi}=\tfrac{\mathcal{R}_{\Delta}^{}\nsp(\alpha)}{2\gamma^{2}_{}}$ and ${\eta}=\tfrac{\pspp \Delta^2_{}}{2\gamma^2_{}}\pspp$, where $\mathcal{R}_{\Delta}^{}\nsp(\alpha)=
\alpha^2_{}-(\pspp[\alpha-\Delta]_{\nspp +}^{})^{ 2}_{}$.
\end{theorem}

\begin{proof}
We again consider the upper bound $\plossupp\nsp(t)$ 
for ${\zeta}_{\pspp d}^{}\nsp(t)\pspp$; from Lemma \ref{lem:zosd}, we have ${\zeta}_{\pspp d}^{}\nsp(T) \uso \plossupp\nsp(T)\pspp$. 
Since stochastic ordering manifests to the ordering of expectations of increasing functions of RVs \cite[eq. (1.A.7)]{shaked2007stochastic},  
we get
\begin{align*}
\mathbb{M}_{{\zeta}_{\pspp d}^{}\nsp(T)}^{}({s}) &= \mathbb{E}\Big[e^{{s}\pspp{\zeta}_{\pspp d}^{}\nsp(T)}\Big] 
\leq \mathbb{E}\Big[e^{{s}\pspp\plossupp\nsp(T)}\Big]
\\ &
= \exp\!\left(\nsp s\tfrac{\mathcal{R}_{\abs{d}}^{}\nsp(\alpha)}{2\gamma^{2}_{}}+s\tfrac{ d^2_{}}{2\gamma^2_{}}\nsp\right)\!\times \mathbb{E}\Big[\exp\!\left(\nsp\tfrac{s\pspp d}{\gamma^2_{}}T\nsp\right)\Big]
\\&
\leq \exp\!\left(\nsp s\tfrac{\mathcal{R}_{\abs{d}}^{}\nsp(\alpha)}{2\gamma^{2}_{}}+s(s+1)\tfrac{\pspp d^2_{}}{2\gamma^2_{}}\nsp\right)\!
\\ &
\leq \exp\!\left(\nsp s\tfrac{\mathcal{R}_{\Delta}^{}\nsp(\alpha)}{2\gamma^{2}_{}}+s(s+1)\tfrac{\pspp \Delta^2_{}}{2\gamma^2_{}}\nsp\right)\!
\nspp\pspp,
\end{align*}
where the penultimate inequality is due to the sub-Gaussianity of $T\sim\mathcal{FH}(\alpha,\gamma^2)$ from Lemma \ref{lem:sub_gau}. Thus, one dimensional flipped Huber mechanism is $\!\left(\nsp \tfrac{\mathcal{R}_{\Delta}^{}\nsp(\alpha)}{2\gamma^{2}_{}},\tfrac{\Delta^2_{}}{2\gamma^2_{}}\nsp\right)\nsp$-zCDP.
\end{proof}
The $L$-fold composition of $({\xi}_l^{},{\eta}_l^{})$-zCDP mechanisms, $l=1,\,2,\,\ldots,\,{L}$, where the output of $l$-th mechanism shall depend on the previous $l-1$ mechanisms' outputs, is $\!\left(\nsp \sum_{l=1}^{L}{\xi}_l^{},  \sum_{l=1}^{L}{\eta}_l^{}\nsp\right)\nsp$-zCDP \cite[Lemma 2.3]{bun2016concentrated}.
In the following, we use this result to account for the privacy of flipped Huber mechanism in an application. 
Since the $K$-dimensional query can be interpreted as a (non-adaptive) composition of $K$ single-dimensional queries, 
the $K$-dimensional flipped Huber mechanism is
$\!\left(\nsp \tfrac{K\mathcal{R}_{\Delta}^{}\nsp(\alpha)}{2\gamma^{2}_{}},\tfrac{\Delta^2_{2}}{2\gamma^2_{}}\nsp\right)\nsp$-zCDP. 

\subsection{Flipped Huber for private coordinate descent} \label{sec:appn_dpcd}

We illustrate the utility of the proposed mechanism in empirical risk minimization through DP coordinate descent (DP-CD) \cite{mangold2022differentially} for training machine learning models, where differential privacy is guaranteed through the perturbation of gradient updates. Coordinate descent typically requires 
fewer data passes to learn the parameter than gradient descent, as learning rates for each of the coordinates in the parameter can be adapted to the coordinate-wise smoothness of the objective function and are typically large.

We perform $L$ batches of coordinate descents, where on each batch, the gradient update on the $i$-th coordinate, $i=1,\,2,\,\ldots,\,K$, which has  sensitivity ${\lambda}_i^{}$, is perturbed by noise. Additional information on the DP-CD algorithm is provided in Appendix \ref{appx:appn}.  
%
%
%
%
%
%
For a given privacy constraints $(\epsilon,\delta)$, zCDP parameters $({\xi},{\eta})$ can be determined using the result \cite[Lemma 3.5]{bun2016concentrated}. By invoking the composition rule, the zCDP parameters for each batch can be selected as ${\xi}_0^{}=\frac{{\xi}}{L}$ and ${\eta}_0^{}=\frac{{\eta}}{L}$; from Theorem \ref{thm:zCDP}, the corresponding noise parameters for the flipped Huber distribution are $\alpha_0^{}=\mathcal{R}_{\Delta}^{-1}\nsp\Big(\frac{\Delta_2^2{\xi}_0^{}}{K{\eta}_0^{}}\Big)\nsp$ and $\gamma_0^{}=\frac{\Delta_2^{}}{\sqrt{2{\eta}_0^{}}}$, where $\Delta=\norm{\boldsymbol{{\lambda}}}_{\infty}^{}$ and $\Delta_2^{}=\norm{\boldsymbol{{\lambda}}}_{2}^{}$.

Among the possible pairs of $({\xi},{\eta})$ resulting in $(\epsilon,\delta)$-DP, we select the pair that results in minimum variance for $\mathcal{FH}(\alpha_0^{},\gamma^2_{0})$ through grid search.
After selecting suitable $({\xi},{\eta})$, we further adapt the noise parameters to each coordinate as done for Gaussian in \cite{mangold2022differentially}. Consider the per coordinate zCDP parameters as ${\xi}_i^{}=\frac{{\xi}_0^{}}{K}$ and ${\eta}_i^{}=\frac{{\eta}_0^{}}{K}, \ i=1,\,2,\,\ldots,\,K$. 
Using Theorem \ref{thm:zCDP}, we can express the per coordinate noise parameters as $\alpha_i^{}=\lambda_i^{}\mathcal{R}_{1}^{-1}\nsp\Big(\frac{{\xi}}{{\eta}}\Big)\nsp$ and $\gamma_i^{}=\lambda_i^{}\sqrt{\frac{KL}{2{\eta}}}$. Thus, when updating the $i$-th coordinate of the parameter to be learnt, we add noise sampled from $\mathcal{FH}(\alpha_i^{},\gamma^2_{i})$.


We perform DP-CD for logistic regression with $\ell_2^{}$ regularization 
and linear regression with $\ell_1^{}$ regularization (i.e., LASSO)  over a wide range of real datasets.
The results are presented in Table \ref{tab:dpcd_FH_results}, where we compare 
our mechanism with Gaussian\footnote{We limited our comparison only to Gaussian since, to our best knowledge, the composition results of other mechanisms considered in Section V are unavailable in amenable form for the derivation of coordinate-wise parameters.}.
Each dataset is partitioned into $80\%$ training data and $20\%$ testing data; the parameter is learned from training data, and we report the following metrics over test data $\{(\widetilde{\mathbf{x}}_n^{},\widetilde{y}_n^{})\}_{n=1}^{\widetilde{N}}$. 
\begin{enumerate}[label=(\roman*)]
\item Normalized mean squared error (NMSE), 
$\frac{{\Vert{\widehat{\boldsymbol{{\theta}}}-\boldsymbol{{\theta}^*_{}}}\Vert_2^{2}}}{\Vert\boldsymbol{{\theta}^*_{}}\Vert_2^2}$, which measures how close the private estimate $\widehat{\boldsymbol{{\theta}}}$ is close to the (non-private) optimum $\boldsymbol{{\theta}^*_{}}$.
\item Test error, which gauges the performance of the estimates on the test data, is computed as 
\begin{enumerate}[label=(\alph*)]
\item Misclassification error, $\frac{1}{\widetilde{N}_{}^{}}\sum_{n=1}^{\widetilde{N}}\mathbb{I}\big\{\widetilde{y}_n^{}\neq\sgn\big(\widetilde{\mathbf{x}}_n^{\top}\widehat{\boldsymbol{{\theta}}}\big)\big\}$ for logistic regression, and
\item Normalized residual sum of squares (RSS), 
%
$\sum_{n=1}^{\widetilde{N}} \big(\widetilde{{y}}_n^{}-\widetilde{\mathbf{x}}_n^{\top}\widehat{\boldsymbol{{\theta}}}\big)^2_{}\Big/ \sum_{n=1}^{\widetilde{N}} \widetilde{{y}}_n^{2} $
for linear regression. 
\end{enumerate} 
\end{enumerate}

For Gaussian and Flipped Huber mechanisms, we perform five random trials for each combination of hyperparameters, and the best NMSE averaged across the trials and the corresponding test errors are reported.
From the results in Table \ref{tab:dpcd_FH_results}, we observe that the flipped Huber mechanism outperforms Gaussian in all the datasets considered, both in terms of NMSE and test error. Particularly, for logistic regression, we observe an $8.29\%$ improvement in NMSE over Gaussian in the Wine quality dataset and misclassification error is reduced by $1.85\%$ for the Heart dataset; for linear regression, flipped Huber offers $10.26\%$ lesser NMSE and $13.57\%$ lesser normalized RSS in the Diabetes dataset compared to Gaussian. 
\begin{table*}[h!]
\centering
\setlength{\tabcolsep}{0pt}
\begin{tabular}{p{\textwidth}}
\begin{equation}\label{eq:flip_hub_inv_cdf}
{G}_{\mathcal{FH}}^{-1}(u;\alpha,\gamma^2) = 
\begin{cases}
\\[-2.75em]
-\frac{\gamma^2_{}}{\alpha}\sgn(2u-1) \log\!\left(\nspp 1-\frac{\alpha\pspp{\omega}}{2\gamma}|2u-1|\psp e^{-\alpha^2/2\gamma^2_{}}\nspp\right)\!
\, , &  
u \wedge (1-u) \geq \frac{\sqrt{2\pi}}{{\omega}} {Q}\!\left(\nsp\frac{\alpha}{\gamma}\nsp\right) \\
\gamma\sgn(2u-1)\psp {Q}^{-1}\!\nsp\left(\nsp\frac{{\omega}(1-|2u-1|)}{2\sqrt{2\pi}}\nsp\right)\!
\, , &
\text{otherwise} \\
\end{cases}
\psp.
\end{equation}
\\
\hline
\end{tabular}
\end{table*}

\subsection{Discussions}

Throughout the article, we have observed that the flipped Huber mechanism provides better utility compared to several other mechanisms in a wide range of settings. 
This shall be attributed to the lighter tails and the sharper centre of the noise density. Note that the truncated Laplace distribution has these traits and is also proven to be an optimal $(\epsilon,\delta)$-DP mechanism for one-dimensional queries in the high-privacy regime \cite{geng2020tighttrunclapl}. However, due to bounded support, truncated Laplace renders the perfect identification of neighbouring datasets possible with non-zero probability. Flipped Huber mechanism does not have this issue, and in Section \ref{sec:priv_1_dim}, we witnessed that it performs as good as truncated Laplace.

Due to the central limit phenomenon, the performance of every mechanism tends to be that of Gaussian in very high levels of composition \cite{sommer2019privacy,dong2022gaussian}. As more and more 
flipped Huber mechanisms get composed, the transition parameter $\alpha$ tends to get smaller, and the noise density approaches that of Gaussian; this is implicit from the 
requirement $\alpha\leq\mathcal{R}_{\Delta}^{-1}\nsp( ({2\gamma^2_{}\epsilon-\Delta_2^2})/{K})$
in Theorem \ref{thm:priv_K_dim}, which also appears if the zCDP guarantee is converted to $(\epsilon,\delta)$-DP (where we would need $\epsilon\geq{\xi}+{\eta}$ \cite{bun2016concentrated}).
However, if the application 
does not require a very large  
number of compositions, flipped Huber mechanism seems to perform significantly better than Gaussian; this was instantiated in the previous subsection for the case of coordinate descent, which typically requires a lesser number of data passes to converge. 

Another limitation of flipped Huber mechanism is that it simultaneously requires several measures of sensitivities. If some of them are not directly available, one can bound the unknown sensitivities with known ones using \eqref{eq:norm_equiv}; however, this might degrade the performance as these bounds are typically loose. In some cases, this can be handled by the clever setting of the problem; for instance, in DP-CD, coordinate-wise clipping enabled us to find all the required sensitivities directly.

\section{Conclusions}\label{sec:conc}
This article introduced the flipped Huber mechanism for differential privacy that perturbs query responses on datasets with noise sampled from a novel density which is a hybrid of Laplace and Gaussian densities and has the desired traits of both. 
Theoretical results had been derived for the flipped Huber mechanism, and we provided a necessary and sufficient condition in one dimension and a sufficient condition in $K$ dimensions for the mechanism to be $(\epsilon,\delta)$-DP. The performance of the flipped Huber mechanism under these conditions was empirically validated, and the results demonstrated that the proposed hybrid density 
provides more accuracy for the given privacy constraints compared to other existing mechanisms in a wide range of settings. 
%
We further demonstrated the utility of flipped Huber mechanism for empirical risk minimization through coordinate descent for learning the regression model parameters on several real-world datasets.


Given the current explosion of literature on differential privacy when differentially private versions of numerous algorithms are being proposed in different fields, we believe our work is critical. 
Construction of the new noise density from the statistical estimation formulation motivates us to explore other hand-crafted noises for differential privacy, which may possibly be better or more competitive to the existing ones, rather than restricting ourselves to well-known distributions.


\appendices

\section{Sampling from the flipped Huber distribution}\label{appx:sampling}
The cumulative distribution function of the flipped Huber distribution $\mathcal{FH}(\alpha,\gamma^2)$ is given  in \eqref{eq:flip_hub_cdf}.  Though the CDF is piecewise, its inverse can be obtained in closed form, and hence, the noise samples can be obtained Smirnov transform (also called the inversion sampling method) \cite{devroye1986non}.

The quantile function 
of 
$\mathcal{FH}(\alpha,\gamma^2)$ is given in \eqref{eq:flip_hub_inv_cdf}. In order to obtain i.i.d. samples  $t_i,\ i=1,\,2,\,\ldots,\,K$ from the flipped Huber distribution $\mathcal{FH}(\alpha,\gamma^2)\pspp$, the i.i.d. samples ${u }_i,\ i=1,\,2,\,\ldots,\,K$ are drawn from the uniform distribution $\mathcal{U}(0,1)\pspp$. Then, the  required samples can be obtained as $t_i={G}_{\mathcal{FH}}^{-1}({u }_i;\alpha,\gamma^2_{})\pspp$.

\section{Variance and Fisher information}\label{appx:var_fish_FH}
{Here, we provide the derivation of variance and Fisher information of the flipped Huber distribution and derive bounds on them. First, we observe that 
\begin{equation}\label{eq:bnd_sinh}
\sinh({a})\leq {a}\cosh({a}) \leq {a}\pspp e^{\pspp {a}}_{}\,\ \forall\, {a}\in\mathbb{R}_{+}^{}
\psp,
\end{equation}
where the first inequality is evident from the Taylor expansion, $ {a}\cosh({a}) = {a} \big(\nspp 1 {+} \tfrac{{a}^2}{2!} {+} \tfrac{{a}^4}{4!} {+} \cdots\nsp\big)\! \geq \!\big(\nspp {a} {+} \tfrac{{a}^3}{3!} {+} \tfrac{{a}^5}{5!} {+} \cdots\nsp\big)\! = \sinh({a}) \,\ \forall\, {a}\in{\mathbb{R}_{+}^{}}\pspp$; 
also, ${a}\pspp e^{\pspp {a}}_{} - {a}\cosh({a}) = {a}\sinh({a}) \geq 0 \,\ \forall\, {a}\in\mathbb{R}\pspp$.
}
Let ${T} \sim \mathcal{FH}(\alpha,\gamma^2)$ be the flipped Huber random variable. Due to the symmetry of the density function, we have $\mathbb{E}[T]=0\pspp$. Hence, the variance of ${T}$ is
\begin{equation*} 
\sigma_{\mathcal{FH}}^2
=\frac{1}{{\kappa}}\nsp\!\left[2\! \int_{0}^{\alpha}\! t^2e^{-\alpha t/\gamma^2_{}} \mathrm{d}t + 2e^{-\alpha^2/2\gamma^2_{}}\!\int_{\alpha}^{\infty}\! t^2e^{-t^2/2\gamma^2_{}} \mathrm{d}t\right]\!\nsp
\psp. 
\end{equation*}
\begin{align*}
\text{Now, } \int_{0}^{\alpha}\! & t^2e^{-\alpha t/\gamma^2_{}} \mathrm{d}t 
= 
-\nspp\!\left[e^{-\alpha t/\gamma^2_{}}\!\left(\nsp t^2\tfrac{\gamma^2_{}}{\alpha}+2t\tfrac{\gamma^4}{\alpha^2}+2\tfrac{\gamma^6}{\alpha^3}\nsp\right)\!\right]_0^{\alpha}
\\&
=\tfrac{4\gamma^6}{\alpha^3} e^{-\alpha^2/2\gamma^2_{}} \sinh\!\left(\nspp\tfrac{\alpha^2}{2\gamma^2_{}}\nspp\right)\!
-e^{-\alpha^2/\gamma^2_{}}\!\left(\nsp\alpha\gamma^2+\tfrac{2\gamma^4}{\alpha}\nsp\right)\!
\end{align*}
\begin{equation*}
\text{and }
\int_{\alpha}^{\infty}\! t^2e^{-t^2/2\gamma^2_{}} \mathrm{d}t
=\gamma^3\sqrt{2\pi}\psp {Q}\!\left(\nspp\tfrac{\alpha}{\gamma}\nspp\right)\!
+\alpha\gamma^2 e^{-\alpha^2/2\gamma^2_{}}\nsp
\psp. \hspace{10em}
\end{equation*}
Hence,
\begin{align*}
\sigma^2_{\mathcal{FH}}
&=\tfrac{2\gamma^2_{}}{\omega}\nsp\!\left[\sqrt{{2}{\pi}} \psp {Q}\!\left(\nspp\tfrac{\alpha}{\gamma}\nspp\right)\!
+\tfrac{4\gamma^3}{\alpha^3} \sinh\!\left(\nspp\tfrac{\alpha^2}{2\gamma^2_{}}\nspp\right)\!
+\tfrac{2\gamma}{\alpha} e^{-\alpha^2/2\gamma^2_{}}
\right]\!
\\&
=\gamma^2 \nspp\!\left[ 1\nspp-\nspp \tfrac{1}{{\omega}}\big(\tfrac{2\gamma}{\alpha}\big)^{\!3}_{}\pspp 
\!\left(\nspp\tfrac{\alpha^2}{2\gamma^2_{}}\cosh\!\left(\nspp\tfrac{\alpha^2}{2\gamma^2_{}}\nspp\right)\! -\sinh\!\left(\nspp\tfrac{\alpha^2}{2\gamma^2_{}}\nspp\right)\!\nspp\right)\!
\right]\!
\psp.
\end{align*}

\begin{table*}[h!]
\centering
\setlength{\tabcolsep}{0pt}
\begin{tabular}{p{\textwidth}}
\begin{equation}\label{eq:zeta_d_flip_hub}
\widetilde{{\zeta}}_{\pspp d}^{}\nsp(t)=
\begin{cases}
\\[-2.75em]
2\sgn(d)\frac{\alpha t}{\gamma^2_{}}\nsp
\, , & \hspace{-0.5em}
|t| \! < \!\left(\alpha \nsp - \nsp \frac{|d|}{2}\nsp\right)\!\wedge\!\frac{|d|}{2}\\[-0.35em]
\frac{\sgn(t)}{\gamma^2_{}}\alpha d\psp\nsp
\, , & \hspace{-0.5em}
|t| \! \in \!\Big[ \frac{|d|}{2}, \alpha \nsp - \nsp \frac{|d|}{2}\nsp\Big)\!\\[-0.35em]
\frac{\sgn(t)}{\gamma^2_{}}\nsp \!\left[\nspp\frac{\sgn(d)}{2}\nspp\!\left(\nspp|t|+\alpha+\frac{|d|}{2}\nspp\right)^{\nsp2}\!-\alpha d\pspp\right]\! \nsp
\, , & \hspace{-0.5em}
|t| \! \in \!\Big[\!\nspp \abs{\nspp\alpha \nsp - \nsp \frac{|d|}{2}\nsp\nspp}\!, \frac{|d|}{2}\nsp \Big)\!\\[-0.35em]
\frac{\sgn(t)}{\gamma^2_{}}\nsp \!\left[\nspp\frac{\sgn(d)}{2}\nspp\!\left(\nspp|t|-\alpha+\frac{|d|}{2}\nspp\right)^{\nsp2}\!+\alpha d\pspp\right]\! \nsp
\, , & \hspace{-0.5em}
|t| \! \in \!\Big[\!\nsp\left(\nspp\alpha \nsp - \nsp \frac{|d|}{2}\nspp\nspp\right)\! \vee \!\frac{|d|}{2}, \alpha \nsp + \nsp \frac{|d|}{2} \nsp\Big)\!\!\! \\[-0.35em]
\frac{td}{\gamma^2_{}}\nsp
\, , & \hspace{-0.5em}
|t| \! \in \!\mathbb{R}_{+}^{}\!\nspp\left\backslash\nspp\Big[\nspp\frac{|d|}{2} \nsp - \nsp \alpha,\frac{|d|}{2} \nsp + \nsp \alpha \nspp\Big)\right.\!\!\\
\end{cases}
\psp.
\end{equation}
\\
\hline
\end{tabular}
\end{table*}

Now, we derive the Fisher information of the flipped Huber distribution  $\mathcal{FH}(\alpha,\gamma^2)\pspp$. 
Let $T$ be 
the flipped Huber 
random variable, $T\sim \mathcal{FH}(\alpha,\gamma^2)$ and let %
${V}={z}+{T}$ be the noisy observation of the 
`location' parameter ${z}\pspp$. Thus, 
the random variable ${V}$ has the density ${g}_{\mathcal{FH}}^{}({v}-\theta;\alpha,\gamma^2)\pspp$. The Fisher information about the location parameter ${z}$ contained in the random observation of %
${V}$ is
\begin{align*}
\mathcal{I}_{\mathcal{FH}}^{}
&
= \mathbb{E}\nsp\!\left[\!\left(\nsp\frac{\partial}{\partial\theta}\nsp\log {g}_{\mathcal{FH}}^{}({V}\!-\nspp\theta;\alpha,\gamma^2)\nsp\right)^{\!\!2}\right]\!\nsp
= \mathbb{E}\nsp\!\left[\frac{(\rho_{\alpha}'\nspp({V}\!-\nspp\theta)\nspp)^2}{\gamma^4}\right]\!
\\&
=\frac{1}{{\kappa}} \!\int_{-\infty}^{\infty}\! \frac{(\rho_{\alpha}'\nspp(t)\nspp)^2}{\gamma^4}
\psp {g}_{\mathcal{FH}}^{}(t;\alpha,\gamma^2)
\, \mathrm{d}t
\psp.
\end{align*}
We know that  
$\rho_{\alpha}'\nspp(t)=\alpha \sgn(t)$ if $|t| \leq \alpha\pspp$, else $\rho_{\alpha}'\nspp(t)=t\pspp$. 
Therefore,
\begin{align*}
\mathcal{I}_{\mathcal{FH}}^{}
&=\frac{1}{{\kappa}}\nsp\!\left[2 \!\int_{0}^{\alpha}\! \frac{\alpha^2}{\gamma^4}e^{-\alpha t/\gamma^2_{}} \mathrm{d}t + 2e^{-\alpha^2/2\gamma^2_{}}\!
\!\int_{\alpha}^{\infty}\! \frac{t^2}{\gamma^4}e^{-t^2/2\gamma^2_{}} \mathrm{d}t\right]\!
\\&
=\tfrac{1}{\gamma^2_{}}
\nsp\!\left[1+\tfrac{4\gamma}{\alpha\pspp{\omega}}\!\left(\nspp\tfrac{\alpha^2}{2\gamma^2_{}}e^{\psp\alpha^2/2\gamma^2_{}}\! -\sinh\!\left(\nspp\tfrac{\alpha^2}{2\gamma^2_{}}\nspp\right)\!\nspp\right)\!
\right]\!
\psp.
\end{align*}

Using 
\eqref{eq:bnd_sinh}, it is evident that $\sigma_{\mathcal{FH}}^2 \leq \gamma^{2}_{}$ and $\mathcal{I}_{\mathcal{FH}}^{} \geq \frac{1}{\gamma^{2}_{}}\nspp\pspp$.
Therefore,  the flipped Huber distribution 
has lesser variance and more Fisher Information compared to the Gaussian with the same scale parameter. 
\section{Centered privacy loss function}
\label{appx:zeta}	

We now provide the centered privacy loss function of the flipped Huber mechanism in closed form. We have
%
$
\widetilde{{\zeta}}_{\pspp d}^{}\nsp(t)
=\tfrac{1}{\gamma^2_{}}\!\Big[\rho_{\alpha}^{}\!\nsp\left(\nspp t\nspp+\nspp\tfrac{d}{2}\nspp\right)\! - \rho_{\alpha}^{}\!\nsp\left(\nspp t\nspp-\nspp\tfrac{d}{2}\nspp\right)\!\!\Big]\!
\pspp$. 
Since the flipped Huber loss function $\rho_{\alpha}^{}\nspp(\cdot)$ is piecewise, obtaining $\widetilde{{\zeta}}_{\pspp d}^{}\nsp(t)$ in closed form is algebraically involved. We need to consider three different cases based on how $\alpha$ and $d$ are related, viz., $\alpha \geq |d|\pspp$, $\frac{|d|}{2} \leq \alpha < |d|$ and $\alpha < \frac{|d|}{2}\pspp$. 
By aggregating 
the centered privacy loss functions corresponding to these three cases, we  get the piecewise closed-form expression for $\widetilde{{\zeta}}_{\pspp d}^{}\nsp(t)$ as in \eqref{eq:zeta_d_flip_hub}. 

\begin{figure}[h]
\begin{subfigure}{0.97\linewidth}
\centering
\includegraphics[width=\linewidth]{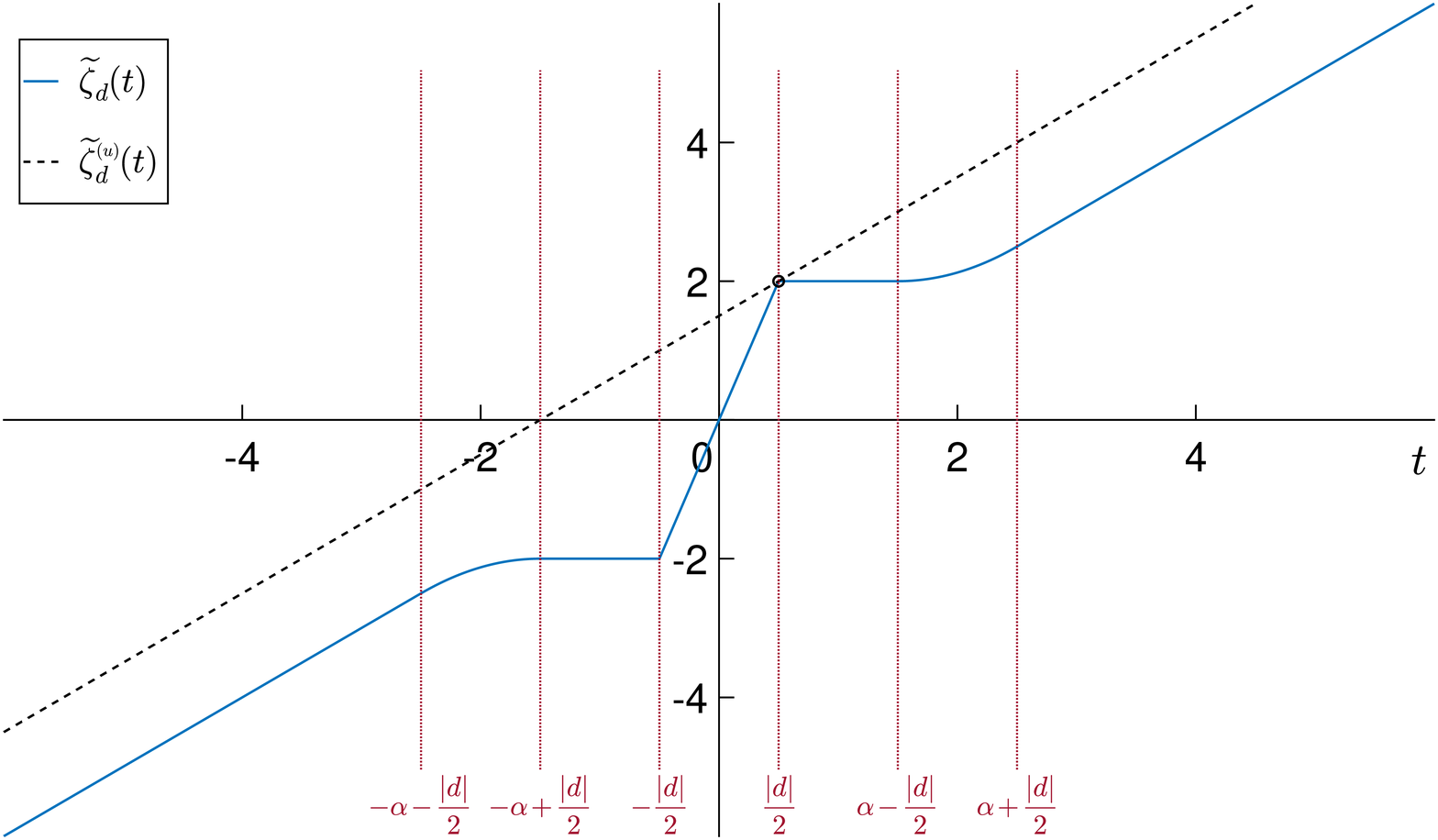}
\caption{case I: $\alpha \geq |d|\pspp$.}
\vspace{1em}
\label{fig:flip_hub_zeta1}
\end{subfigure}
\begin{subfigure}{0.97\linewidth}
\centering
\includegraphics[width=\linewidth]{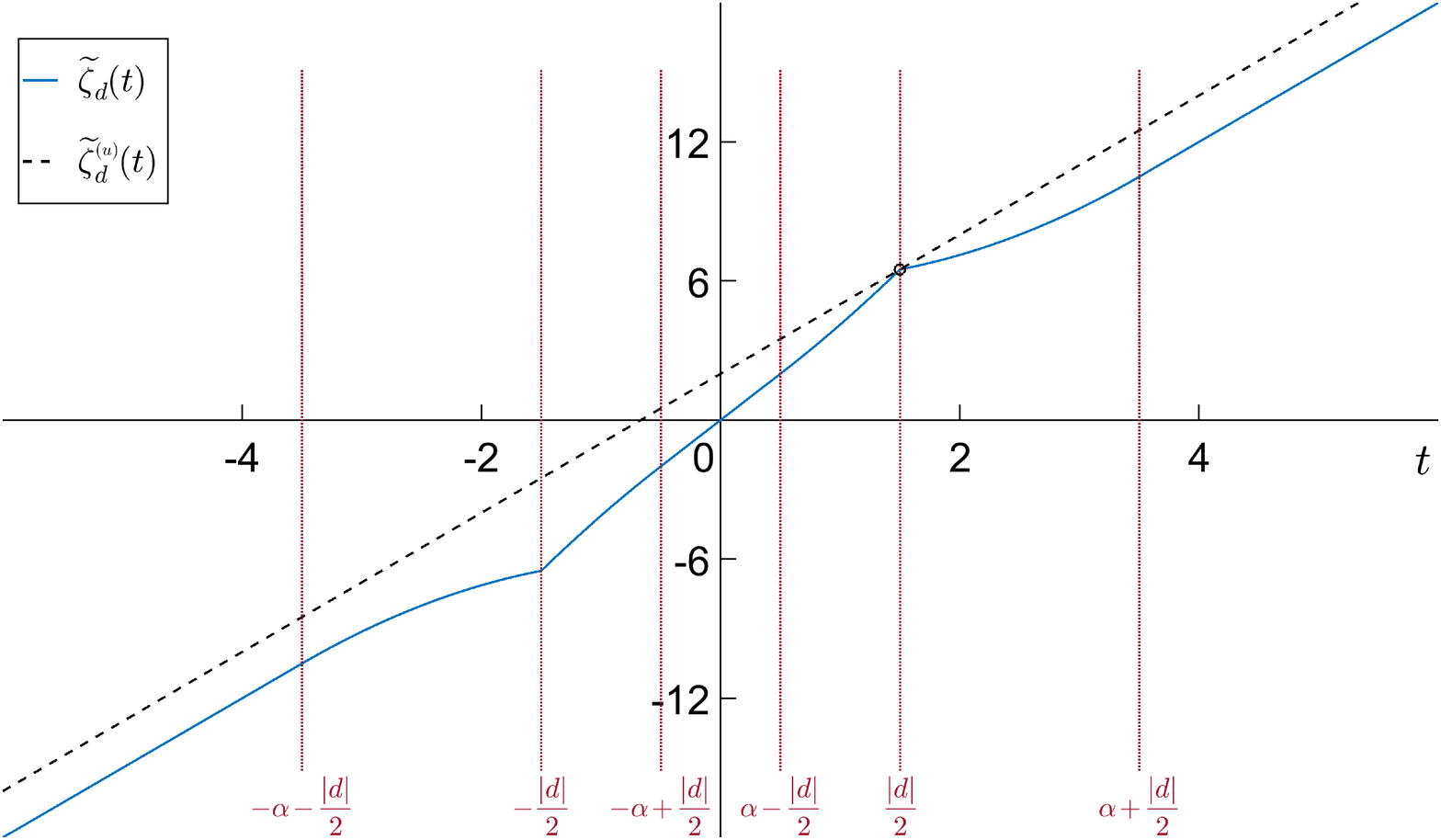}
\caption{case II: $\frac{|d|}{2} \leq \alpha < |d|\pspp$.}
\vspace{1em}
\label{fig:flip_hub_hub_zeta2}
\end{subfigure}
\begin{subfigure}{0.97\linewidth}
\centering
\includegraphics[width=\linewidth]{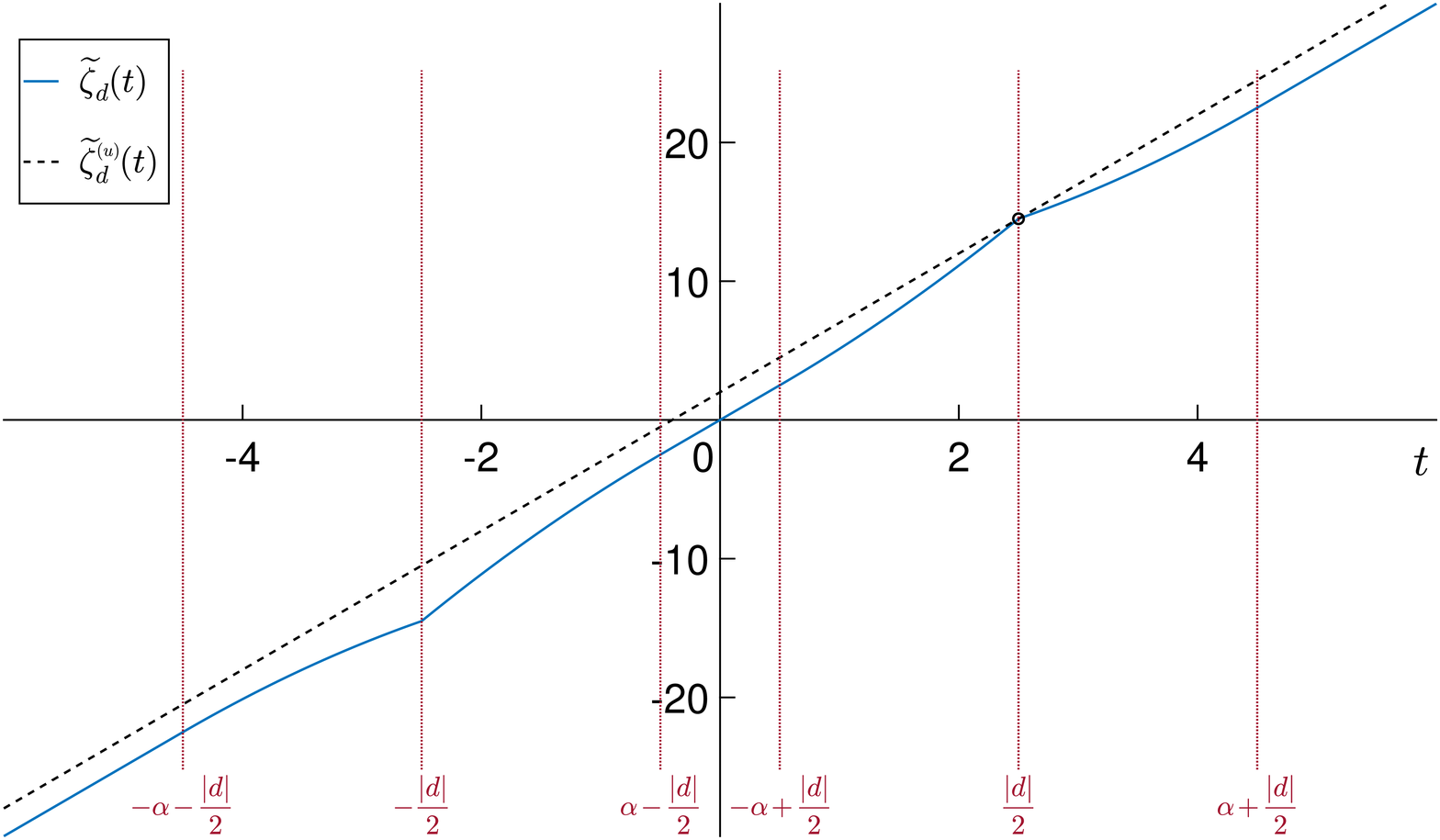}
\caption{case III: $\alpha<\frac{|d|}{2}\pspp$.}
\vspace{1em}
\label{fig:flip_hub_hub_zeta3}
\end{subfigure}
\caption{Centered privacy loss functions of 
$\mathcal{FH}(\alpha,\gamma^2_{})$ with $\alpha=2$ and $\gamma=1$ and their upper bounds for 
(a) $d=1\pspp$, (b) $d=3$ and (c) $d=5\pspp$.}
\label{fig:flip_hub_hub_zeta}
\end{figure}
%

Fig.  \ref{fig:flip_hub_hub_zeta} illustrates the centered privacy loss 
for the three cases. 
{Note that $\widetilde{{\zeta}}_{\pspp d}^{}\nsp(t)$ is the monotonic function in the direction of $\sgn(d)$ and exhibits odd symmetry over $t$ as well as $d\pspp$, i.e., $\widetilde{{\zeta}}_{\pspp -d}^{}\nsp(t)=-\widetilde{{\zeta}}_{\pspp d}^{}\nsp(t)=\widetilde{{\zeta}}_{\pspp d}^{}\nsp(-t)\pspp$; this is because the flipped Huber density is symmetric and log-concave. }
%
%
Also, in all the 
cases, $\widetilde{{\zeta}}_{\pspp d}^{}\nsp(t)$ is 
a \textit{corrugated} linear function that is crumpled at the middle; 
the tail of $\widetilde{{\zeta}}_{\pspp d}^{}\nsp(t)$ is 
linear with slope $\frac{d}{\gamma^2_{}}\nspp\pspp$, 
i.e., 
$\widetilde{{\zeta}}_{\pspp d}^{}\nsp(t)=\frac{td}{\gamma^2_{}}
\pspp$, 
whenever $|t|\geq \alpha+\frac{|d|}{2}\nspp\pspp$. 
These tails coincide with the centered privacy loss function of the Gaussian $\mathcal{N}(0,\gamma^2_{})\pspp$, 
given by $\frac{td}{\gamma^2_{}} \pspp \,\ \forall\, t\in \mathbb{R}\pspp$. 
This is a direct consequence of the flipped Huber density having the tail of 
$\mathcal{N}(0,\gamma^2_{})\pspp$.
%
\\

\subsubsection{Upper bound of $\widetilde{{\zeta}}_{\pspp d}^{}\nsp(t)\pspp$}

We now deduce a function that upper bounds 
$\widetilde{{\zeta}}_{\pspp d}^{}\nsp(t)$ 
that 
is useful in the derivation of sufficient condition for $(\epsilon,\delta)$-DP in higher dimensions and the characterization of zCDP. 
As noted 
before, $\widetilde{{\zeta}}_{\pspp d}^{}\nsp(t)$ is a \textit{zigzag} 
function which is linear at the tails, 
and it has hinges at $\pm \tfrac{d}{2}\nspp\pspp$: a ridge at $\tfrac{d}{2}$  and a groove at $-\tfrac{d}{2}\nspp\pspp$. 

The intrinsic 
choice for the bounding function is the affine function passing through the ridge that has the same slope of $\frac{d}{\gamma^2_{}}\nspp\pspp$ as that of tail segments of $\widetilde{{\zeta}}_{\pspp d}^{}\nsp(t)$. 
The ridge points for all three cases can be collectively expressed as $
\!\left(\nsp \frac{d}{2},\frac{\alpha^2_{}+d^2_{}-(\pspp[\alpha-|d|]_{\nspp +}^{})^{ 2}_{}}{2\gamma^2_{}}\nsp\right)\!\pspp\pspp$, 
and hence, $\tfrac{td}{\gamma^2_{}}+\tfrac{\alpha^2_{}-(\pspp[\alpha-|d|]_{\nspp +}^{})^{ 2}_{}}{2\gamma^2_{}}\nspp\pspp$ is the aforementioned straight line through the ridge. 
This function is shown as the dashed line in Fig.  \ref{fig:flip_hub_hub_zeta} for each case, alongside $\widetilde{{\zeta}}_{\pspp d}^{}\nsp(t)\pspp$, where the ridge points are marked as $\circ\pspp$. 
%
Since the above function is monotonic in $|d|\pspp$, we arrive at the bound
$
\widetilde{{\zeta}}_{\pspp d}^{}\nsp(t)
\leq \tfrac{td}{\gamma^2_{}}+\tfrac{\alpha^2_{}-(\pspp[\alpha-\Delta]_{\nspp +}^{})^{ 2}_{}}{2\gamma^2_{}}
\triangleq\widetilde{{\zeta}}_{\pspp {d}}^{_{(\nspp u\nspp)}}\nsp
\nsp(t)
\pspp$.
%

Consequently, for the $K$-dimensional privacy loss function, the upper bound can be obtained as ${\zeta}_{\pspp \mathbf{d}}^{}\nsp(\mathbf{t})= \sum_{i=1}^K \widetilde{{\zeta}}_{\pspp d_i}^{}\nsp\!\left(\nsp t_i^{} \nsp +\nsp \frac{d_i}{2}\nsp\right)\!
\leq \plossup\nsp(\mathbf{t})\pspp$, where
$
\plossup\nsp(\mathbf{t}) =
\tfrac{\mathbf{t}^{\top}\mathbf{d}}{\gamma^2_{}}+ \tfrac{\norm{\mathbf{d}}^2_{2}}{2\gamma^2_{}}+\tfrac{K(\alpha^2_{}-(\pspp[\alpha-\Delta]_{\nspp +}^{})^{ 2}_{})}{2\gamma^2_{}}
\nspp\pspp$. 
The affine nature of this bound plays a significant role in simplifying the analysis; it facilitates the usage  sub-Gaussianity and stochastic ordering in higher dimensions and enables us to obtain a simpler sufficient condition. 

\section{Additional details 
on private coordinate descent 
} 
\label{appx:appn}	
Let us consider the empirical risk minimization problem, 
\begin{equation*}\label{eq:erm}
\underset{\boldsymbol{{\theta}}\in\mathbb{R}^K_{}}\min\ 
\frac{1}{n}\nspp\summ_{n=1}^N J(\boldsymbol{{\theta}}; \mathcal{D}_n^{})+ {\psi}(\boldsymbol{{\theta}})
\psp,
\end{equation*}
where $\boldsymbol{{\theta}}\in\mathbb{R}^K_{}$ is the model parameter to be optimized, 
$\mathcal{D}=(\mathcal{D}_1^{},\,\mathcal{D}_2^{},\,\ldots,\,\mathcal{D}_N^{})\in\mathcal{X}$ 
is the dataset of $N$ samples, and $\mathcal{D}_n^{}=(\mathbf{x}_n^{},y_n^{})$ is the tuple of $n$-th user's attribute and label. Also, $J:\mathbb{R}^K_{} \times \mathcal{X} \to \mathbb{R}$ is a convex and smooth loss function, and ${\psi}:\mathbb{R}^K_{} \to \mathbb{R}$ is a convex and separable regularizing function, ${\psi}(\boldsymbol{{\theta}})=\sum_{i=1}^{K}{\psi}_i^{}(\theta_i^{})$. We assume the coordinate-wise smoothness constants of the objective $\{{M}_i^{}\}_{i=1}^{K}$ to be given (For generalized linear models, they can be obtained from the data - see \cite[Section 5.2]{mangold2022differentially}).
The proximal operators corresponding to the regularizers are
\begin{equation*}
\operatorname{prox}_{{\tau}_i^{}{\psi}_i^{}}^{} \nsp (\boldsymbol{{\theta}})  =
\underset{\mathbf{w}\in\mathbb{R}^K_{}}{\operatorname{argmin}} \ \tfrac{1}{2}\norm{\mathbf{w} - \boldsymbol{{\theta}}}_2^2 + {\tau}_i^{}{\psi}_i^{}(w_i^{})
\psp, 
\end{equation*}
where ${\tau}_i^{}=\tfrac{{\tau}}{{M}_i^{}}$ is the learning rate for the $i$-th coordinate. 
We consider the least squares and logistic regression losses and $\ell_1^{}$  and $\ell_2^{}$ regularizations.

The procedure is summarized in Algorithm \ref{alg:dpcd}. We perform $L$ batches of coordinate descents.
Before updating the coordinate, the gradient corresponding to that coordinate is perturbed with noise to guarantee DP. In order to calibrate the noise, gradients should be bounded; the $i$-th coordinate gradients corresponding to each user are clipped to have a maximum absolute value of $C_i^{}$ and averaged. Hence, the sensitivity of $i$-th coordinate update is ${\lambda}_i^{}=\frac{2C_i^{}}{N}$. The clipping constants are adaptively chosen as $C_i^{}=C\sqrt{\frac{M_i^{}}{\sum_{j=1}^{K} {M}_j^{}}}$. The hyperparameters 
$L, {\tau}$ and ${C}$ are 
tuned as described in \cite{mangold2022differentially}.
\begin{algorithm}[h]
\caption{Differentially Private Coordinate Descent (DP-CD).}
\label{alg:dpcd}
\textbf{Input}: Dataset $\mathcal{D}$, privacy parameters $\epsilon\in\mathbb{R}_{++}^{}$ and $\delta\in(0,1)$, iteration budget $L\in\mathbb{N}$, initial point $\boldsymbol{{\theta}}^{(0)}_{}\in \mathbb{R}^K_{}$, Clipping constants $\{{C}_i^{}\}_{i=1}^K$, and step sizes $\{{\tau}_i^{}\}_{i=1}^K$
\begin{algorithmic}[1]
\State Determine noise parameters $\{\sigma_i^{}\}_{i=1}^K$ for Gaussian and $\{\alpha_i^{},\gamma_i^{}\}_{i=1}^K$ for flipped Huber from the privacy parameters.
\For{$l=1,\,2,\,\ldots,\,L$}
\State $\boldsymbol{{\theta}}^{(l)}_{} \gets \boldsymbol{{\theta}}^{(l-1)}_{}\pspp$.
\For{$i=1,\,2,\,\ldots,\,K$}
\State Sample $t_i^{(l)}\sim \mathcal{N}(0,\sigma_j^2)$ or $t\sim\mathcal{FH}(\alpha_i^{},\gamma_i^2)$
\State ${\nu}^{(l)}_i=\frac{1}{n}\nspp\sum_{n=1}^N \operatorname{clip}_{}^{}\left(\nabla_i^{}\pspp J(\boldsymbol{{\theta}}_{}^{(l)}; \mathcal{D}_n^{});\pspp 
C_i^{}\right)+t^{(l)}_i$ 
\State ${\theta}^{(l)}_i \gets \operatorname{prox}_{{\tau}_i^{}{\psi}_i^{}}^{} \!\nspp \left( {\theta}^{(l)}_i- {\tau}_i^{} {\nu}^{(l)}_i \right)$
\EndFor
\EndFor
\end{algorithmic}
\textbf{Output}: $\widehat{\boldsymbol{{\theta}}} = \boldsymbol{{\theta}}^{(L)}_{}$
\end{algorithm}

\section{Auxiliary results and analyses}\label{appx:addn_res} 
This section provides some results in addition to those given in Section V. 

\subsection{Privacy profiles in one dimension}
We present the privacy profiles of flipped Huber, OSGT and Gaussian mechanisms in Fig.  \ref{fig:priv_prof}, where we plot the smallest $\delta$ that the mechanism could afford  at the same variance for various $\epsilon$. The simulation setup is the same as that of Section V-A. We set the noise variance as 16. Thus, the scale parameter for the Gaussian mechanism is 4, and we set the scale parameters of OSGT and flipped Huber as 7; the other parameter ($\alpha$ and ${\vartheta}$ for flipped Huber and OSGT, respectively) is set to retain the variance as 16.

It is evident that the flipped Huber mechanism under consideration could afford smaller $\delta$ for a given $\epsilon$ than OSGT and Gaussian, or equivalently, afford smaller $\epsilon$ for a given $\delta\pspp$. However, we remark that this performance is for the given set of parameters, and in practice, the parameters are chosen to meet the given specification of $(\epsilon,\delta)$. 

\subsection{Flipped Huber vs OSGT}\label{appx:addn_res_osgt}

\begin{figure*}[h!]
	\centering
	\begin{minipage}{0.45\linewidth}
		\centering
		\includegraphics[width=0.88\linewidth]{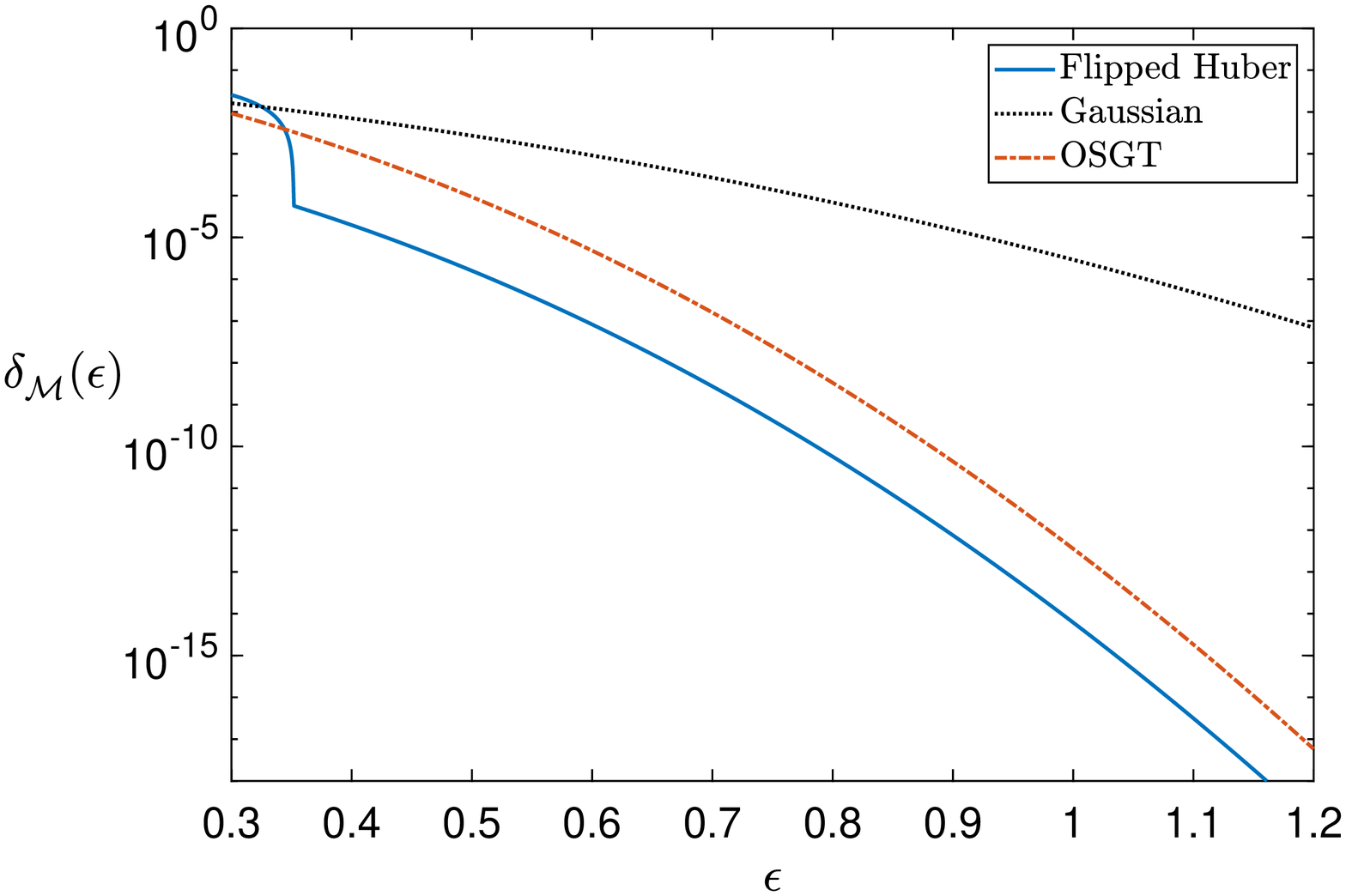}
		\caption{Privacy profiles of various noise mechanisms in one dimension for the noise variance of 16 and $\Delta=1\pspp$.}
		\label{fig:priv_prof}
	\end{minipage}%
	\hspace{2em}
	\begin{minipage}{0.45\linewidth}
		\centering
		\includegraphics[width=0.88\linewidth]{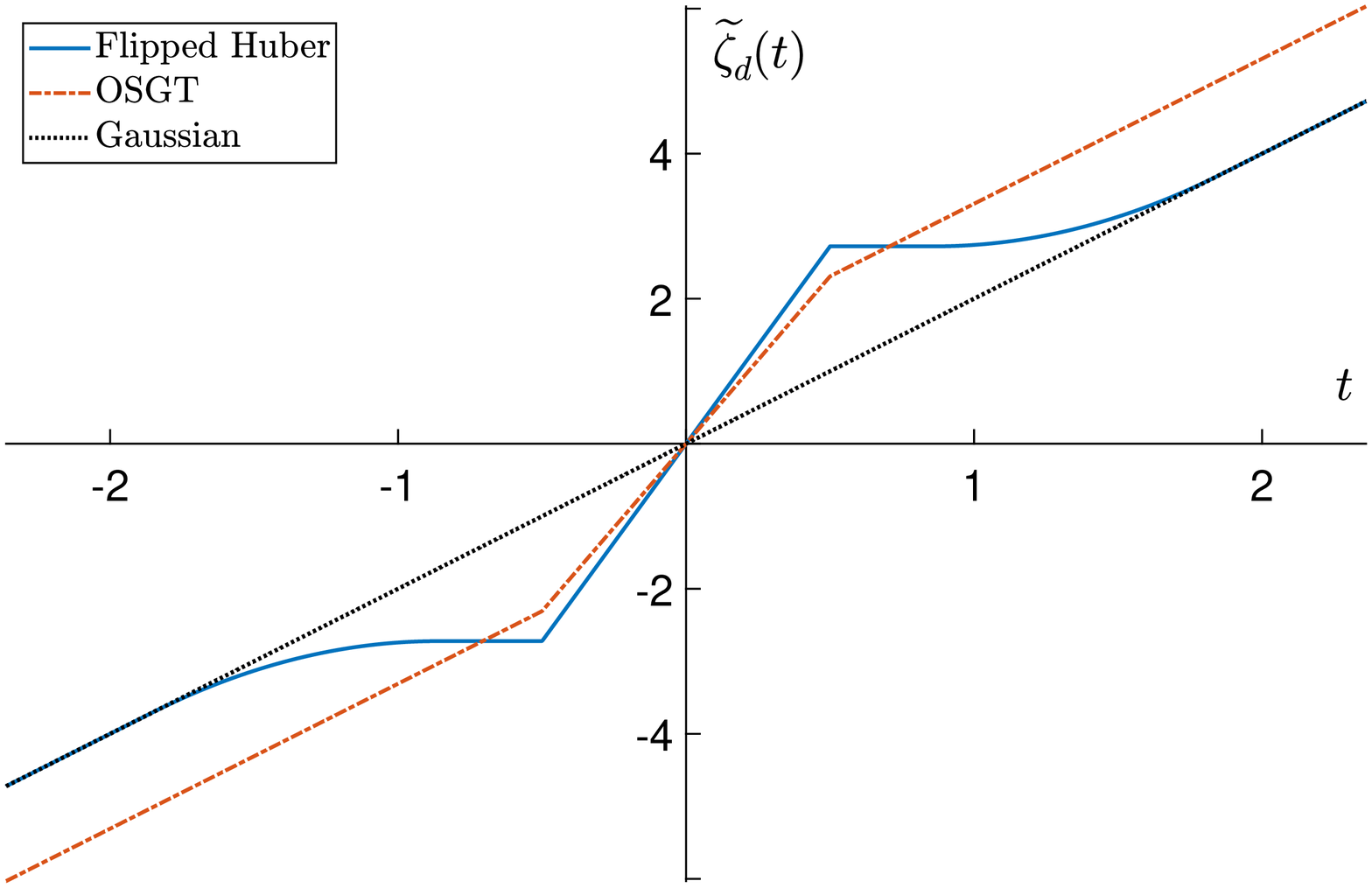}
		\caption{Centred privacy loss functions for various noise mechanisms for the same scale parameter of $\tfrac{1}{\sqrt{2}}\nspp\pspp$ and $d=1\pspp$.}
		\label{fig:ploss_compare}
	\end{minipage}
\end{figure*}

Note that the centred loss privacy functions of Gaussian and Laplace mechanisms in one dimension are respectively 
$
\tfrac{td}{\sigma^2_{}}
$ 
and $
\tfrac{2}{\beta}\sgn(t)\sgn(d)
\nspp\!\left(\nspp|t|\wedge\nspp\tfrac{|d|}{2}\nspp\right)\!\nsp\psp
\pspp$, 
whereas, for the  OSGT mechanism, the centred loss privacy function is given as 
$
\tfrac{td}{{\varrho}^2_{}}+\tfrac{2{\vartheta}}{{\varrho}^2_{}}\sgn(t)\sgn(d)
\nspp\!\left(\nspp|t|\wedge\nspp\tfrac{|d|}{2}\nspp\right)\!\nsp\psp
\pspp$, 
Thus, for OSGT, the centered privacy loss function and consequently, the privacy loss of the OSGT mechanism is the sum of those of $\mathcal{N}(0,{\varrho}^2_{})$ and $\mathcal{L}\nspp\big(0,\tfrac{{\varrho}^2_{}}{\vartheta}\big)\nspp\pspp$.
But, for flipped Huber, as noted  in Appendix \ref{appx:zeta}, the centered privacy loss function coincides with that of Gaussian with the same scale parameter in the tails. Thus, the left tail of the privacy loss function of OSGT is with when the scale parameters of both are the same. This phenomenon is illustrated in Fig.  \ref{fig:ploss_compare}.


Consequently, the level set $\{t:{\zeta}_{\pspp d}^{}\nsp(t) \geq \epsilon\}$ are `larger' for OSGT relative to the flipped Huber. Thus, intuitively, the term $\mathbb{P}\{ {\zeta}_{d}^{}\nsp({T}) > \epsilon \}
-e^{\epsilon}_{}\psp \mathbb{P}\{{\zeta}_{\pspp-{d}}^{}\nsp({T}) < -\epsilon\}$ is larger when $T\sim\mathcal{O}({\vartheta},{\varrho}^2_{})$ than when $T\sim\mathcal{FH}(\alpha,{\varrho}^2_{})$, unless when $\epsilon$ is very small and the second term dominates (which could be observed in Fig.  \ref{fig:priv_prof}, where OSGT offers smaller $\delta$ for very small $\epsilon$ compared to flipped Huber with the same scale parameter). This explains why Flipped Huber performs better than OSGT in one dimension.

\section{Necessary and sufficient conditions for $(\epsilon,\delta)$-DP}\label{appx:ns_cond}

This section provides some known conditions that are necessary and sufficient for $(\epsilon,\delta)$-DP; the conditions relevant to this work, along with their derivations, have been provided 
in the notational convention followed in this article.

A generic condition 
based on the tail probabilities of the privacy loss 
is given in the following lemma. The result appeared as Theorem 5 in \cite{balle2018improving}, and here we re-derive it for additive noise mechanism in terms of the noise density function and our redefined privacy loss.

\begin{lemma}\label{lem:balle}
	The additive noise mechanism $\mathcal{M}:\mathcal{X}\rightarrow\mathbb{R}^{K}_{}$ that adds noise from a distribution with density function ${g}_{\pspp\mathbf{T}}^{}\nspp(\cdot)$ is $(\epsilon,\delta)$-DP if and only if 
	\begin{equation}\label{eq:balle0_K}
		\int_{\mathcal{E}_{*}^{}} \!\left({g}_{\pspp\mathbf{T}}^{}\nspp(\mathbf{t})-e^{\epsilon}_{}{g}_{\pspp\mathbf{T}}^{}\nspp(\mathbf{t}\nspp+\nspp \mathbf{d})\nsp\right)\! \psp \mathrm{d}\mathbf{t}  \leq \delta
		\ \,\,\ \forall \ \neighbdsets
		\psp,
	\end{equation}
	where 
	$\mathcal{E}_{*}^{}=\{ \mathbf{t} \in \mathbb{R}^{K}_{} |\, {\zeta}_{\pspp \mathbf{d}}^{}\nsp(\mathbf{t}) \geq \epsilon \}=\{ \mathbf{t} \in \mathbb{R}^{K}_{} |\, {g}_{\pspp\mathbf{T}}^{}\nspp(\mathbf{t}) \geq e^\epsilon_{} {g}_{\pspp\mathbf{T}}^{}\nspp(\mathbf{t}\nspp+\nspp \mathbf{d})\}$ 
	and $\mathbf{d}={f}(\mathcal{D})-{f}(\widecheck{\mathcal{D}})\pspp$. 
	This can be equivalently written as
	\vspace{-0.85em}
	\begin{equation}\label{eq:balle_K}
		\supoverneighb \,
		\mathbb{P}\{{\zeta}_{\pspp \mathbf{d}}^{}\nsp(\mathbf{T})\geq\epsilon\}-e^{\epsilon}_{}\psp \mathbb{P}\{{\zeta}_{\pspp -\mathbf{d}}^{}\nsp_{}(\mathbf{T})\leq-\epsilon\} \leq \delta
		\psp.	
	\end{equation}
\end{lemma}
\begin{proof}
	Let $\mathcal{E}$ be any measurable subset of $\mathbb{R}^{K}_{}\nspp\pspp$. We have  $\mathbb{P}\{\mathcal{M}(\mathcal{D})\in\mathcal{E}\}=\int_{\mathcal{E}}^{} {g}_{\pspp\mathbf{T}}^{}\nspp(\mathbf{t}) \psp \mathrm{d}\mathbf{t}$ and
	$\mathbb{P}\{\mathcal{M}(\widecheck{\mathcal{D}})\in\mathcal{E}\}=\int_{\mathcal{E}}^{} {g}_{\pspp\mathbf{T}_{}}^{}\nspp(\mathbf{t}\nspp+\nspp \mathbf{d}) \psp \mathrm{d}\mathbf{t}\pspp$. Thus, 
	$\mathbb{P}\{\mathcal{M}(\mathcal{D})\in\mathcal{E}\} \leq e^{\epsilon}_{}\psp \mathbb{P}\{\mathcal{M}(\widecheck{\mathcal{D}})\in\mathcal{E}\}+\delta$ if and only if 
	$\int_{\mathcal{E}}^{} \!\big( {g}_{\pspp\mathbf{T}}^{}\nspp(\mathbf{t}) - e^{\epsilon}_{} {g}_{\pspp\mathbf{T}_{}}^{}\nspp(\mathbf{t}\nspp+\nspp \mathbf{d}) \nsp\big) \psp \mathrm{d}\mathbf{t} \leq \delta\pspp$.
	
	Let ${q}(\mathbf{t})={g}_{\pspp\mathbf{T}}^{}\nspp(\mathbf{t}) -e^{\epsilon}_{}{g}_{\pspp\mathbf{T}_{}}^{}\nspp(\mathbf{t}\nspp+\nspp \mathbf{d})\pspp$. 
	Hence, 
	$ \int_{\mathcal{E}}^{}{q}(\mathbf{t})\psp \mathrm{d}\mathbf{t} =\int_{\mathcal{E}\,\cap\,\mathcal{E}_{*}^{}}^{} {q}(\mathbf{t})\psp \mathrm{d}\mathbf{t}
	+\int_{\mathcal{E}\,\cap\,\mathcal{E}_{*}^{\comp}}^{} {q}(\mathbf{t})\psp \mathrm{d}\mathbf{t}
	\leq\int_{\mathcal{E}\,\cap\,\mathcal{E}_{*}^{}}^{} {q}(\mathbf{t})\psp \mathrm{d}\mathbf{t}
	\leq\int_{\mathcal{E}_{*}^{}}^{} {q}(\mathbf{t})\psp \mathrm{d}\mathbf{t}\pspp$, 
	where the first inequality is because ${q}(\mathbf{t})$ is negative over $\mathcal{E}\pspp \cap\pspp \mathcal{E}_{*}^{\comp}\pspp$, and  the second inequality follows from  $\mathcal{E}\,\cap\,\mathcal{E}_{*}^{} \subseteq \mathcal{E}_{*}^{}\nspp\pspp$. The above inequalities are tight for $\mathcal{E}=\mathcal{E}_{*}^{}\nspp\pspp$. Since the definition of DP is for any measurable $\mathcal{E}$ (including $\mathcal{E}^{*}_{}$), $\int_{\mathcal{E}_{*}^{}}^{} {q}(\mathbf{t})\psp \mathrm{d}\mathbf{t}\leq \delta$ is a necessary and sufficient condition for $(\epsilon,\delta)$-DP.  Thus, \eqref{eq:balle0_K} has been derived.
	
	Now, $\int_{\mathcal{E}_{*}^{}}^{} {g}_{\pspp\mathbf{T}}^{}\nspp(\mathbf{t}) \psp \mathrm{d}\mathbf{t}
	=\int_{\mathbb{R}^{K}_{}} \mathbb{I}\pspp[\pspp{\zeta}_{\pspp \mathbf{d}}^{}\nsp(\mathbf{T}) \geq \epsilon\pspp]\psp  {g}_{\pspp\mathbf{T}}^{}\nspp(\mathbf{t})\psp \mathrm{d}\mathbf{t} 
	=\mathbb{P}\{{\zeta}_{\pspp \mathbf{d}}^{}\nsp(\mathbf{T}) \geq \epsilon\}\pspp$, where $\mathbb{I}\pspp[\pspp\cdot\pspp]$ is the indicator function. 
	Similarly,
	$\int_{\mathcal{E}_{*}^{}}^{} {g}_{\pspp\mathbf{T}_{}}^{}\nspp(\mathbf{t}\nspp+\nspp \mathbf{d})\psp \mathrm{d}\mathbf{t} =\int_{\mathbb{R}^{K}_{}} \mathbb{I}\pspp[\pspp{g}_{\pspp\mathbf{T}}^{}\nspp(\mathbf{t}) \geq e^\epsilon_{} {g}_{\pspp\mathbf{T}}^{}\nspp(\mathbf{t}\nspp+\nspp \mathbf{d})]\psp {g}_{\pspp\mathbf{T}_{}}^{}\nspp(\mathbf{t}\nspp+\nspp \mathbf{d}) \psp \mathrm{d}\mathbf{t}
	=\int_{\mathbb{R}^{K}_{}} \mathbb{I}\pspp[\pspp{g}_{\pspp\mathbf{T}}^{}\nspp(\mathbf{r}) \leq e^{-\epsilon}_{} {g}_{\pspp\mathbf{T}}^{}\nspp(\mathbf{r}-\mathbf{d})]\psp {g}_{\pspp\mathbf{T}_{}}^{}\nspp(\mathbf{r}) \psp \mathrm{d}\mathbf{r}
	= \mathbb{P}\{{\zeta}_{\pspp -\mathbf{d}}^{}\nsp(\mathbf{T}) \leq -\epsilon\}\pspp$. 
	Hence, from \eqref{eq:balle0_K}, $\mathbb{P}\{{\zeta}_{\pspp \mathbf{d}}^{}\nsp(\mathbf{T}) \geq \epsilon\}-e^{\epsilon}_{}\psp\mathbb{P}\{{\zeta}_{\pspp -\mathbf{d}}^{}\nsp(\mathbf{T}) \leq -\epsilon\} < \delta$ is the necessary and sufficient condition for $(\epsilon,\delta)$-DP.
\end{proof}

\begin{remark}\label{rem:bound_p_zeta}
	The {pDP} sufficient condition, $\mathbb{P}\{{\zeta}_{\pspp \mathbf{d}}^{}\nsp(\mathbf{T})\geq\epsilon\} \leq \delta\pspp$, can be obtained from \eqref{eq:balle_K} by trivially lower bounding the 
	term $\mathbb{P}\{{\zeta}_{\pspp -\mathbf{d}}^{}\nsp(\mathbf{T})\leq-\epsilon\}$ by zero. However, this would typically result in  noise variance much larger than what is truly needed for ensuring privacy. 
	%
	Tighter sufficient conditions can be obtained by using tight lower bounds for $\mathbb{P}\{{\zeta}_{\pspp -\mathbf{d}}^{}\nsp(\mathbf{T})\leq-\epsilon\}\pspp$. 
\end{remark}
Though condition \eqref{eq:balle_K} is powerful, it might be difficult to characterize the event $\mathcal{E}_{*}^{}$ or the tail probabilities of ${\zeta}_{\pspp \mathbf{d}}^{}\nsp(\mathbf{T})$ and ${\zeta}_{\pspp -\mathbf{d}}^{}\nsp(\mathbf{T})$ in high dimensions, thereby limiting its applicability. 
%
\\

\subsubsection{Condition for symmetric, log-concave noise densities:} 

Log-concave densities have been extensively studied in the past, 
and most of the noise densities proposed for differential privacy fall under this category \cite{vinterbo2022differential}. A simple necessary and sufficient condition for $(\epsilon,\delta)$-DP, based 
on the survival function of the noise distribution 
with symmetric log-concave density, is provided below.


\begin{lemma}[Lemma 1 in \cite{vinterbo2022differential}]\label{lem:log_concave}
	A one-dimensional mechanism $\mathcal{M}:\mathcal{X}\rightarrow\mathbb{R}$ that adds noise 
	from a distribution with symmetric log-concave density function is $(\epsilon,\delta)$-DP 
	if and only if
	\begin{equation}\label{eq:log_concave}
		\overline{{G}}_{\pspp T}^{}\nsp\!\left(\nspp \widetilde{{\zeta}}^{-1}_{\Delta}\nsp(\epsilon) - \tfrac{\Delta}{2}\nsp\right)\!
		- e^{\epsilon}_{}\psp \overline{{G}}_{\pspp T}^{}\nsp\!\left(\nspp \widetilde{{\zeta}}^{-1}_{\Delta}\nsp(\epsilon) + \tfrac{\Delta}{2}\nsp\right)\! \leq \delta
		\psp.
	\end{equation}
\end{lemma}
\begin{proof}
	For the single-dimensional case, condition \eqref{eq:balle0_K} becomes
	$\int_{\mathcal{E}_{*}^{}} \!\left({g}_{\pspp T}^{}\nspp(t)-e^{\epsilon}_{}{g}_{\pspp T}^{}\nspp(t\nsp+\nsp d)\nsp\right)\! \psp \mathrm{d}t \leq \delta\pspp$, where $\mathcal{E}_{*}^{}=\{ t \in \mathbb{R} \,|\, {\zeta}_{\pspp d}^{}\nsp(t) \geq \epsilon \}\pspp$.
	Without loss of generality, let us assume $d>0$ (otherwise, the datasets' labels can be interchanged). Using Proposition 2.3 (b) in \cite{saumard2014log}, we conclude that ${\zeta}_{\pspp d}^{}\nsp(t)$ is a monotonic increasing function in $t\pspp$. Let 
	${\nu} 
	=\widetilde{{\zeta}}^{-1}_{d}\nsp(\epsilon)=\sup\nspp\big\{t\,\big|\,\widetilde{{\zeta}}^{}_{d}\nsp(t)\leq \epsilon\big\}\pspp$. 
	Clearly, $\widetilde{{\zeta}}^{}_{d}\nsp({\nu})=\epsilon$ and hence, 
	${g}_{\pspp T}^{}\nspp\!\left(\nspp {\nu} \nsp-\nsp \tfrac{d}{2}\nspp\right)\nsp-e^{\epsilon}_{}{g}_{\pspp T}^{}\nspp\!\left(\nspp {\nu} \nsp+\nsp \tfrac{d}{2}\nspp\right)\nsp =0 \pspp$.	
	Also, $\int_{\mathcal{E}_{*}^{}} {g}_{\pspp T}^{}\nspp(t) \psp \mathrm{d}t = \overline{{G}}_{\pspp T}^{}\nspp\!\left(\nspp {\nu} \nsp-\nsp \tfrac{d}{2}\nspp\right)\nsp$ and 
	$\int_{\mathcal{E}_{*}^{}} {g}_{\pspp T}^{}\nspp(t\nsp+\nsp d) \psp \mathrm{d}t  = \overline{{G}}_{\pspp T}^{}\nspp\!\left(\nspp {\nu} \nsp+\nsp \tfrac{d}{2}\nspp\right)\nsp \pspp$. 
	The necessary and sufficient condition now becomes 
	${{\mathcal{B}}}(\epsilon;d)\leq \delta\pspp$, where 
	${{\mathcal{B}}}(\epsilon;d)= \overline{{G}}_{\pspp T}^{}\nspp\!\left(\nspp {\nu} \nsp-\nsp \tfrac{d}{2}\nspp\right)\nsp
	-e^{\epsilon}_{} \overline{{G}}_{\pspp T}^{}\nspp\!\left(\nspp {\nu} \nsp+\nsp \tfrac{d}{2}\nspp\right)\nsp\pspp$. 
	Now, 
	$\frac{\partial {{\mathcal{B}}}}{\partial d}
	= -\big[{g}_{\pspp T}^{}\nspp\!\left(\nspp {\nu} \nsp-\nsp \tfrac{d}{2}\nspp\right)\nsp-e^{\epsilon}_{}{g}_{\pspp T}^{}\nspp\!\left(\nspp {\nu} \nsp+\nsp \tfrac{d}{2}\nspp\right)\nsp\nsp\big]\nsp \frac{\partial {\nu}}{\partial d}
	+\frac{1}{2}\!\big[{g}_{\pspp T}^{}\nspp\!\left(\nspp {\nu} \nsp-\nsp \tfrac{d}{2}\nspp\right)\nsp+e^{\epsilon}_{}{g}_{\pspp T}^{}\nspp\!\left(\nspp {\nu} \nsp+\nsp \tfrac{d}{2}\nspp\right)\nsp\nsp\big]\!$
	$=\frac{1}{2}\!\big[{g}_{\pspp T}^{}\nspp\!\left(\nspp {\nu} \nsp-\nsp \tfrac{d}{2}\nspp\right)\nsp +e^{\epsilon}_{}{g}_{\pspp T}^{}\nspp\!\left(\nspp {\nu} \nsp+\nsp \tfrac{d}{2}\nspp\right)\nsp\nsp\big]\!
	\geq 0\pspp$. Thus, ${{\mathcal{B}}}(\epsilon;d)$ is a monotonic increasing function in $d\pspp$; replacing $d$ by $\Delta\pspp$, we get the necessary and sufficient condition \eqref{eq:log_concave}. 
\end{proof}

\subsection{Conditions for some existing mechanisms}\label{appx:ns_cond_eg}
		In the following, necessary and sufficient conditions for $(\epsilon,\delta)$-DP for various additive noise mechanisms are listed; whenever trivial, the conditions for $K$-dimensions are also provided. 
\begin{enumerate}[label*=(\alph*)]
	\item \textbf{Gaussian mechanism: } When the noise from $\mathcal{N}(0,\sigma^2)$ is added, the mechanism ensures $(\epsilon,\delta)$-DP if and only if
	\begin{equation}\label{eq:gau_K}
		{Q}\!\left(\tfrac{\sigma\epsilon}{\Delta_2^{}}-\tfrac{\Delta_2^{}}{2\sigma}\right)\! -
		e^{\epsilon}_{} {Q}\!\left(\tfrac{\sigma\epsilon}{\Delta_2^{}}+\tfrac{\Delta_2^{}}{2\sigma}\right)\!
		\leq \delta
		\psp.
	\end{equation}
	It can be noted that $\sigma$ satisfying \eqref{eq:gau_K} cannot be expressed in closed form. The smallest $\sigma$ satisfying the condition for any given $(\epsilon,\delta)$ can be obtained by binary search 
	\cite{balle2018improving}. 
	%
	
	\item \textbf{Laplace mechanism:} 
	This mechanism adds noise sampled from $\mathcal{L}(0,\beta)\pspp$ and is capable of warranting pure DP, i.e., DP with $\delta=0\pspp$, when $\beta\geq\tfrac{\Delta_1^{}}{\epsilon}\nspp\pspp$.
	The privacy profile of this mechanism in one dimension is given as $ \!\left[\nspp1-\exp\!\left(\nspp\tfrac{1}{2}(\epsilon-\Delta_{}^{}/\beta)\nspp\right)\!\right]_{\nsp +}^{}\!\nspp\pspp$, and the necessary and sufficient condition for $(\epsilon,\delta)$-DP \cite{balle2020privacy,vinterbo2022differential} in the single-dimensional case is  
	$
	\beta\geq\tfrac{\Delta_{}^{}}{\epsilon-2\log(1-\delta)}
	\pspp$.
	%
	
	\item \textbf{OSGT mechanism\cite{sadeghi2022offset}:}
	OSGT mechanism adds noise sampled from the density function $g_{\mathcal{O}}^{}(t;{\vartheta},{\varrho}^2_{})=
	\frac{1}{2{Q}\nspp\left(\nspp\frac{{\vartheta}}{{\varrho}}\nspp\right)\nsp}\psp\phi(\abs{t};-{\vartheta},\,{\varrho}^2_{})\nspp\pspp$,
	where $\phi(\psp\cdot\,;{\upsilon},\sigma^2_{})$ is the density function of $\mathcal{N}({\upsilon},\sigma^2_{})\pspp$. This distribution can be perceived as the \textit{modulation} of Laplace density by Gaussian density, and hence, it is also a hybrid of Laplace and Gaussian densities. In the single-dimension case, 
	the OSGT mechanism guarantees $(\epsilon,\delta)$-DP if and only if
	$\delta_{\mathcal{O}}^{}(\epsilon) \leq \delta\pspp$, 
	where
	\begin{equation*}\label{eq:del_el_net}
		\hspace{-2em}
		\delta_{\mathcal{O}}^{}\nsp(\epsilon) \! = \! \begin{cases}
			\\[-3em]
			\nsp 1\nsp -\nsp\frac{1}{2{Q}\nsp\left(\nsp\frac{{\vartheta}}{{\varrho}}\nsp\right)\nsp} \nspp\!\left[{Q}\!\left(\nsp\frac{{\varpi}}{2{\varrho}}\nsp - \nsp\frac{{\varrho}\epsilon}{{\varpi}}\nsp\right)\! \nsp +\nsp
			e^{\epsilon}_{} {Q}\!\left(\nsp\frac{{\varpi}}{2{\varrho}}\! + \!\frac{{\varrho}\epsilon}{{\varpi}}\nsp\right)\!\right]\!\!\!
			\, , & \hspace{-0.8em}
			\epsilon\nsp \leq\nsp \frac{{\varpi}\Delta}{2{\varrho}^2} \\[-0.35em]
			\nsp \frac{1}{2{Q}\nsp\left(\!\frac{{\vartheta}}{{\varrho}}\nsp\right)\nsp}\nspp \!\left[{Q}\!\left(\nsp\frac{{\varrho}\epsilon}{\Delta}\nsp - \nsp\frac{\Delta}{2{\varrho}}\nsp\right)\! \nsp-\nsp
			e^{\epsilon}_{} {Q}\!\left(\nsp\frac{{\varrho}\epsilon}{\Delta}\! + \!\frac{\Delta}{2{\varrho}}\nsp\right)\!\right]\!\!\!
			\, , & \hspace{-0.8em}
			\epsilon\nsp >\nsp \frac{{\varpi}\Delta}{2{\varrho}^2}
		\end{cases}\!\nsp
		\pspp, \hspace{-1.9em}
	\end{equation*}
	and ${\varpi}=\Delta+2{\vartheta}\pspp$. 
	In high dimensions, the necessary and sufficient condition is not available in closed form, and hence we have to perform iterative numerical integration to determine the parameters that could meet the privacy constraints. 
	%
\end{enumerate}

\section*{Acknowledgment}
\addcontentsline{toc}{section}{Acknowledgment}
The authors thank the anonymous reviewers and the associate editor for their constructive feedback and suggestions, which helped to improve the article.
{
\bibliographystyle{IEEEtran}
\bibliography{ref_flip_hub}
}

\begin{IEEEbiography}[{\includegraphics[width=1in,height=1.25in,clip,keepaspectratio]{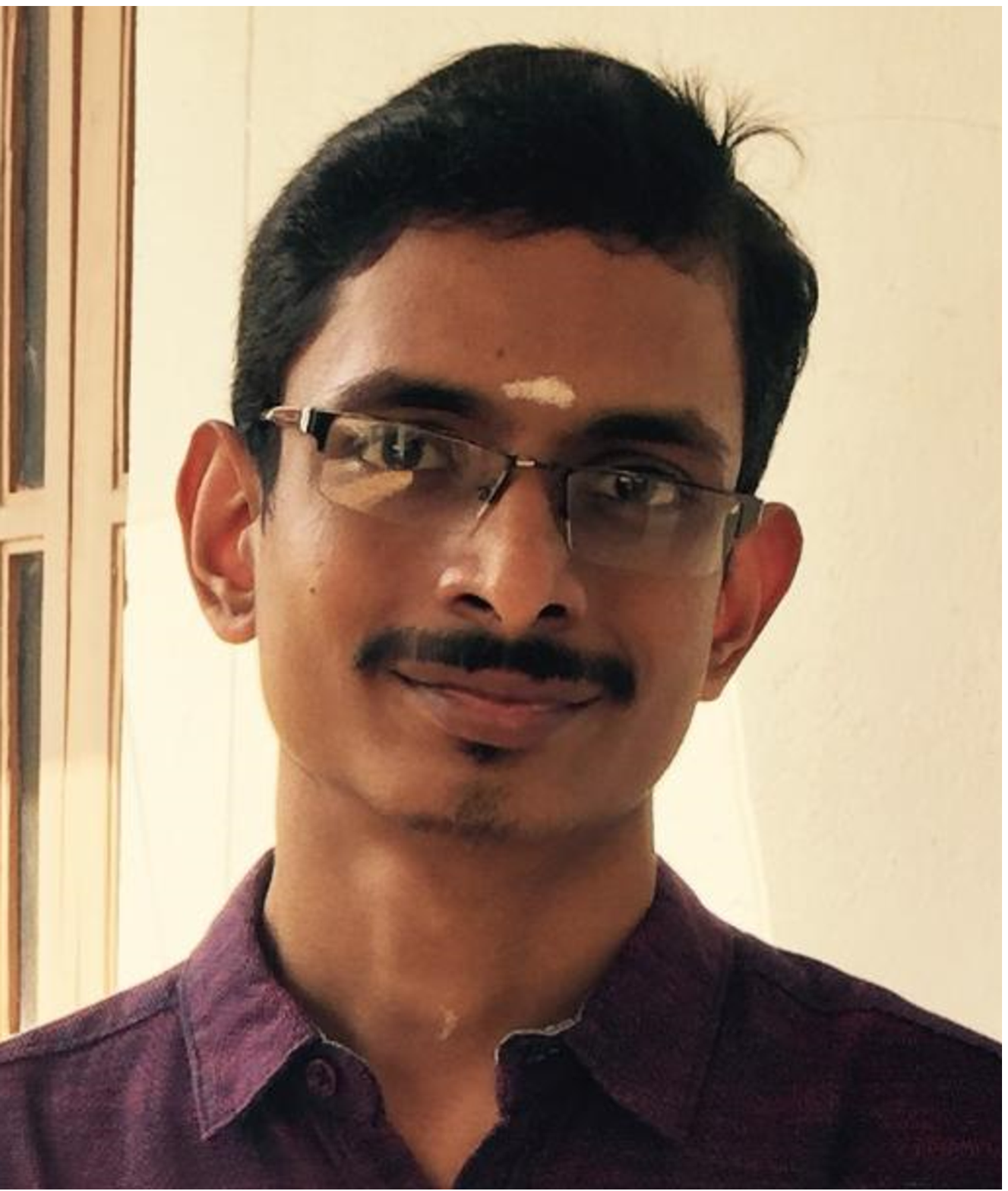}}]
{Gokularam Muthukrishnan} received his B.E. degree in electronics and communication engineering in 2017 from PSG College of Technology, Coimbatore, India. He is currently a Ph.D. Research Scholar at the Department of Electrical Engineering, Indian Institute of Technology Madras, Chennai, India. His current research interests are differential privacy, statistical signal processing, robust estimation, signal processing for distributed radar systems, non-parametric algorithms and wireless communication.
\end{IEEEbiography}


\begin{IEEEbiography}[{\includegraphics[width=1in,height=1.25in,clip,keepaspectratio]{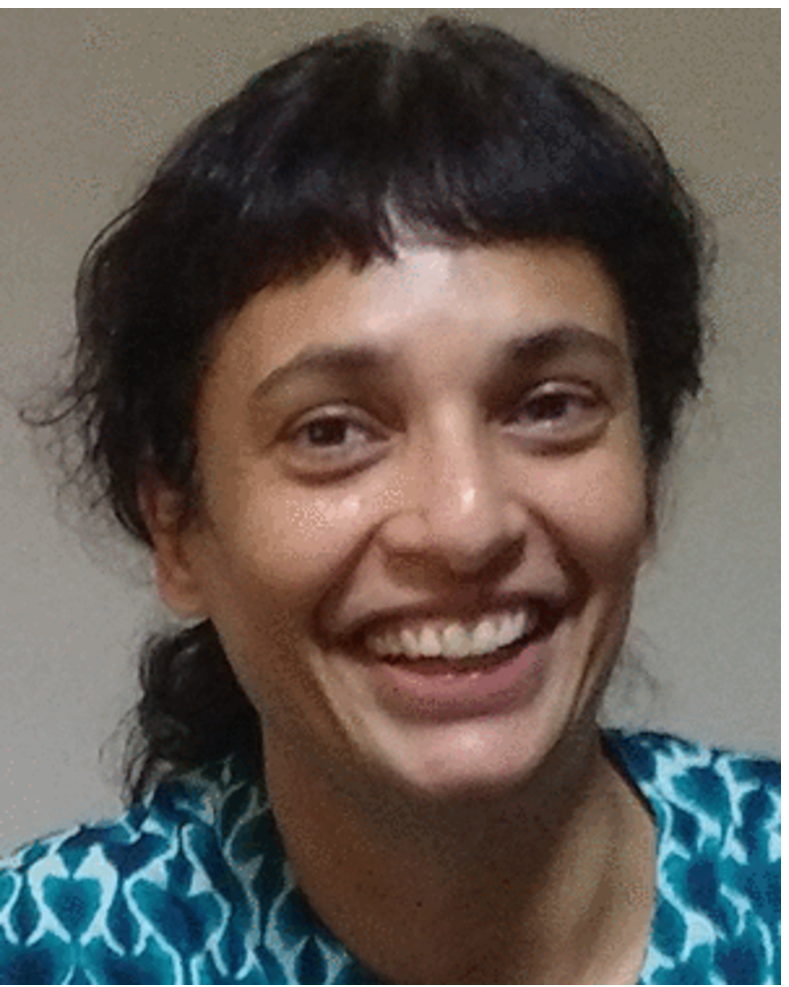}}]
{Sheetal Kalyani} received the B.E. degree in electronics and communication engineering from Sardar Patel University, Anand, India, in 2002, and the Ph.D. degree in electrical engineering from the Indian Institute of Technology Madras (IIT Madras), Chennai, India, in 2008. From 2008 to 2012, she was a Senior Research Engineer with the Centre of Excellence in Wireless Technology, Chennai. She is currently a Professor with the Department of Electrical Engineering, IIT Madras. Her research interests include extreme value theory, generalized fading models, hypergeometric functions, performance analysis of wireless systems/networks, compressed sensing, machine learning, deep learning for wireless applications and differential privacy.
\end{IEEEbiography}

\vspace{96ex}
\end{document}